\documentclass[manuscript]{aastex}

\slugcomment{Accepted for publication in Astrophys J.}

\shorttitle{Neutrino transfer in 3D}
\shortauthors{Sumiyoshi et al.}

\begin{document}

\title{Neutrino Transfer in Three Dimensions \\
for Core-Collapse Supernovae. I. Static Configurations}

\author{K. Sumiyoshi}
\affil{Numazu College of Technology, 
       Ooka 3600, Numazu, Shizuoka 410-8501, Japan \\
       \& \\
       Theory Center, 
       High Energy Accelerator Research Organization (KEK), \\
       Oho 1-1, Tsukuba, Ibaraki 305-0801, Japan}
\email{sumi@numazu-ct.ac.jp}

\and

\author{S. Yamada}
\affil{Science and Engineering
       \& 
       Advanced Research Institute for Science and Engineering, \\
       Waseda University, 
       Okubo, 3-4-1, Shinjuku, Tokyo 169-8555, Japan}
\email{shoichi@heap.phys.waseda.ac.jp}


\begin{abstract}
We develop a numerical code to calculate the neutrino transfer 
with multi-energy and multi-angle 
in three dimensions (3D) for the study of core-collapse supernovae.  
The numerical code solves the Boltzmann equations for neutrino 
distributions by the discrete-ordinate (S$_{n}$) method 
with a fully implicit differencing for time advance.  
The Boltzmann equations 
are formulated in the inertial frame 
with collision terms being evaluated to the zeroth order of $v/c$.  
A basic set of neutrino reactions for three neutrino species 
is implemented together with a realistic equation of state of dense matter.  
The pair process is included approximately 
in order to keep the system linear.  
We present numerical results for a set of test problems 
to demonstrate the ability of the code.  
The numerical treatments of 
advection and collision terms are validated first 
in the diffusion and free streaming limits.  
Then we compute steady neutrino distributions for a background 
extracted from a spherically symmetric, general relativistic simulation 
of 15M$_{\odot}$ star and compare them with the results 
in the latter computation.  
We also demonstrate multi-D capabilities of 
the 3D code solving 
neutrino transfers for artificially deformed supernova cores 
in 2D and 3D.  
Formal solutions along neutrino paths are utilized 
as exact solutions.  
We plan to apply this code 
to the 3D neutrino-radiation hydrodynamics 
simulations of supernovae.  
This is the first article in a series of reports on 
the development.  
\end{abstract}


\section{Introduction}

Elucidating the explosion mechanism of core-collapse supernovae 
is a grand challenge in astrophysics, 
nuclear and particle physics as well as computing science.  
It requires the numerical computations of hydrodynamics, 
neutrino transfer and electromagnetism 
that incorporate 
detailed microphysics at extreme conditions.  
Delicate interplays of these multiple elements in various physics 
have to be described quantitatively in numerical simulations 
in order to determine unambiguously 
the outcome of collapse and bounce 
of central cores and to obtain 
the magnificent optical display, 
neutrino burst, and ejected material as well as 
remnants that fit observations appropriately.  
Despite the extensive studies with ever increasing 
theoretical achievement 
and computational resources 
for more than four decades, the critical element of supernova explosion 
has been elusive \citep{bet90,kot06,jan07}.  

One of the key issues is the neutrino transfer.  
Interactions of neutrinos with material play 
important roles in a couple of ways.  
The lepton fraction stored in the bouncing core is determined 
by the neutronization and neutrino trapping;  
the core bounce launches a shock wave, which 
stalls on the way out 
owing to energy losses through the dissociation of nuclei and 
the neutrino cooling; 
the fate of the stagnated shock wave is eventually determined by 
the flux of neutrinos 
that are emitted copiously by a proto-neutron star and 
interact with the material below the shock wave.  
Neutrino signals detected at terrestrial detectors are 
an important probe into 
what is happening in the deep interior of thickly veiled supernova cores, 
which was indeed vindicated for SN1987A \citep{hir87}.  

Among viable scenarios for supernova explosions, 
the neutrino heating mechanism \citep{bet85} is currently the most promising.  
The heating of material below the stalled shock wave 
by the absorption of neutrinos emitted from deeper 
inside the core is supposed to revive the stagnated shock wave.  
Since neutrinos carry away most of 
the liberated energy of $\sim$10$^{53}$ erg, 
tapping a hundredth the energy transferred by neutrinos 
is sufficient to obtain the observed explosion energy of $\sim$10$^{51}$ erg.  
Even if the neutrino heating is not the main cause of explosions, 
it will be still critical to accurately evaluate the energy exchanges 
between neutrinos and matter.  
For this purpose, 
one has to solve the neutrino transfer equations and obtain 
fluxes, energy spectra, emissions and absorptions.  

The numerical treatment of neutrino transfer is a longstanding problem though.  
One of the difficulties is the wide range of opacity in the supernova core.  
Deep inside, interactions are so frequent that thermal and chemical equilibrium 
is established in a quasi-static manner and neutrinos spread in a diffusive way.  
As the density and temperature become lower in the outer region, 
the neutrino interactions 
become less frequent and the statistical equilibrium is not achieved.  
Further outside, neutrinos propagate freely through transparent material.  
It is the intermediate regime between diffusion and free-streaming 
that the neutrino heating mechanism operates.  
It is hence mandatory in the computation of neutrino transfer to treat 
all these regimes.  

It is computationally demanding, however, 
as one often sees in the applications of radiation transfer to other subjects.  
Even in spherical symmetry, the neutrino transfer is a three-dimensional problem 
(or four-dimensional if one includes time) 
since the neutrino distribution is 
a function of neutrino energy, angle of neutrino momentum 
with respect to the radial direction as well as 
radial position.  
Since the neutrino interactions are strongly energy-dependent, 
the multi-energy group treatment is indispensable.  
Without any symmetry, the neutrino transfer 
becomes a six dimensional problem: 
the neutrino distribution is then 
as a function of three spatial coordinates 
and three neutrino momentum (energy and two angles that specify 
the direction of neutrino momentum are commonly used as independent variables).  
With limited computational resources, 
this fact makes the numerical computation of neutrino transfer 
such a formidable task and various levels of approximations have been employed 
(See \S \ref{nu-transfer} for details) so far.  

The numerical treatment of neutrino transfer under spherical symmetry 
has been sophisticated over the years - so much so that the Boltzmann equation 
is solved without any approximation.  
Unfortunately, it has been consistently demonstrated 
with scrutinized microphysics \citep{ram00,lie01,mez01,tho03,sum05} 
that no explosion occurs under spherical symmetry.  
On the other hand, recent multi-dimensional simulations have revealed 
the critical role of hydrodynamical instabilities with/without possible 
neutrino assistance for successful supernova explosions \citep{bur06a,mar09,bru10,suw10}.  
Core-collapse simulations in axial symmetry have adopted different 
approximations and only recently the Boltzmann equations were directly 
solved by the discrete-ordinate method 
for a handful of post-bounce evolutions \citep{ott08,bra11}.  
In three spatial dimensions (3D) without any symmetry, 
most of the simulations done so far focused on the hydrodynamical instabilities 
with very simple treatments of neutrinos.  
Very recently the first 3D simulations 
with the neutrino transport being treated with the isotropic diffusion source 
approximation have been performed \citep{tak11}. 
Unfortunately, their results are inconclusive owing to limited numerical resolutions. 
3D computations of neutrino transfer have just begun and further development 
is highly awaited.

Exploring the explosion mechanism in three dimensions is indeed 
the current focus in the supernova society.  
This is mainly because 
marginal explosions have been observed in many 2D simulations so far 
and it is expected that 3D will give a boost: 
increased degrees of freedom may give fluid elements 
more time to hover around the heating region.  
It was indeed demonstrated by a systematic numerical experiment \citep{nor10} that 
the critical neutrino luminosity is lowered and explosions occur 
more easily as the spatial dimension increases \citep[See also][]{han11}.  
This needs confirmation by more realistic simulations with detailed neutrino transfer. 
It is interesting to assess quantitatively 
whether the observed rotation and/or magnetic fields of pulsars can be induced 
by 3D hydrodynamical instabilities as suggested by some authors \citep{blo07a,ran11}.  
Moreover, gamma ray bursts may be produced by 
anisotropic neutrino radiations \citep{har10a,har10b}.  
3D neutrino transfer in these contexts has not been 
studied yet and will be the target in our investigations to come.  

Under these circumstances, we have developed a numerical code 
to treat the time-dependent neutrino transfer 
in 3D space with no symmetry.  
We adopt a discrete-ordinate method
(S$_n$) and deploy multi-angle, multi-energy groups 
to directly solve the Boltzmann equation 
for the neutrino distribution function in six dimensional phase space.  
The fully implicit finite differencing in time is employed.  
The neutrino reactions and EOS of dense matter 
that are suitable for supernova simulations have 
been implemented.  

As far as the authors know, this is the first study 
on the 3D neutrino transfer for core-collapse supernovae.  
The purpose of this article is to report the first results 
on the performance of the new 3D code.  
As a first step, 
we study the neutrino transfer in static backgrounds in this paper.  
Starting with some basic tests in idealized settings, 
we then examine the code performance for exemplary core profiles 
both before and after bounce with the appropriate microphysics inputs.  
By these tests and applications, we demonstrate that 
the 3D simulation of neutrino transfer is now feasible 
in all the opacity regimes.  
It is true that 
the problem size is very large and the resolution is 
still severely limited by the current computing resources.  
We discuss the computing power required for the full 3D core-collapse simulations 
based on the current numerical experiments.  

We arrange the article as follows.  
We briefly describe the recent developments of neutrino transfer 
computations for supernovae in \S \ref{nu-transfer}.  
In \S \ref{formulation}, we give the basic equations of neutrino transfer 
together with the microphysics included in the current version of the code 
and explain their numerical implementation.  
We report in \S \ref{basic} the results of the basic tests to validate 
the code: 
the neutrino transfer in the diffusion and free-streaming limits are 
tested in \S \ref{advection}; 
the implementations of neutrino reactions as collision terms 
are examined in \S \ref{collision}.  
After these basic tests, 
we proceed to some applications of the code to more realistic 
background models adopted from 
supernova simulations in \S \ref{applications}: 
we first compute 
spherically symmetric neutrino transfers in \S \ref{1Dconfig} 
and compare them with the results published in other papers.  
We then demonstrate some basic features of 
2D and 3D neutrino transfers using artificially deformed core profile 
in \S\S \ref{2Dconfig} and \ref{3Dconfig}.  
The summary is given in \S \ref{summary} with some discussions 
on further extensions of the current code.  

\section{Developments on Neutrino transfer}\label{nu-transfer}

We describe briefly the recent developments of neutrino transfer 
in core-collapse supernovae in order to grasp the status of 
prescriptions to handle the neutrino transfer in the previous studies.  
Further references to overview the historical and modern developments, 
we refer the review articles by \citet{suz94,kot06,jan07,ott09} 
on core-collapse supernovae as well as the books by \citet{per02,cas04} 
on the radiation transfer.  

Under spherical symmetry, the development of neutrino transfer 
has reached a level of sophistication by direct solutions of equations, 
thanks to enough computing resources recently.  
After longstanding efforts on the treatment of neutrino transfer 
by invoking approximations \citep[See][for example]{suz94}, 
the numerical solution of 
the neutrino transfer with multi-energy and multi-angle 
has become possible under spherical symmetry \citep{ram00,tho03,mez01} 
in general relativity \citep{lie01,lie04,sum05}.  
Through the developments of the frameworks, 
the method of numerical solutions both in inertial and co-moving frames 
have been established to handle the neutrino transfer 
with the Lorentz transformation \citep{bura06}.  
The comparison of the methods to solve the neutrino transfer 
by direct solutions of the Boltzmann equations 
and by iterations of the moment equations with a variable Eddington factor 
has been also made to test the approaches of 
neutrino-radiation hydrodynamics \citep{lie05}.  
The microphysics of the equation of state and the neutrino reactions 
have been implemented to examine their influence in detail 
by having the advance of nuclear physics in supernovae.  
The most updated set of neutrino reactions in dense matter \citep{bur06} 
as well as electron capture rates of nuclei \citep{lan03} have been adopted 
to test the impact on the explosion mechanism \citep{lan03a,hix03,bura06}.  
The role of the equation of state based on the unstable nuclei 
has been examined through the delicate connections with 
the neutrino transfer \citep{sum05}.  
Even with the exact treatment of the neutrino transfer and 
the updated sets of microphysics, 
the numerical studies of the gravitational collapse of 
massive stars have shown that there is no successful case 
of supernova explosion under spherical symmetry 
for sets of stellar models.  

In the developments of the numerical methods of neutrino transfer 
under spherical symmetry, there has been progress on 
clarifying the role of neutrino transfer in core-collapse supernovae.  
The proper treatment of neutrino transfer is crucial to determine 
the amount of neutrino trapping in the collapsing core, 
the energy and flux of neutrinos emitted from the cooling region, 
the heating behind the stalled shock wave and 
the prediction of the spectrum of supernova bursts.  
The progress from the early methods such as the light-bulb approximation, 
the leakage scheme and the flux limited diffusion method 
to the exact treatment of neutrino transfer has clarified 
the influence of approximations to those quantities.  
Among others, the neutrino heating is sensitive to the neutrino transfer, 
especially, to the flux factor of neutrino distributions \citep{jan96}.  
The accurate evaluation by the neutrino transfer is necessary 
since the flux limited diffusion method may overestimate 
the flux factor, resulting underestimation of the heating \citep{yam99}.  
It is to be noted that the proper treatment of neutrino 
transfer enables ones to predict the energy spectrum of 
the supernova bursts for terrestrial observations \citep{tho03,sum05,sum07,fis09}.  
The energy spectra predicted by the neutrino-radiation hydrodynamics 
are used to evaluate the event numbers for supernova explosions and 
black hole formations near the Galaxy in future 
by taking into account the neutrino oscillations and 
the specification of neutrino detectors \citep{tot98,and05,nak10a,kee10}.  
%

In two dimensions, the approximate treatments of the neutrino transfer 
have been adopted in most of the recent studies by state-of-the-art 
calculations.  
This is because the exact treatment of neutrino transfer 
becomes much harder than the case of spherical symmetry 
due to the increase of dimension of phase space from three to five.  
There are several categories in the approximate methods, 
depending on the degree of accuracy regarding transfer and 
dimensional assumption.  
Putting the emphasis on the two dimensional behavior, 
the flux limited diffusion method has been adopted for 
the neutrino-radiation hydrodynamics 
to reveal the explosion mechanism \citep{liv04,wal05,bur06a}.  
The diffusion equations of neutrino energy and flux distributions 
with multi-energy groups are solved with the flux-limiter 
to handle the transition to the free-streaming limit. 
Although this approach is advantageous to describe 
the lateral transport of neutrinos in supernova cores, 
the intermediate regime from diffusion to transparent 
regimes are handled by the prescribed flux-limiter.  
Attempting to describe better the transition of neutrino transfer 
in the intermediate regime, 
the method of "ray-by-ray plus" approximation 
has been adopted for the neutrino-radiation hydrodynamics \citep{bura06,mar09}
utilizing the developed code of neutrino transfer 
for spherical symmetry \citep{ram02}.  
The equation of neutrino transfer along each radial ray 
is solved independently for many directions to cover 
the whole region of the supernova core.  
By using the exact treatment of the 1D neutrino transfer 
for multi-energy groups, 
it is advantageous to describe the whole regime of neutrino 
transfer.  
By assuming the spherical symmetry on the transport along each ray, 
the transport of neutrinos for lateral directions 
is neglected except for the advection with material and 
a partial contribution of pressure.  
This prescription along independent rays overestimates 
the angular dependence of neutrino quantities and 
enhances the neutrino fluxes along the radial directions.  
Some of the recent studies adopt the mixed approach of 
the ray-by-ray plus approximation together with 
the approximation of flux limited diffusion \citep{bru06} 
or the isotropic diffusion source approximation (IDSA) \citep{lie09,suw10}.  
The ray-tracing method has been utilized 
for the analysis of anisotropic neutrino radiations \citep{kot09}.  

Recently, the 2D numerical simulations by the multi-angle, 
multi-energy group neutrino-radiation hydrodynamics have been done 
to study the postbounce phase of core-collapse supernovae.  
The neutrino transfer is solved by the discrete-ordinate 
(S$_n$) method to describe the whole regime of neutrino transfer \citep{liv04}.  
In order to save computational load and to have large time steps, 
the flux limited diffusion approximation is adopted for the central 
part of supernova core for the study on the time evolution 
for a long period \citep{ott08}.  
The basic behavior of neutrino quantities in 2D 
has been reported through the examination of angle dependence, 
moments of energy and angle of neutrino distributions 
in realistic profiles of supernova dynamics.  
The advantage to describe the neutrino transfer in 2D 
has been demonstrated 
through comparisons with the counterpart 
by the flux limited diffusion approximation.  
The multi-angle treatment of neutrino transfer has revealed 
the substantial effects such as 
asymmetries in neutrino fluxes between pole and equator, 
enhancements of the neutrino heating 
through the integral of neutrino sources over many angles \citep{ott08,bra11}.  
With those levels of approximations and sophistications described above, 
the mechanism of core-collapse supernovae has been explored 
by neutrino radiation hydrodynamics in 2D 
to find a handful of successful models with new mechanisms, 
which are still under debate.  

In three dimensions, the treatment of neutrino transfer 
is still in its infancy.  
In most of the numerical simulations in 3D \citep{blo03,ohn06,blo07b,iwa08}, 
the neutrino transfer has been neglected or simplified 
by assuming the given rate of 
neutrino cooling and/or heating \citep{nor10,ran11}.  
The central part of supernova core was often omitted to 
avoid the treatment of neutrino trapping and emission 
in the early stage of researches.  
These simplifications in the numerical studies are made 
to explore the frontier of hydrodynamical instabilities 
and to seek the favorable conditions of explosion 
in 3D beyond the assumption of axial symmetry.  
Moreover, the computational cost of neutrino transfer 
in 3D has been formidable to handle the time evolution of neutrino 
distributions in six dimensions.  
The 3D numerical simulations by the ray-by-ray plus approximations 
have been reported recently by adopting the flux limited diffusion \citep{bru10} 
or the isotropic diffusion source approximation (IDSA) \citep{tak11} 
for radial transport.  
However, the direct solution of the neutrino transfer in 3D 
has not been implemented in the numerical simulations so far.  

There have been long-term efforts on the numerical method 
to solve directly the neutrino transfer in multi-dimensions 
besides the approximative approaches 
mentioned above \citep[See also,][]{swe09}.  
General forms of the neutrino transfer in general relativity 
have been studied by \cite{car03,car05}.  
A time-dependent Boltzmann transport scheme in multi-energy 
and multi-angle has been recently developed for neutrino-radiation 
hydrodynamics in one and two dimensions \citep{liv04}.  
The Boltzmann equation in 2D axisymmetric geometry 
is discretized in conservative form 
using the discrete-ordinates (S$_n$) method 
by dropping the velocity dependent terms.  
The 2D transport method incorporated to the neutrino radiation hydrodynamics 
is applied to a time-dependent 2D test of a post-bounce supernova core.  
The neutrino radiation hydrodynamics code (VULCAN-2D) has been utilized 
for the 2D supernova simulations 
with a variant of the flux limited method 
as mentioned above \citep{ott08,bra11}.  
More recently, 
a new algorithm to solve the neutrino transfer in two dimensions 
has been developed to conform the Lorentz transformation 
in the transport equation \citep{hub07}.  
They derive the formulation using the mixed-frame approach 
by evaluating the collision term in the comoving frame 
with a Taylor expansion regarding Lorentz shifts.  
The new formalism has been applied to one-dimensional tests 
of stationary solutions and proto-neutron star cooling.  

Our study here is to establish the numerical solver of 
the Boltzmann equation for neutrinos in three dimensions, 
for the first time, 
beyond the previous developments in two dimensions.  
We develop a numerical code to solve the Boltzmann equation 
for multi-energy and multi-angle group in 3D spatial coordinates.  
We take an approach to solve the Boltzmann equation 
in the inertial frame, on which we report below, as a basis 
for our developments.  
We extend the formulation and its numerical implementation 
by evaluating the collision term 
according to the Lorentz transformation as a next step, 
which will be reported separately elsewhere.  


\section{Formulations}\label{formulation}

\subsection{Boltzmann Equation}\label{Boltzmann}
\begin{figure}
\epsscale{.50}
\plotone{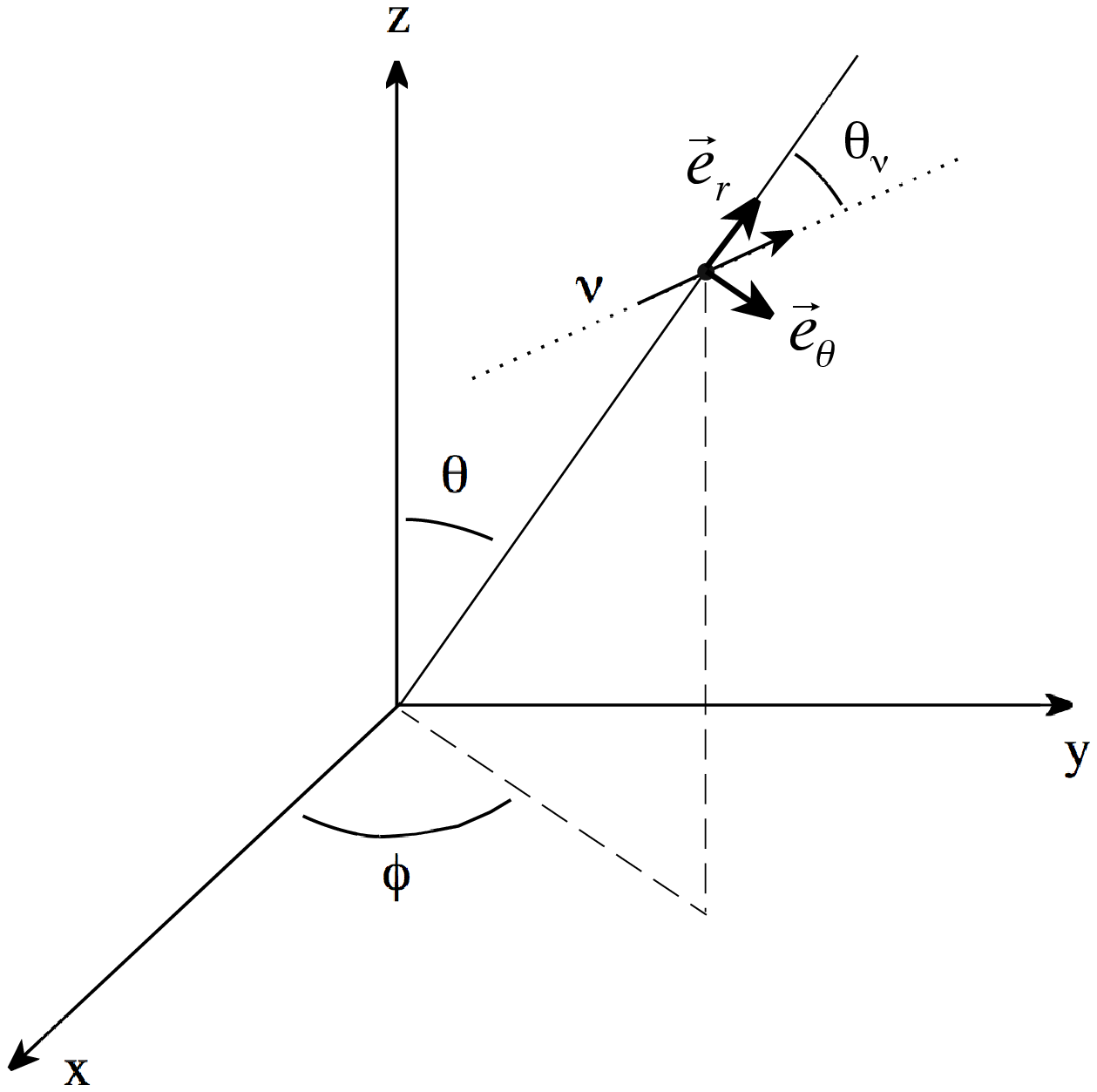}
\plotone{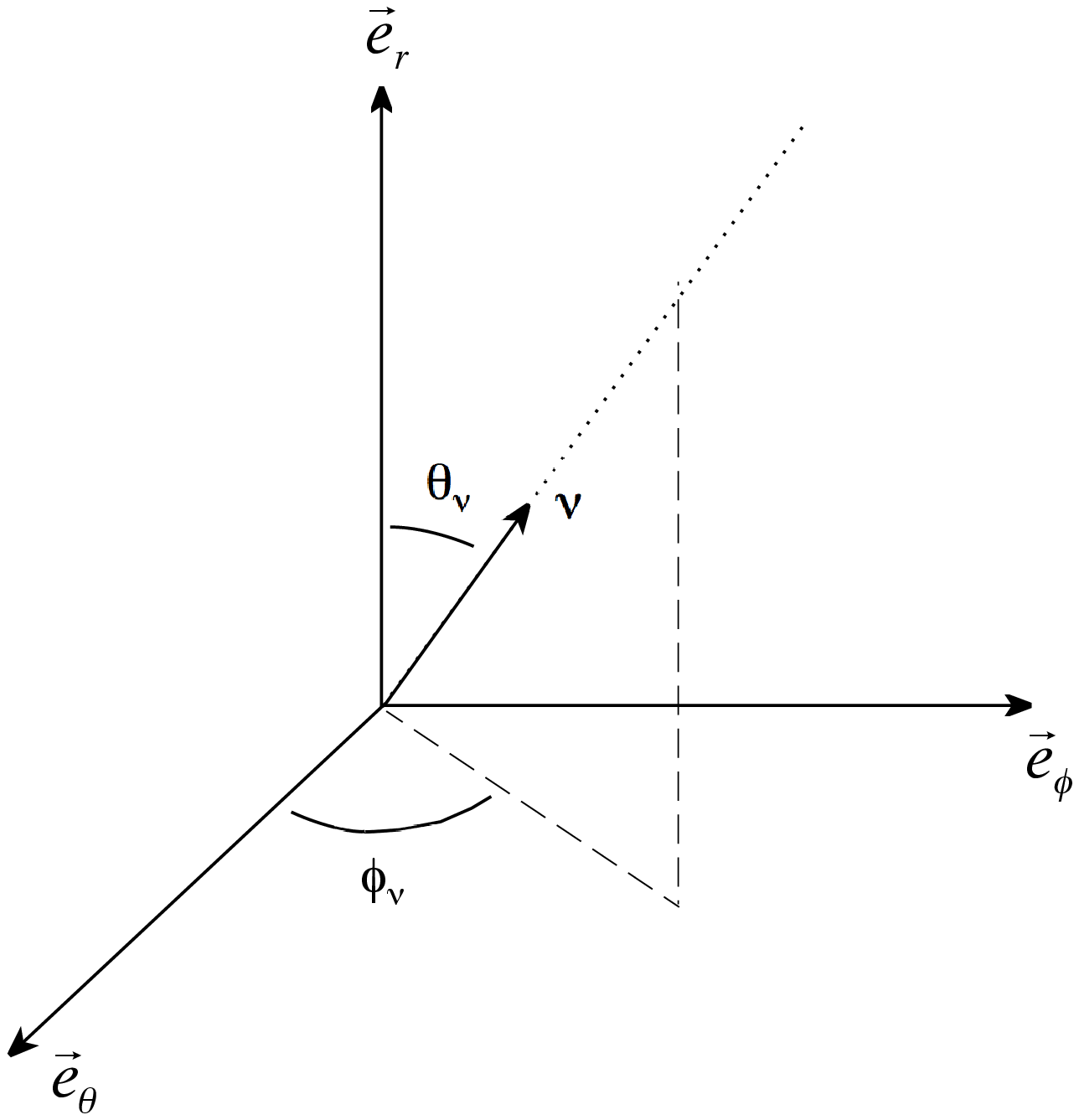}
\caption{Definition of the neutrino direction.  
The direction of neutrino propagation 
is specified by the neutrino angle variables, 
$\theta_{\nu}$ and $\phi_{\nu}$,  
in the spherical coordinate system.  
The neutrino angle, $\theta_{\nu}$, is measured 
from the radial direction.  
The unit vectors along the radial and theta directions 
are depicted in the spherical coordinate (top).  
The unit vector of the phi direction is defined 
in the right-handed system (bottom).  
The neutrino angle, $\phi_{\nu}$, is measured 
from the theta direction.  }
\label{fig:coordinate}
\end{figure}

In our numerical code for the neutrino transfer,  
we solve the Boltzmann equation for the neutrino distribution 
by a discrete ordinate (S$_n$) method.  
Our starting point is the Boltzmann equation, 
\begin{equation}
\label{eqn:eqtransfin}
\frac{1}{c}\frac{\partial f}{\partial t} 
+ \frac{\partial f}{\partial s}
= \left[ \frac{1}{c} \frac{\delta f}{\delta t} \right]_{collision},
%
%
\end{equation}
for the neutrino distribution function, 
$f(\mbox{\boldmath $r$},t;\varepsilon,\mbox{\boldmath $n$}) $, 
at position, $\mbox{\boldmath $r$}$, and time, $t$, 
along path length, $s$.  
The right hand side is the collision term, 
which expresses the time rate of change 
due to the neutrino reactions such as emissions, absorptions and scatterings.  
We prepare the neutrino distributions,
\begin{equation}
\label{eqn:fin}
f^{in}(r,\theta,\phi,t;\varepsilon^{in},\mbox{\boldmath $n$}^{in}) ,
\end{equation}
where $\varepsilon^{in}$ and {\boldmath $n$}$^{in}$ are 
the neutrino energy and the unit vector of neutrino momentum, 
respectively, in the inertial frame.  
We adopt spatial variables, $r$, $\theta$, $\phi$, in the spherical coordinate system.  
The unit vector of neutrino momentum is defined with respect to the radial direction 
along the coordinate $r$ as in Fig. \ref{fig:coordinate}.  
We adopt the neutrino angles, $\theta_{\nu}$, $\phi_{\nu}$, 
and the neutrino energy, $\varepsilon^{in}$, to designate the neutrino momentum.  

We take an approach in the inertial (laboratory) frame 
to write down the equation of neutrino transfer 
and to handle the neutrino quantities.  
The way of solutions of neutrino transfer 
differs very much depending on the frame \citep{mih99}.  
The two major ways in the comoving and inertial frames 
have both easiness and difficulty in the procedures of solution.  
On the one hand, the form of left hand side of Eq. (\ref{eqn:eqtransfin}) 
is simple in the inertial frame, 
while the derivative terms in the left hand side are complicated 
with velocity dependent terms in the comoving frame \citep{bura06}.  
On the other hand, the collision term can be calculated 
easily in the comoving frame, where the neutrino reactions 
occur in the moving fluid.  
The collision term in the inertial frame requires tedious 
procedures through the Lorentz transformation of reaction 
rates from the comoving frame in principle.  
In our strategy, we take the simplicity of the transfer equation 
in three dimensions and will make numerical efforts to handle 
the collision term in a next step of the development.  

Fixing the framework in the inertial frame, 
the Boltzmann equation, Eq. (\ref{eqn:eqtransfin}), 
in the spherical coordinate system is expressed as
\begin{eqnarray}
\label{eqn:eqtransfin-spherical}
\frac{1}{c}\frac{\partial f^{in}}{\partial t} 
+ {\rm cos}~\theta_{\nu} \frac{\partial f^{in}}{\partial r} 
+ \frac{{\rm sin}~\theta_{\nu}~{\rm cos}~\phi_{\nu}}{r} 
  \frac{\partial f^{in}}{\partial \theta} 
+ \frac{{\rm sin}~\theta_{\nu}~{\rm sin}~\phi_{\nu}}{r {\rm sin}~\theta} 
  \frac{\partial f^{in}}{\partial \phi} \nonumber \\
+ \frac{{\rm sin}^{2}~\theta_{\nu}}{r} 
  \frac{\partial f^{in}}{\partial {\rm cos}~\theta_{\nu}} 
- \frac{{\rm sin}~\theta_{\nu}~{\rm sin}~\phi_{\nu}}{r} 
  \frac{{\rm cos}~\theta}{{\rm sin}~\theta}
  \frac{\partial f^{in}}{\partial \phi_{\nu}} 
= \left[ \frac{1}{c} \frac{\delta f^{in}}{\delta t} \right]_{collision},
\end{eqnarray}
with the definition of the neutrino direction angles \citep{pom73}.  
We remark that there is neither velocity-dependent term nor 
energy derivative in the equation in the inertial frame, 
being different from that in the comoving frame.  
Choosing the angle variable $\mu_{\nu}={\rm cos}~\theta_{\nu}$ 
instead of $\theta_{\nu}$, the equation can be written by 
\begin{eqnarray}
\label{eqn:eqtransfin-spherical2}
\frac{1}{c}\frac{\partial f^{in}}{\partial t} 
+ \mu_{\nu} \frac{\partial f^{in}}{\partial r} 
+ \frac{\sqrt{1-\mu_{\nu}^{2}}~{\rm cos}~\phi_{\nu}}{r} 
  \frac{\partial f^{in}}{\partial \theta} 
+ \frac{\sqrt{1-\mu_{\nu}^{2}}~{\rm sin}~\phi_{\nu}}{r {\rm sin}~\theta} 
  \frac{\partial f^{in}}{\partial \phi} \nonumber \\
+ \frac{1-\mu_{\nu}^{2}}{r} 
  \frac{\partial f^{in}}{\partial \mu_{\nu}} 
- \frac{\sqrt{1-\mu_{\nu}^{2}}~{\rm sin}~\phi_{\nu}}{r} 
  \frac{{\rm cos}~\theta}{{\rm sin}~\theta}
  \frac{\partial f^{in}}{\partial \phi_{\nu}} 
= \left[ \frac{1}{c} \frac{\delta f^{in}}{\delta t} \right]_{collision}.
\end{eqnarray}
For the numerical calculation, we rewrite the equation in the conservative form as, 
\begin{eqnarray}
\label{eqn:eqtransfin-spherical2a}
\frac{1}{c}\frac{\partial f^{in}}{\partial t} 
+ \frac{\mu_{\nu}}{r^{2}} \frac{\partial}{\partial r} (r^{2} f^{in})
+ \frac{\sqrt{1-\mu_{\nu}^{2}}~{\rm cos}~\phi_{\nu}}{r {\rm sin}~\theta} 
  \frac{\partial}{\partial \theta} ({\rm sin}~\theta f^{in})
+ \frac{\sqrt{1-\mu_{\nu}^{2}}~{\rm sin}~\phi_{\nu}}{r {\rm sin}~\theta} 
  \frac{\partial f^{in}}{\partial \phi} \nonumber \\
+ \frac{1}{r} 
  \frac{\partial}{\partial \mu_{\nu}} [(1-\mu_{\nu}^{2}) f^{in}]
- \frac{\sqrt{1-\mu_{\nu}^{2}}}{r} 
  \frac{{\rm cos}~\theta}{{\rm sin}~\theta}
  \frac{\partial}{\partial \phi_{\nu}} ({\rm sin}~\phi_{\nu} f^{in})
= \left[ \frac{1}{c} \frac{\delta f^{in}}{\delta t} \right]_{collision}.
\end{eqnarray}
We adopt this equation as the basis for our numerical code.  
We remark that the neutrino distribution function is a function of 
time and six variables in the phase space as written by,
%
\begin{equation}
\label{eqn:fin3d}
f^{in}(r,\theta,\phi,t; \mu_{\nu}, \phi_{\nu},\varepsilon^{in}) .
\end{equation}
In the above expressions, the angle variables, $\mu_{\nu}$ and $\phi_{\nu}$ 
are those measured in the inertial frame.  

\subsection{Neutrino Reactions}\label{reaction}

We implement the rate of neutrino reactions with the composition of dense matter 
as contributions to the collision term.  
We take here several simplifications to make the neutrino transfer 
in 3D feasible.  

As the first step of 3D calculations, 
we treat mainly the case of static background of material 
or the case where the motion is very slow so that $v/c$ is very small.  
In the current study, we evaluate the collision term of the Boltzmann equation 
to the zeroth order of $v/c$ 
by neglecting the terms due to the Lorentz transformation.  
For dynamical situations in general, 
this drastic approximation 
will be studied carefully by evaluating 
the effects from the Lorentz transformation in future.  
We plan to implement such effects 
in all orders of $v/c$ 
in our formulation by taking into account 
the energy shift by the Doppler effects and the angle shifts by the aberration 
in the collision term.  

In addition, we limit ourselves within a set of neutrino reactions 
to make the solution of Boltzmann equation possible 
in the current supercomputing facilities.  
In order to avoid the energy coupling in the collision term, 
we do not take into account energy-changing scatterings 
such as the neutrino-electron scattering \citep{bur06a}.  
This makes the size of the block matrix due to the collision term 
smaller and the whole matrix tractable in the system of equations.  
As a further approach, 
we linearize the collision term for the pair process 
to avoid the non-linearity in equations and 
to guarantee the convergence.  

In future, having enough supercomputing resources, 
we will be able to include the energy-changing reactions 
by enlarging the size of block matrices.  
We also may be able to solve the full reactions 
by the Newton iteration, which requires the complicated 
matrix elements by derivatives, 
as have been accomplished in the spherical calculations \citep{sum05}.  

In the numerical study under the assumptions above, 
we implement the collision term in the following way.  
We utilize directly the neutrino distribution function in the inertial frame 
to evaluate the collision term.  
We use the energy and angle variables in the inertial frame 
in the calculation of the collision term by dropping the shifts.  
We drop the superscript {\it in} for the inertial frame 
in the following expressions.  
For the emission and absorption of neutrinos, 
the collision term for the energy, $\varepsilon$, and 
the angles, $\mu_{\nu}$ and $\phi_{\nu}$, is expressed as,
\begin{equation}
\label{eqn:colemisabs}
\left[ \frac{1}{c} \frac{\delta f}{\delta t} \right]_{emis-abs} = 
- R_{abs}(\varepsilon, \Omega) f(\varepsilon, \Omega)  
+ R_{emis}(\varepsilon, \Omega) ( 1 - f(\varepsilon, \Omega)).  
\end{equation}
Hereafter we suppress the spatial variables and 
use $\Omega$ to denote the two angle variables 
for the compactness of equations.  
The emission rate is related with the absorption rate through 
the detailed balance as,
\begin{equation}
\label{eqn:ratebalance}
R_{emis}(\varepsilon, \Omega) = 
R_{abs}(\varepsilon, \Omega) e^{-\beta (\varepsilon - \mu_{\nu})},
\end{equation}
where $\beta=1/k_B T$ is the inverse of temperature 
and $\mu_{\nu} = \mu_{p} + \mu_{e} -\mu_{n}$ is 
the chemical potential for neutrinos.  
The collision term for the scattering is expressed by,
\begin{eqnarray}
\label{eqn:colscat}
\left[ \frac{1}{c} \frac{\delta f}{\delta t} \right]_{scat} = 
- \int \frac{d \varepsilon' \varepsilon'^2}{(2 \pi)^3}~
\int d \Omega'~R_{scat}(\varepsilon, \Omega; \varepsilon', \Omega') 
f(\varepsilon, \Omega)  
[ 1 - f(\varepsilon', \Omega')  ] \nonumber \\
+ \int \frac{d \varepsilon' \varepsilon'^2}{(2 \pi)^3}~
\int d \Omega' ~R_{scat}(\varepsilon', \Omega'; \varepsilon, \Omega)
f(\varepsilon', \Omega')  
[ 1 - f(\varepsilon, \Omega)  ] , 
\end{eqnarray}
where $\Omega'$ denotes the angle variables after/before the scattering.  
The energy integration can be done by assuming the iso-energetic 
scattering.  
The expression can be reduced as 
\begin{equation}
\label{eqn:colscat2}
\left[ \frac{1}{c} \frac{\delta f}{\delta t} \right]_{scat} = 
- \frac{\varepsilon^2}{(2 \pi)^3}~
\int d \Omega'~R_{scat}(\Omega; \Omega') 
[ f(\varepsilon, \Omega)  - f(\varepsilon, \Omega') ]  ,
\end{equation}
with the relation, $R_{scat}(\Omega'; \Omega)=R_{scat}(\Omega; \Omega')$.  
The collision term for the pair-process is expressed by, 
\begin{eqnarray}
\label{eqn:colpair}
\left[ \frac{1}{c} \frac{\delta f}{\delta t} \right]_{pair} = 
- \int \frac{d \varepsilon' \varepsilon'^2}{(2 \pi)^3}~
\int d \Omega'~
R_{pair-anni}(\varepsilon, \Omega; \varepsilon', \Omega') 
f(\varepsilon, \Omega)  
\overline{f}(\varepsilon', \Omega') \nonumber \\
+ \int \frac{d \varepsilon' \varepsilon'^2}{(2 \pi)^3}~
\int d \Omega' ~
R_{pair-emis}(\varepsilon, \Omega; \varepsilon', \Omega')
[ 1 - f(\varepsilon, \Omega) ] 
[ 1 - \overline{f}(\varepsilon', \Omega') ] ,
\end{eqnarray}
where $\overline{f}(\varepsilon', \Omega')$ denotes 
the distribution of anti-neutrinos.  
From the detailed balance, the following relation holds;  
\begin{equation}
\label{eqn:ratebalance2}
R_{pair-anni}(\varepsilon, \Omega; \varepsilon', \Omega') = 
R_{pair-emis}(\varepsilon, \Omega; \varepsilon', \Omega')
e^{\beta (\varepsilon + \varepsilon')}.
\end{equation}
We linearize the collision term, Eq. (\ref{eqn:colpair}), 
by assuming that the distribution for anti-neutrinos is 
given by that in the previous time-step or the equilibrium distribution.  
This is a good approximation since the pair-process is dominant 
only in high temperature regions, where neutrinos are 
in thermal equilibrium.  
We adopt the approach with the distribution in the previous time-step 
in all of the numerical calculations with pair processes in the current study.  
We utilize further the angle average of the distribution 
when we take the isotropic emission rate as we will state.  
We have also tested that the approach with the equilibrium distribution 
determined by the local temperature and chemical potential works equally well.  

As for the reaction rates, we take mainly 
from the conventional set by \cite{bru85} with some extensions \citep{sum05}.  
We implement the neutrino reactions, 
\begin{eqnarray}
\label{eqn:nureaction1}
e^- + p  \longleftrightarrow  \nu_e + n       \ \ [ecp]  ,\\
\label{eqn:nureaction2}
e^+ + n  \longleftrightarrow  \bar{\nu}_e + p \ \ [aecp] ,\\
\label{eqn:nureaction3}
e^- + A  \longleftrightarrow  \nu_e + A'      \ \ [eca]  ,
\end{eqnarray}
for the absorption/emission,
\begin{eqnarray}
\label{eqn:nureactionscat1}
\nu + N  \longleftrightarrow  \nu + N         \ \ [nsc]  ,\\
\label{eqn:nureactionscat2}
\nu + A  \longleftrightarrow  \nu + A         \ \ [csc]  ,
\end{eqnarray}
for the iso-energetic scattering.  
We do not take into account the neutrino-electron scattering.  
It is well known that the influence of this reaction is minor 
although it contributes to the thermalization \citep{bur06a}.  
As for the pair-process, we take the electron-positron process 
and the nucleon-nucleon bremsstrahlung as follows,
\begin{eqnarray}
\label{eqn:nureactionpair1}
e^- + e^+  \longleftrightarrow    \nu_i + \bar{\nu}_i  \ \ [pap] , \\
\label{eqn:nureactionpair2}
N + N \longleftrightarrow N + N + \nu_i + \bar{\nu}_i  \ \ [nbr] .
\end{eqnarray}
For these pair processes, we take the isotropic emission 
rate as an approximation, which avoids complexity but 
describes the essential roles.  
We remark that the set of the reaction rates adopted 
in the current study is the minimum, which describes sufficiently 
the major role of neutrino reactions in the supernova mechanism.  
Further implementation of other neutrino reactions 
and more sophisticated description of reaction rates 
in the modern version \citep{bura06,bur06} will be done once we have enough 
computing resources in future.  
%

\subsection{Equation of State}\label{EOS}

We utilize the physical equation of state (EOS) of dense matter 
to evaluate the rates of neutrino reactions.  
It is necessary to have the composition of dense matter 
and the related thermodynamical quantities such as 
the chemical potentials and the effective mass of nucleon.  
We implement the subroutine for the evaluation of quantities 
from the data table of EOS as used in the other simulations 
of core-collapse supernovae \citep{sum05,sum07}.  
We adopt the table of the Shen equation of state \citep{she98a,she98b,she11} 
in the current study.  
Other sets of EOS can be used by simply replacing the data table.  

\subsection{Numerical Scheme}\label{scheme}

We describe the numerical scheme employed in the numerical code 
for the neutrino transfer in three dimensions.  
The method of the discretization is based on the approach 
by \citet{mez93,cas04}.  
We refer also the references by \citet{swe09, sto92c} 
for the other methods of discretization of neutrino transfer 
and radiation transfer.  

We define the neutrino distributions at the cell centers and 
evaluate the advection at the cell interfaces and the collision terms 
at the cell centers.  
We describe the neutrino distributions in the space coordinate 
with radial $N_r$-, polar $N_{\theta}$- and azimuthal $N_{\phi}$-grid points 
and in the neutrino momentum space with energy $N_{\varepsilon}$-grid points 
and angle $N_{\theta_{\nu}}$- and $N_{\phi_{\nu}}$-grid points.  
We explain the detailed relations to define the numerical grid 
in \S \ref{scheme-grid}.  

We discretize the Boltzmann equation, Eq. (\ref{eqn:eqtransfin-spherical2a}), 
for the neutrino distribution, $f_{i}^{n}$, in a finite differenced form on the grid points.  
Here we assign the integer indices, $n$ and $n+1$, for the time steps 
and, $i$, for the grid position.  
We adopt the implicit differencing in time to ensure the numerical stability 
for stiff equations and to have long time steps for supernova simulations.  
We solve the equation for $f_{i}^{n+1}$ 
by evaluating the advection and collision terms at the time step $n+1$ 
in the following form, 
\begin{eqnarray}
\label{eqn:boltzmann-implicit}
\frac{1}{c}\frac{f_{i}^{n+1} - f_{i}^{n}}{\Delta t} 
+ \left[ \frac{\mu_{\nu}}{r^{2}} \frac{\partial}{\partial r} (r^{2} f) \right]^{n+1}
+ \left[ \frac{\sqrt{1-\mu_{\nu}^{2}}~{\rm cos}~\phi_{\nu}}{r {\rm sin}~\theta} 
  \frac{\partial}{\partial \theta} ({\rm sin}~\theta f) \right]^{n+1} \nonumber \\
+ \left[ \frac{\sqrt{1-\mu_{\nu}^{2}}~{\rm sin}~\phi_{\nu}}{r {\rm sin}~\theta} 
  \frac{\partial f}{\partial \phi} \right]^{n+1} 
+ \left[ \frac{1}{r} 
  \frac{\partial}{\partial \mu_{\nu}} [(1-\mu_{\nu}^{2}) f] \right]^{n+1} \nonumber \\
+ \left[ - \frac{\sqrt{1-\mu_{\nu}^{2}}}{r} 
  \frac{{\rm cos}~\theta}{{\rm sin}~\theta}
  \frac{\partial}{\partial \phi_{\nu}} ({\rm sin}~\phi_{\nu} f) \right]^{n+1} 
= \left[ \frac{1}{c} \frac{\delta f}{\delta t} \right]_{collision}^{n+1},
\end{eqnarray}
where we schematically express the advection terms for the cell containing $f_{i}^{n+1}$.  
We evaluate the advection at the cell interface 
by the upwind and central differencing for free-streaming and diffusive limits, 
respectively.  
The two differencing methods are smoothly connected by a weighting factor 
in the intermediate regime between the free-streaming and diffusive limits.  
We describe the numerical scheme for the evaluation of the advection terms 
in \S \ref{scheme-adv}.  
We express the collision terms by the summation of the integrand 
using the neutrino distributions at the cell centers.  

\subsection{Solution of Linear Equation}
We arrange the discretized neutrino distribution as a vector 
for the system of linear equations.  
The length of the vector is $N_{vector} = N_r N_{\theta} N_{\phi} N_{\varepsilon} N_{\theta_{\nu}} N_{\phi_{\nu}}$.  
We advance time-step by the implicit method for a time step, $\Delta t$, 
by the relation,
\begin{equation}
\label{eqn:implicit}
\frac{\mbox{\boldmath $f$}^{n+1} - \mbox{\boldmath $f$}^{n}}{\Delta t} = F[\mbox{\boldmath $f$}^{n+1}], 
\end{equation}
for the vector of the neutrino distribution, $\mbox{\boldmath $f$}^{n+1}$, 
at the time step, $n+1$.  
By linearizing the collision term as described above, 
we rearrange the Boltzmann equation as a set of linear equations, 
$A \mbox{\boldmath $f$}^{n+1} = \mbox{\boldmath $b$}$.  
The matrix, $A$, contains the terms from time-advance, advection and absorption terms.  
The source vector contains the old vector and emission terms.  
This large sparse matrix ($N_{vector} \times N_{vector}$) contains 
block diagonal matrices of the size, 
$N_{\theta_{\nu}} N_{\phi_{\nu}} \times N_{\theta_{\nu}} N_{\phi_{\nu}}$, 
which comes mainly from the angle-coupling due to scatterings, 
together with the lines of none-zero elements due to the advection in the three directions.  
We remark that the size of the block matrices becomes huge as 
$N_{\varepsilon} N_{\theta_{\nu}} N_{\phi_{\nu}} \times N_{\varepsilon} N_{\theta_{\nu}} N_{\phi_{\nu}}$, 
if we take the energy changing reactions and the Lorentz energy shifts fully into account.  

We solve this system of equations by the matrix solver using the iterative method \citep{saa03}.  
We use the Bi-CGSTAB method by utilizing a program in the Templates library \citep{bar94} 
with the point-Jacobi method as a pre-conditioner.  
We set the allowable convergence measure to be 10$^{-8}$ and 
get the convergence typically within 20 iterations for the current numerical studies.  

We solve the evolution of neutrino distributions for multi-species ($N_f$).  
For the basic tests, we treat the one species of neutrino ($N_f=1$).  
For the applications of supernova cores, we treat the three species 
of neutrinos, $\nu_e$, $\bar{\nu}_e$ and $\nu_{\mu}$ ($N_f=3$).  
We treat $\nu_{\mu}$ as a representative of the group for four species 
$\nu_{\mu}$, $\bar{\nu}_{\mu}$, $\nu_{\tau}$ and $\bar{\nu}_{\tau}$.

\section{Basic Tests}\label{basic}

We performed the series of basic tests on the advection and collision terms, 
in order to validate the numerical code to solve the neutrino transfer 
in three dimensions.  
We report here the two tests on the advection in the diffusion and 
free-streaming limits, where we can compare with the analytic behavior.  
These tests are performed to validate the advection part of the Boltzmann solver.  
In addition, we report the tests on the stationary solution, 
the time evolution toward the equilibrium 
and the comparison with the spherical calculations.  
These tests are used to check the collision term as a source term and 
to examine the neutrino reactions with dense matter in supernova cores.  
The method of basic tests in the current study are based on 
the standard tests described in \citep{sto92c,swe09}.  

\subsection{Advection Term}\label{advection}

\subsubsection{Diffusion Limit}\label{diffusion}
In order to show that the new code can properly handle 
diffusions of neutrinos in opaque material, 
we compute the diffusion of a Gaussian packet in a uniform background.  
Taking a scattering 
as a sole contribution to the collision term, 
we assume that it is isotropic and isoenergetic and 
its rate is independent of incident neutrino energy.  
We keep multiple energy bins to confirm that the neutrino 
energy spectrum is unchanged during the simulation.  

Analytic solutions are available for this test.  
The diffusion of a Gaussian packet 
with its central position located at $\mbox{\boldmath $r$}_0$ 
and its width, $d_0$, being $(4 D t_0)^{1/2}$ initially is described by 
\begin{equation}
\label{eqn:gauss}
f(\mbox{\boldmath $r$},t) = f_0 \left( \frac{t_0}{t_0 + t} \right)^{\alpha}
\exp \left\{ -\frac{| \mbox{\boldmath $r$} - \mbox{\boldmath $r$}_0 |^2}
                   {4 D (t_0 + t)}  \right\},
\end{equation}
where $f(\mbox{\boldmath $r$},t)$ is the neutrino distribution function 
at the position, \mbox{\boldmath $r$}, 
and time, $t$, after the initial time, $t_0$, \citep{swe09}.  
The diffusion coefficient, $D$, is related with the mean free path 
for the scattering, $\lambda$, as $D=c \lambda / 3$.  
The parameter, $f_0$, is the initial height of the packet.  
The value of $t_0=d_0^2/4D$ also gives the time scale of diffusion.  
The power index, $\alpha$, is related with the space dimension, $N_D$, 
as $\alpha=N_D/2$.  

We first describe 
the spherical spreading of the Gaussian packet 
located at the center. 
The computation was done in 3D. 
This is a very crude approximation to what happens in the opaque region 
in the supernova core.  
We additionally examine non-radial diffusions of the Gaussian packet 
in the box geometry.  
Although this has no counterpart in the supernova explosion, 
it is a common benchmark test for neutrino (or radiation) 
transfer codes \citep{swe09,sto92c}.  

For the first test 
we place the Gaussian packet at the center of the sphere 
with the radial coordinate extending from $r=0$ to $r=5\times10^5$ cm 
and the polar and azimuthal coordinates from $\theta=0$ to $\pi$ and 
from $\phi=0$ to $2\pi$.  
The initial width of the Gaussian packet is set to $d_0=1\times10^5$ cm.  
We consider only the isotropic scattering with a mean free path 
of $\lambda=10^{3}$~cm.  
This corresponds to a diffusion time scale of $2.5\times10^{-4}$ sec. 
We deploy $N_r=$20, 40, 80 and 160 radial grid points 
and $N_{\theta}=18$ polar grid points and $N_{\phi}=36$ 
azimuthal grid points as spatial grids.  
For the momentum space, on the other hand, we employ 
$N_{\varepsilon}=2$ for energy grid points 
and $N_{\theta_{\nu}}=12$, $N_{\phi_{\nu}}=12$ for angle grid points.  

\begin{figure}
\epsscale{.70}
\plotone{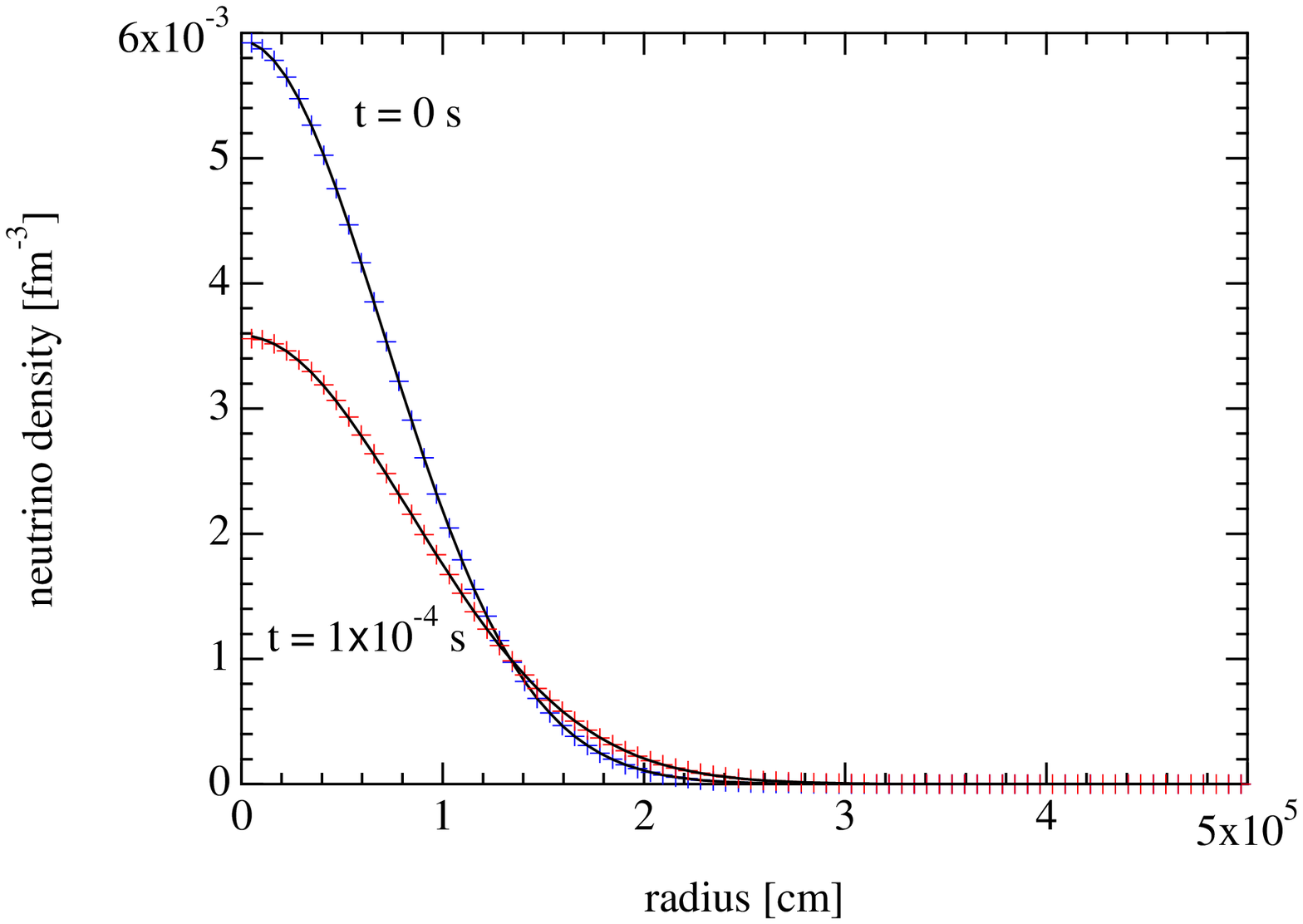}
\plotone{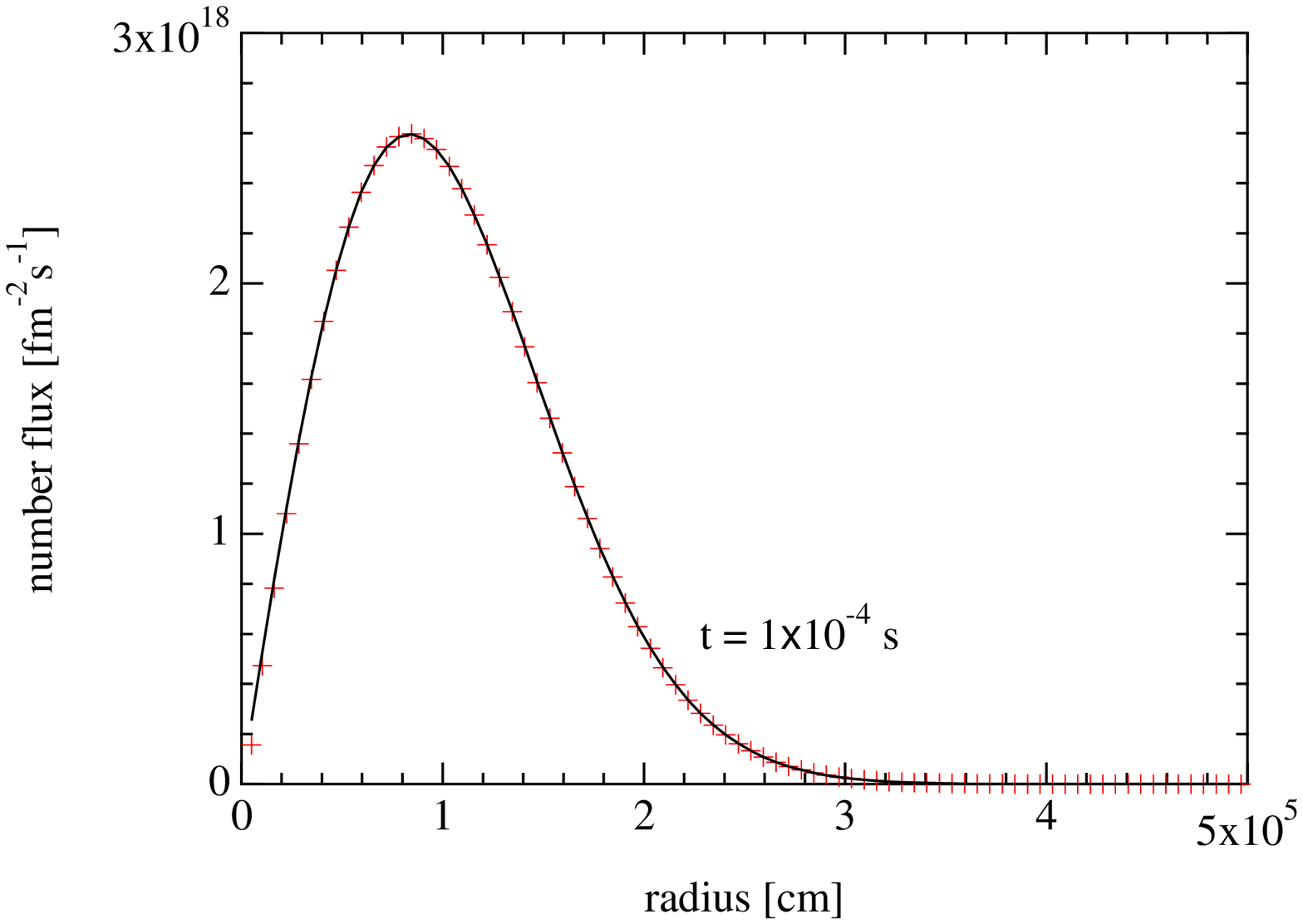}
\caption{Diffusion of the Gaussian packet under spherical symmetry.  
The radial profiles of the neutrino density 
at the initial and final steps are shown in the top panel.  
The radial profile of the number flux at the final time step 
is shown in the lower panel.  
The solid lines and the cross symbols denote 
the analytic solution and the numerical results, respectively.  }
\label{fig:cgauss}
\end{figure}

\begin{figure}
\plotone{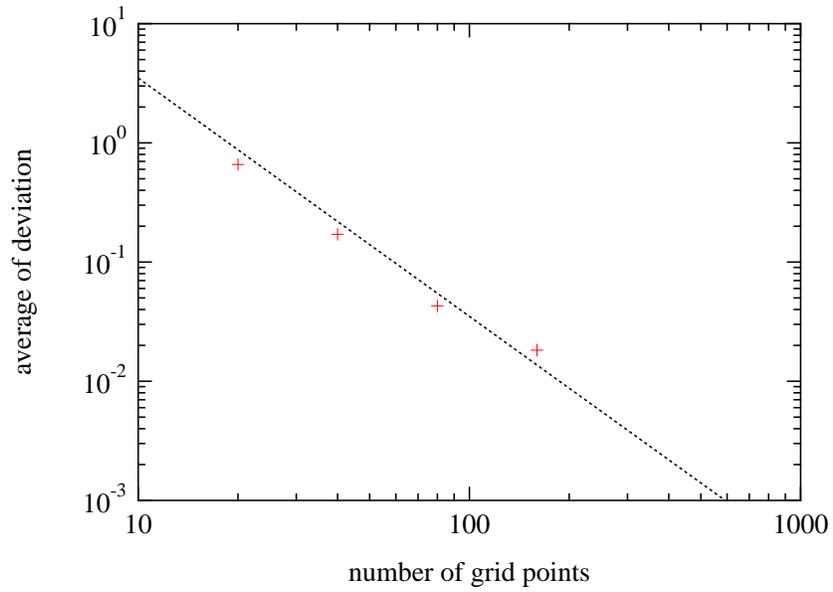}
\caption{Relative deviations of the numerical results by the 3D code 
for the diffusion of Gaussian packet under spherical symmetry 
as a function of the number of radial grid points.  
The relative deviations of the neutrino densities 
from the analytic solutions are averaged over all radial grid points.  
The dotted line shows reference proportional to $N_r^{-2}$.  }
\label{fig:cgauss-error}
\end{figure}

We show in Fig. \ref{fig:cgauss} the radial profiles of 
neutrino density at the initial ($t=0$ s) and 
final ($t=1\times10^{-4}$ s) time steps 
(10$^{4}$ steps) for the case of $N_r=80$.  
The radial profile of number flux is 
also shown for the final time step.  
We compare the numerical results (cross symbols) 
with the analytic solutions (solid lines).  
It is evident that they agree well with each other.  
In Figure \ref{fig:cgauss-error}, we give relative deviations of 
the numerical results from the analytic solutions 
averaged over all the radial grid points 
for different grid sizes.  
As the number of radial grid points increases, 
the deviation is reduced as $N_r^{-2}$ 
as predicted for the second order central differencing scheme.  

\begin{figure}
\epsscale{.65}
\plotone{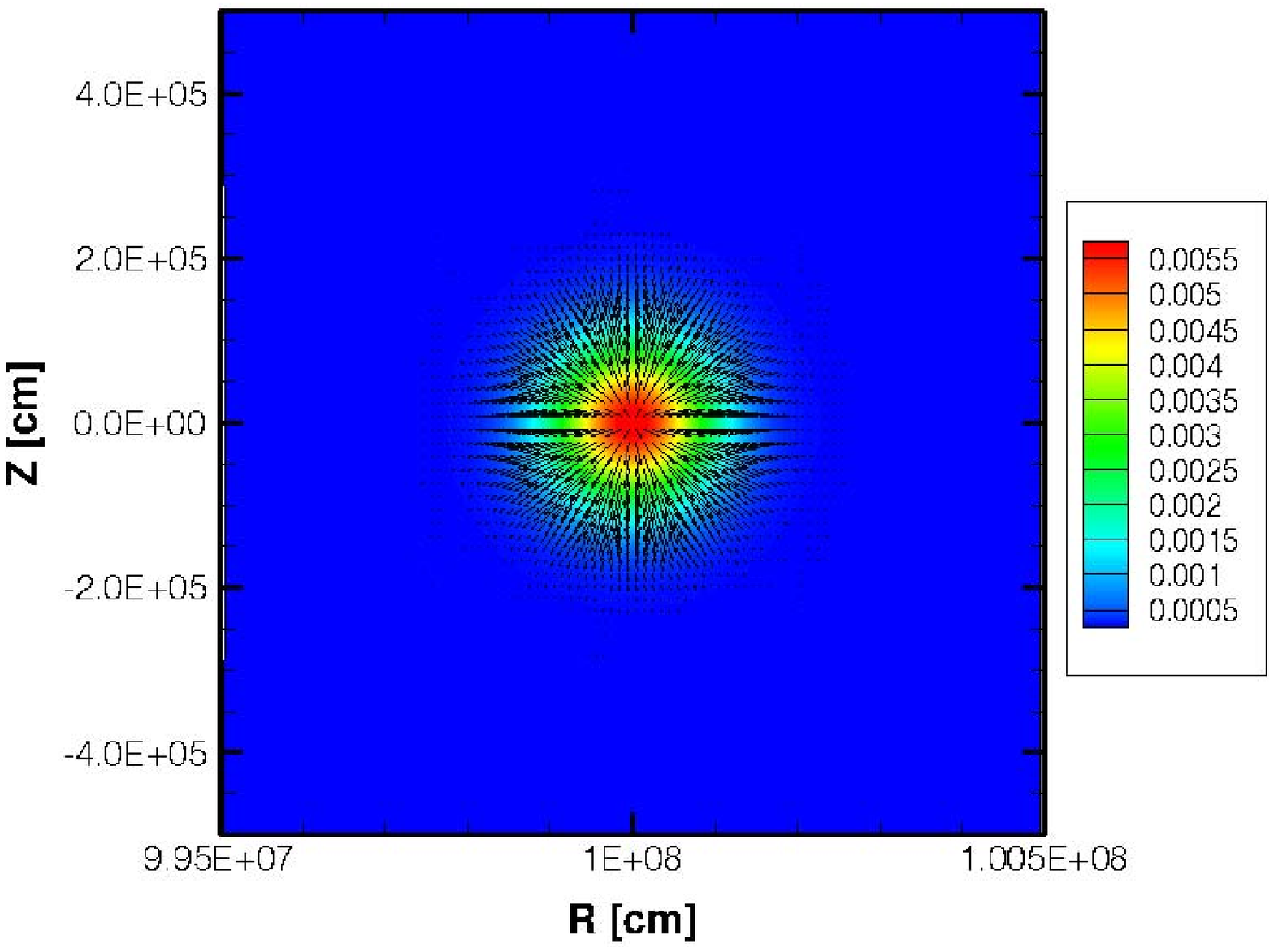}
\plotone{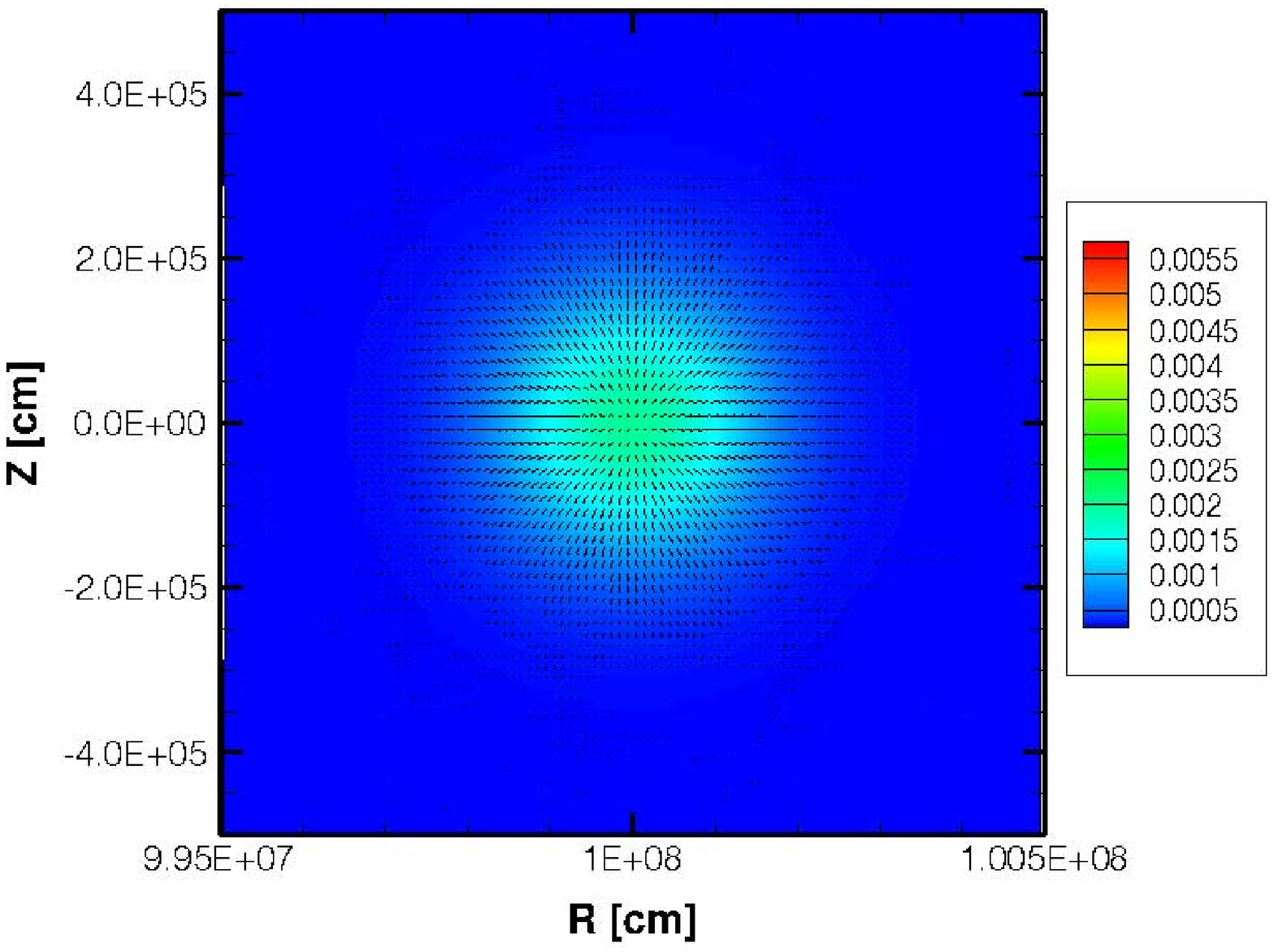}
\caption{Early and final distributions of neutrinos 
in the diffusion of 2D Gaussian packet 
by contour plot on the 2D plane.  
The number density in fm$^{-3}$ is expressed by the color code. 
The arrows express the neutrino fluxes.  }
\label{fig:gauss2d}
\end{figure}

\begin{figure}
\plotone{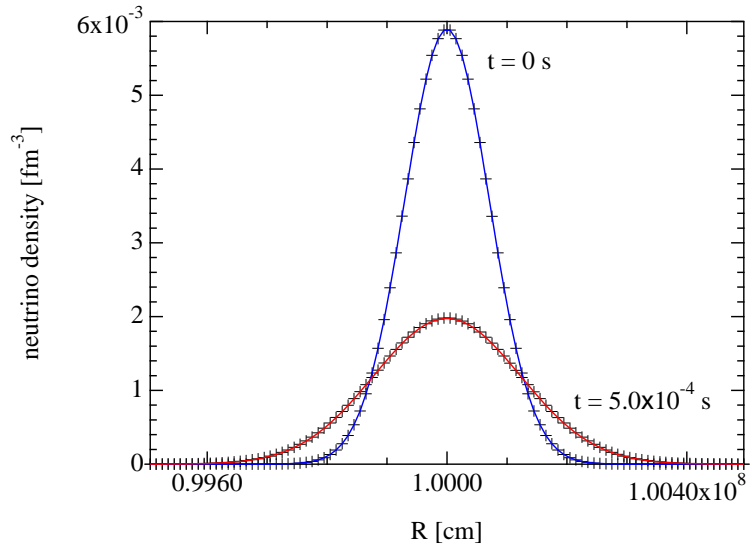}
\caption{Neutrino distributions in the diffusion of 2D Gaussian packet.  
The cross symbols show the initial and final profiles in the numerical results 
along $Z=-8.1\times10^{3}$ cm (near the center of square).  
The two solid lines show the analytic solutions.  }
\label{fig:gauss2dana}
\end{figure}

The second test for non-radial diffusion is done in 2D and 3D.  
In 2D, we place the Gaussian packet in the square 
that has a side of $10^{6}$~cm and is located at $r=10^8$ cm.  
In this small area, coordinates are approximately Cartesian.  
We consider again only the isotropic scattering with the same mean free path 
of $\lambda=10^{3}$~cm.  
We take $N_r=100$, $N_{\theta}=96$ for the spatial grid 
and $N_{\theta_{\nu}}=12$, $N_{\phi_{\nu}}=12$ for the angle grid points 
in the momentum space.  
We deploy $N_{\varepsilon}=4$ energy grid points despite 
the calculation is independent of neutrino energy in this test.  
We show in Fig. \ref{fig:gauss2d} 
the neutrino density at an early time ($t=5.0\times10^{-6}$ s) 
and the final time ($t=5.0\times10^{-4}$ s).  
The polar axis is denoted as Z ($Z=r \cos \theta$) and the distance 
from the polar axis as R ($R=r \sin \theta$) in the plot.  
We compare the numerical result with the analytic solution 
along $Z=-8.1\times10^{3}$ cm 
(near the center of square) in Fig. \ref{fig:gauss2dana}.  
The deviation of the numerical result from the analytic solution is less than 0.5 \% 
in the central region after 10$^{4}$ time-steps ($\Delta t=5\times10^{-8}$ s).  

\begin{figure}
\plotone{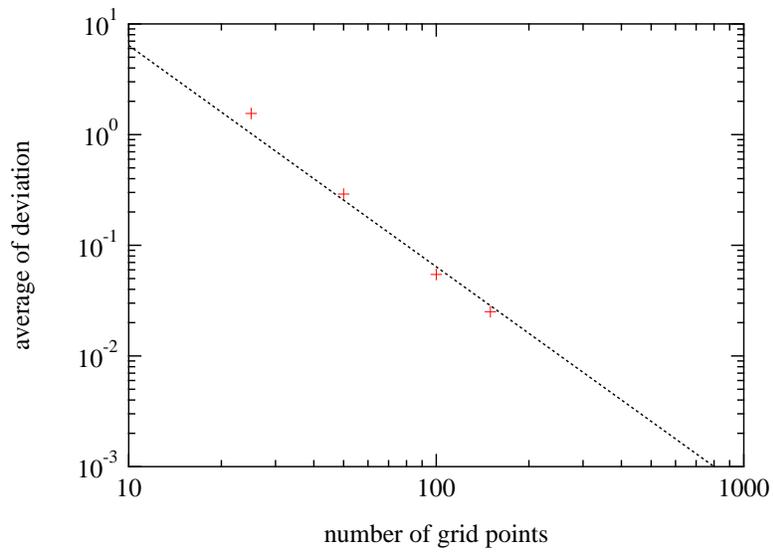}
\caption{Relative deviations of the numerical results 
for the diffusion of 2D Gaussian packet 
as a function of the number of radial grid points.  
The relative deviations of neutrino densities 
from the analytic solutions are averaged over grid points 
along $Z\sim0$ (near the center of square).  
The dotted line shows reference proportional to $N_r^{-2}$.  }
\label{fig:gauss2derror}
\end{figure}

In order to see the convergence of the numerical results quantitatively, 
we repeat the same computation with different resolutions: 
($N_r$, $N_{\theta}$)=(25,24), (50,48), (100,96) and (150,144); 
for the momentum space the grid points are essentially the same 
$N_{\theta_{\nu}}=12$, $N_{\phi_{\nu}}=12$ except 
the number of energy grid points is reduced to $N_{\varepsilon}=2$ 
to save the memory size.  
In Figure \ref{fig:gauss2derror}, we show the relative 
deviations of the neutrino density averaged over the radial grid points along $Z\sim0$.  
The deviation decreases quadratically as the number of grid points increases 
as expected for the central differencing scheme.  

\begin{figure}
\epsscale{.60}
\plotone{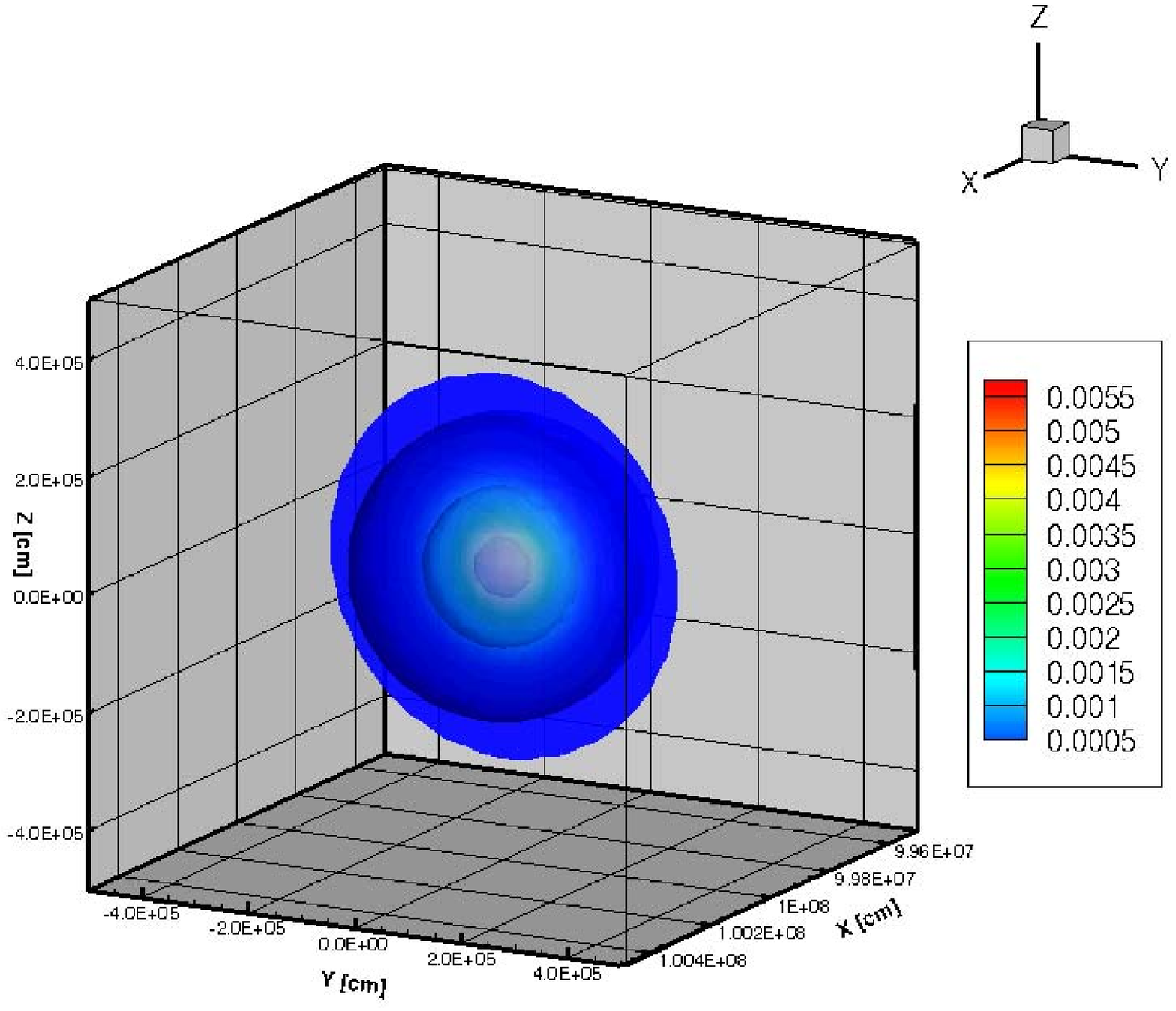}
\plotone{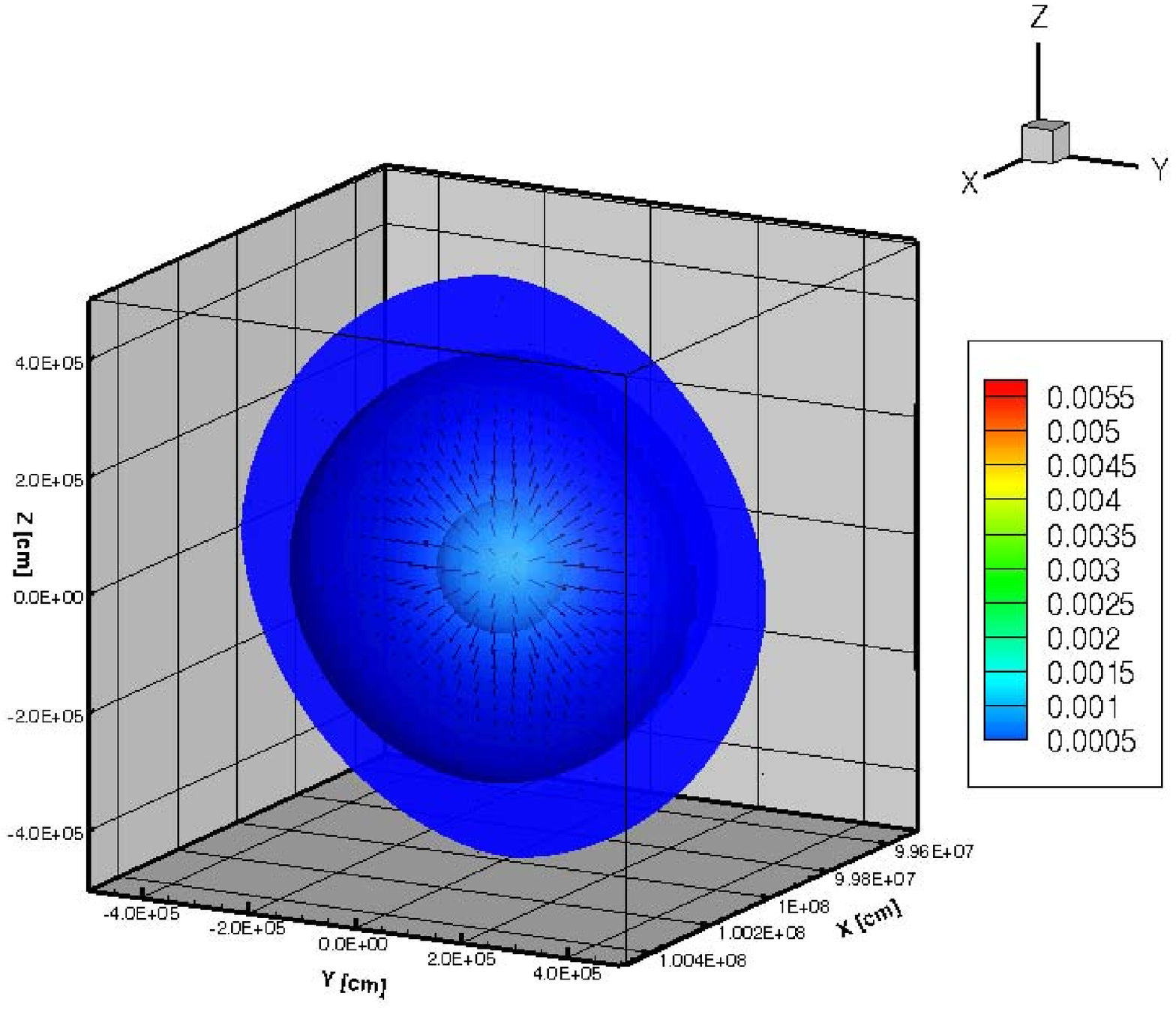}
\caption{Initial and final distributions of neutrinos 
in the diffusion of 3D Gaussian packet.  
The neutrino densities in fm$^{-3}$ 
are shown by color contour plot on the 2D plane 
along $x=1.0\times10^{8}$ cm.  
The isosurfaces of the neutrino densities 
($1.0\times10^{-5}$, $1.0\times10^{-3}$, $4.5\times10^{-3}$ fm$^{-3}$) 
are also shown by the color code.  
The arrows express the neutrino fluxes.  }
\label{fig:gauss3d}
\end{figure}

\begin{figure}
\plotone{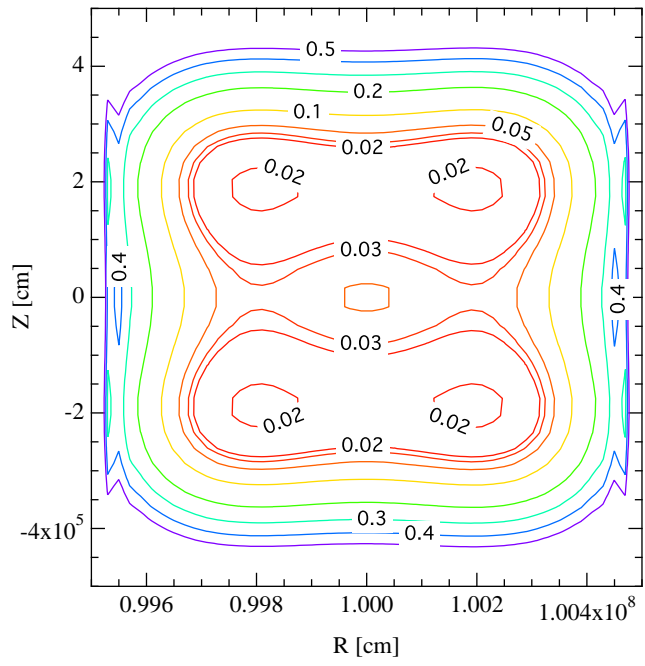}
\caption{Magnitude of relative deviations in the diffusion of 3D Gaussian packet.   
The relative deviations of the neutrino densities by the numerical calculation 
from the analytic solutions are shown by contour plot 
on the plane along $y=-1.6\times10^{4}$ cm.  }
\label{fig:gauss3dana}
\end{figure}

The test is also done in 3D.  
We put a cubic box with a side of $10^{6}$~cm at a radial position of $r=10^{8}$~cm.  
The box is small enough to regard the patch of spherical coordinates as Cartesian.  
The mean free path of the isotropic scattering is again $\lambda=10^{3}$~cm.  
The numbers of grid points employed in this test are 
$N_r=50$, $N_{\theta}=48$, $N_{\phi}=48$ for the spatial grid 
and $N_{\theta_{\nu}}=12$, $N_{\phi_{\nu}}=12$, $N_{\varepsilon}=4$ 
for the momentum space.  
We show in Fig. \ref{fig:gauss3d} the neutrino density on the plane $x=10^{8}$~cm 
for the initial (0 s) and final ($t=3.3\times10^{-4}$ s) times as a contour map.  
The isosurfaces are also shown in color.  
We compare the numerical result with the analytic formula, Eq. (\ref{eqn:gauss}).  
The relative deviation is shown in Fig. \ref{fig:gauss3dana}.  
The final distribution obtained by the numerical computation agrees 
very well with the analytic one.  
The deviation is typically less than 5 \% in the central region 
for the case with $N_r=50$, $N_{\theta}=48$, $N_{\phi}=48$ 
at $t=5.0\times10^{-4}$ s 
after 10$^{4}$ time-steps ($\Delta t=5\times10^{-8}$ s). 
Large deviations near the edge of the box are mainly due to 
the fixed boundary condition we use for this test.  

In order to see the convergence of numerical solutions, 
we perform another test computation with a higher spatial resolution, 
$N_r=75$, $N_{\theta}=72$, $N_{\phi}=72$, and a smaller time step 
of $\Delta t=2\times10^{-8}$ s.  
The results for the different grid sizes are compared 
at the same timing $t=2.0\times10^{-4}$ s.  
The accuracy is improved by a factor of $\sim$2 over the whole region.  
Unfortunately, further computations with even higher resolutions are impossible 
owing to the limitation of available memory space (See \S \ref{CompSize}) 
and the power index of convergence cannot be determined at present. 
It is encouraging, however, that the numerical solution is indeed 
converging to the exact solution in a way that is consistent 
with the 1D and 2D counterparts.  
We note incidentally that the total number of neutrinos in the box is conserved 
within the accumulation of round-off errors as expected for the conservative scheme.  

\subsubsection{Free Streaming Limit}\label{free}
\begin{figure}
\epsscale{.55}
\plotone{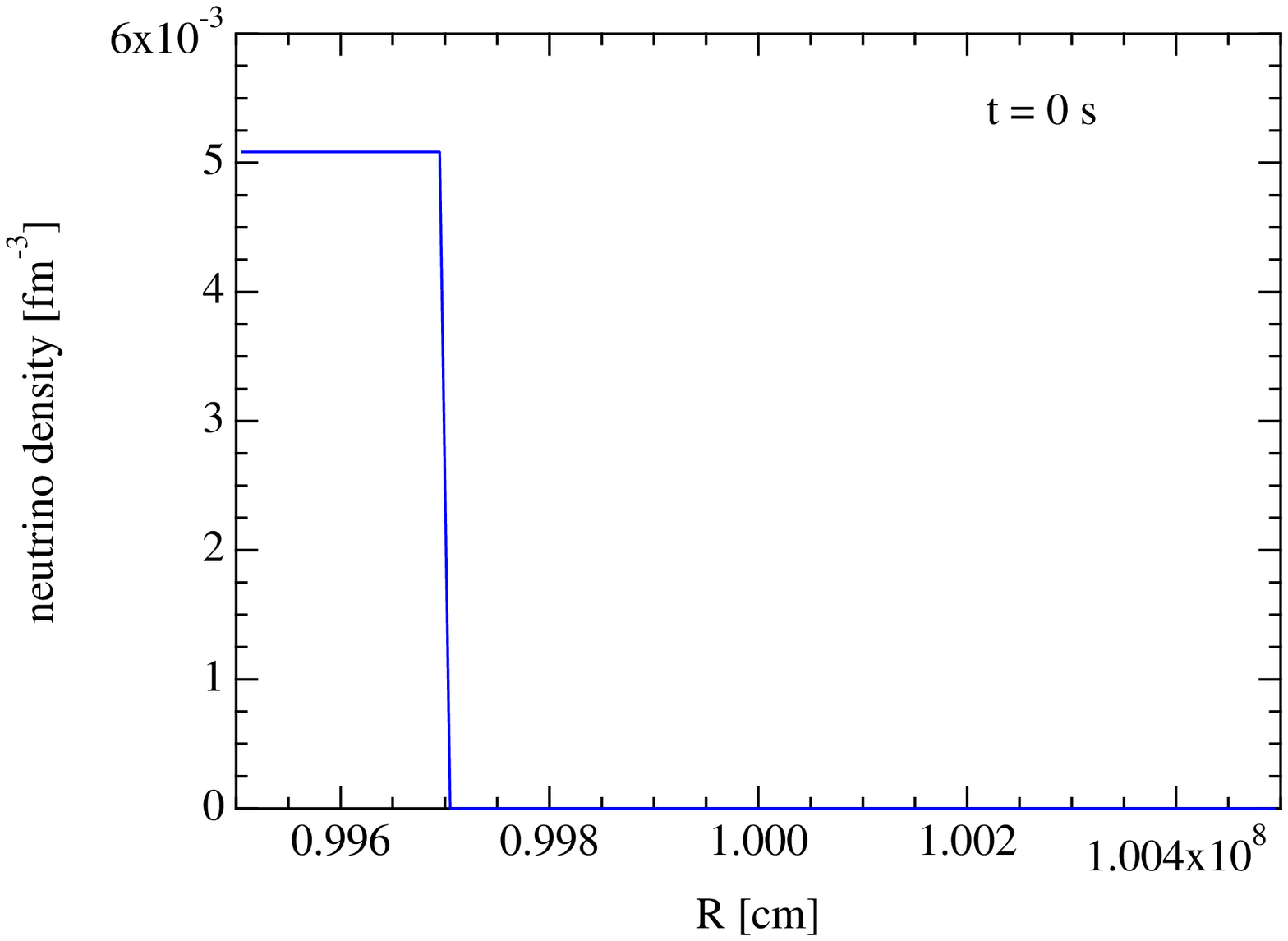}
\plotone{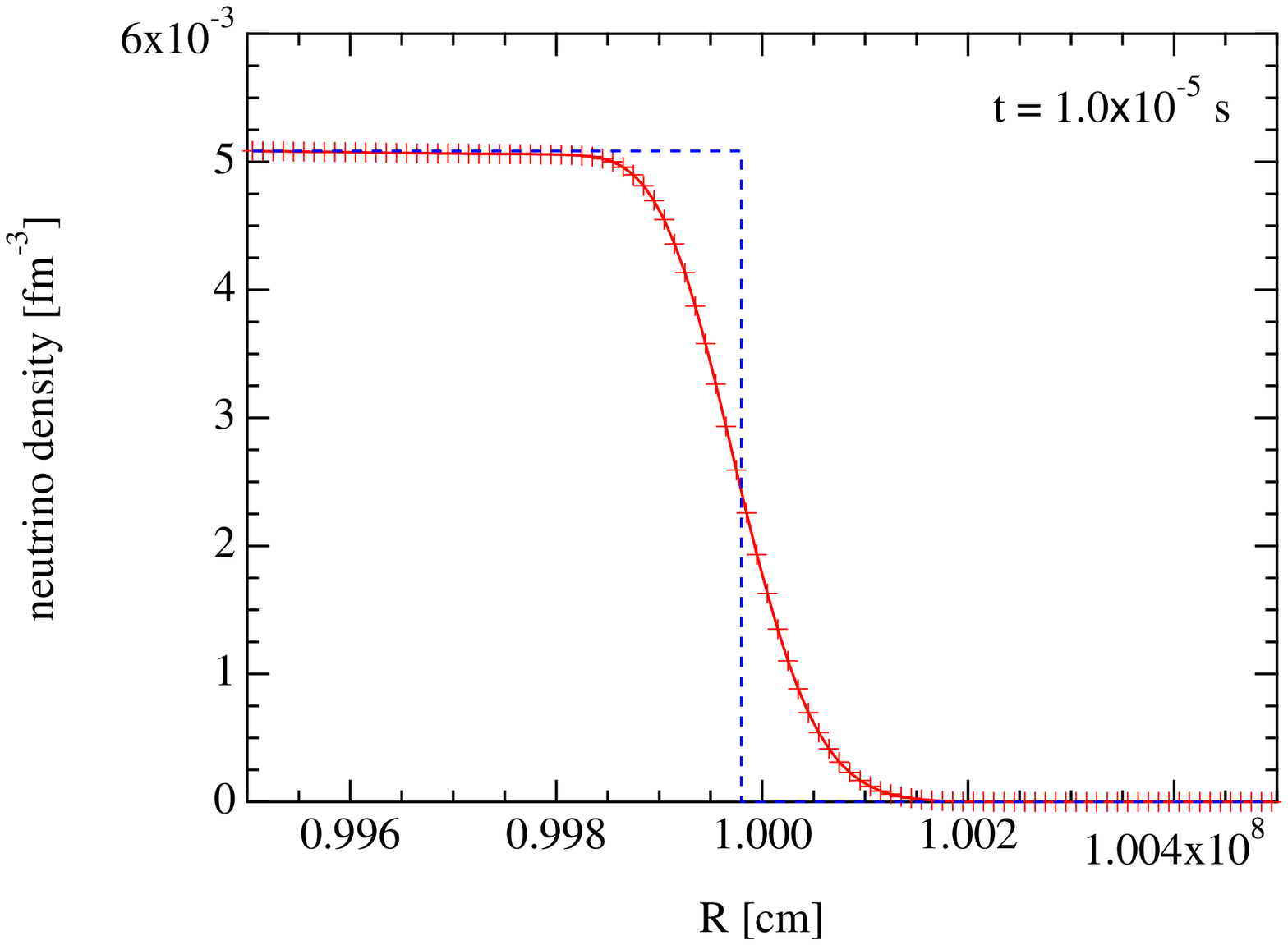}
\plotone{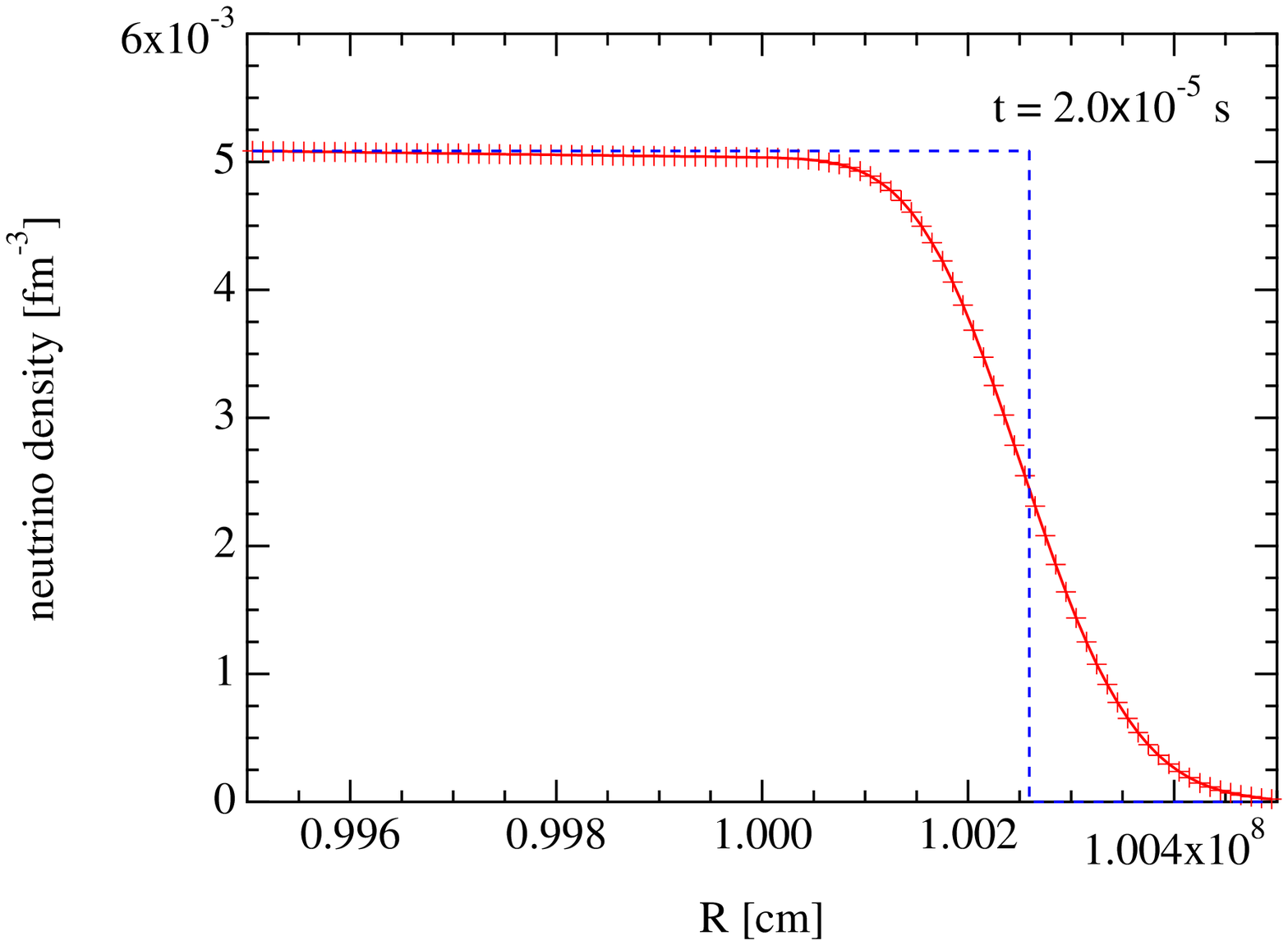}
\caption{1D advection of a step-like distribution.  
The neutrino density as a function of radial coordinate 
is shown by solid lines with cross symbols for the initial time (top), 
the time at $1.0\times10^{-5}$ s (middle), 
$2.0\times10^{-5}$ s (bottom).  
The analytic position of the wave front is shown by dashed lines.  }
\label{fig:1d-adv}
\end{figure}

We next examine the free streaming limit.  
We switch off all neutrino reactions in the following.  
As the simplest test, we compute a 1D advection of neutrinos.  
We utilize a small radial interval located at a large distance.  
We set up a step-like distribution initially (top panel of Fig. \ref{fig:1d-adv}).  
All neutrinos have $\mu_{\nu}=0.93247$, 
which corresponds to the forward grid point for the polar coordinate 
of neutrino momentum whereas they are uniformly distributed 
in the azimuthal direction.  
The numbers of mesh points are 
$N_r=100$, $N_{\theta}=3$, $N_{\phi}=3$ for the spatial grid 
and $N_{\theta_{\nu}}=6$, $N_{\phi_{\nu}}=6$, $N_{\varepsilon}=4$ 
for the momentum space.  
This means that we employ the full 3D code 
to solve the 1D advection in space. 
The time step of $1.0\times10^{-7}$ s is adopted to follow the evolution 
over $2.0\times10^{-5}$ s.  

As shown in Fig. \ref{fig:1d-adv}, 
the step in the neutrino distribution propagates 
radially at the projected light speed, $\mu_{\nu}c$.  
The edge of the step 
smears out gradually owing to numerical diffusions as it propagates.  
The upwind scheme is adopted for the current test (See \ref{scheme-adv}).  

\begin{figure}
\epsscale{.55}
\plotone{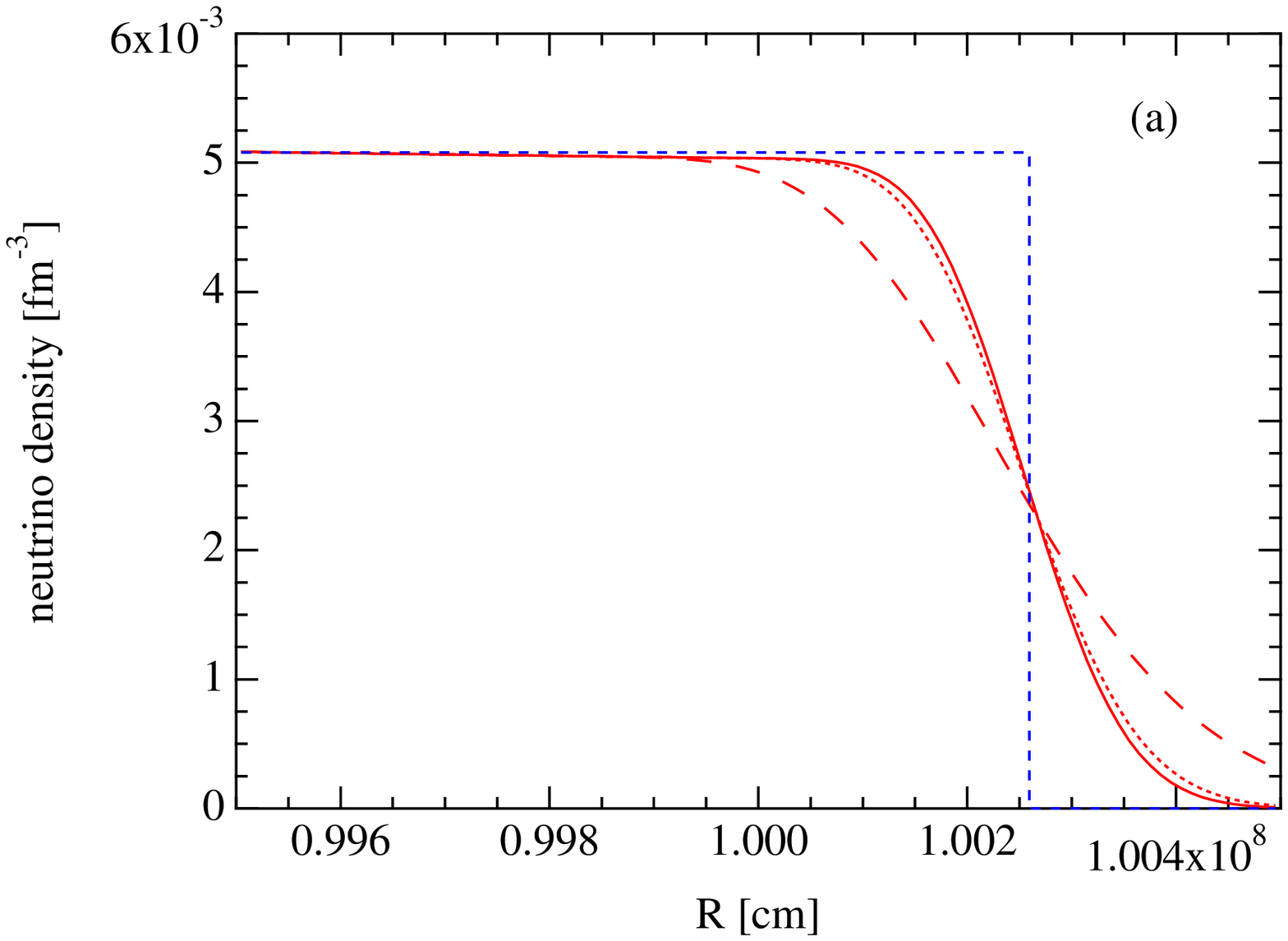}
\plotone{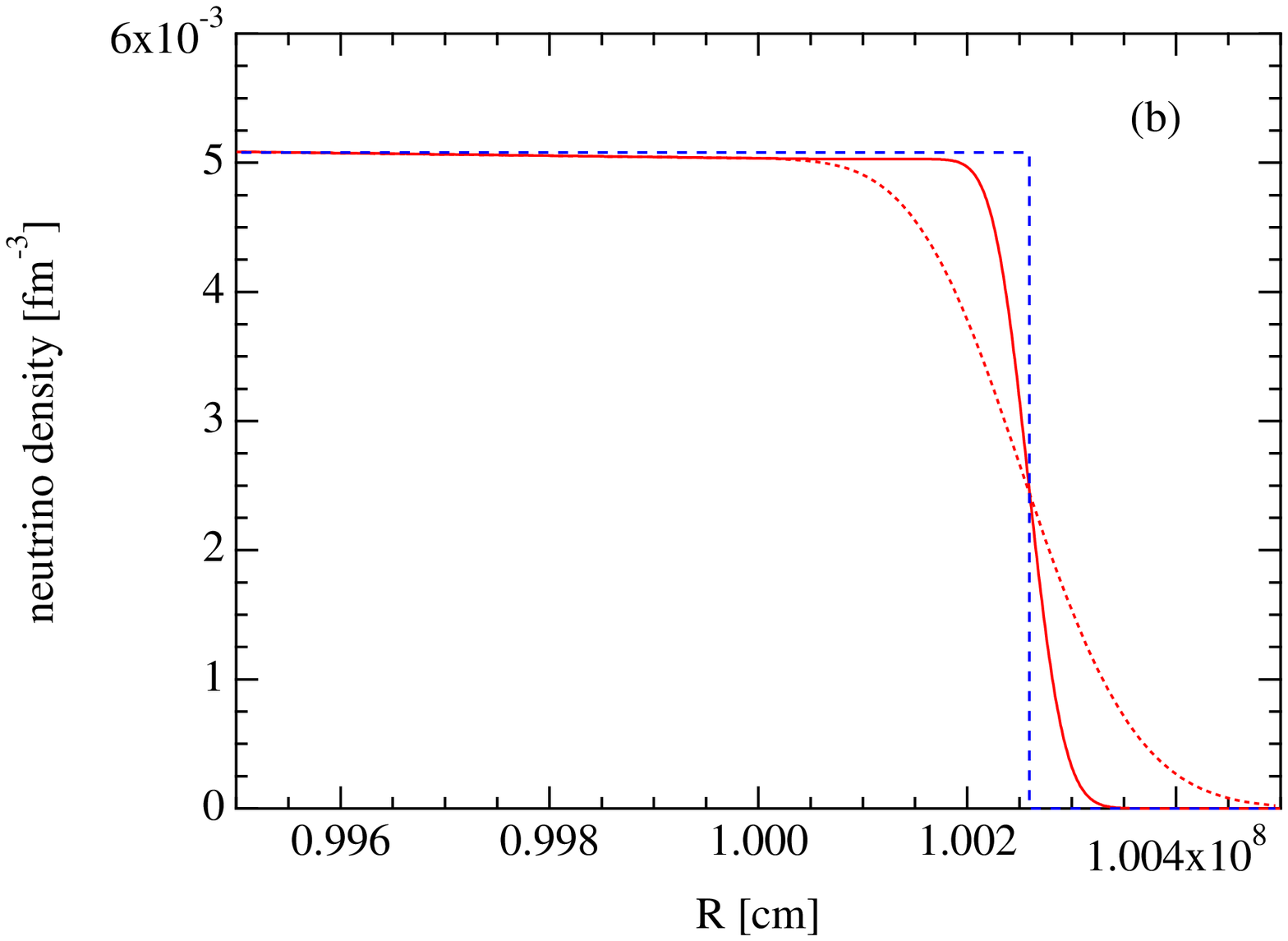}
\caption{Neutrino densities as a function of radial coordinate 
for the time at $2.0\times10^{-5}$ s by the calculations 
with different time steps and resolutions.  
(a) The upper panel shows the different cases of 
$\Delta$t=10$^{-6}$ s (long-dashed line), 
          10$^{-7}$ s (dotted line) and  
          10$^{-8}$ s (solid line) 
for the radial grid with N$_{r}$=100.  
(b) The lower panel shows the cases for the different resolutions 
with 
N$_{r}$=100,  $\Delta$t=10$^{-7}$ s (dotted line) and 
N$_{r}$=1000, $\Delta$t=10$^{-8}$ s (solid line).  
The analytic position of the wave front is shown by dashed lines.  }
\label{fig:1d-adv-check}
\end{figure}

We next confirm that this numerical diffusion is reduced 
when we deploy finer spatial grids and take smaller time steps.  
We show in Fig. \ref{fig:1d-adv-check} the neutrino distributions 
at the final time ($2.0\times10^{-5}$ s, 
corresponding to the bottom panel in Fig. \ref{fig:1d-adv}) 
for different spatial and temporal resolutions.  
In the upper panel, we show the cases for different time steps, 
$\Delta$t=10$^{-6}$ s, 10$^{-7}$ s and 10$^{-8}$ s 
(corresponding to the Courant numbers, 3, 0.3 and 0.03, respectively) 
for the same number of radial grid points of $N_r$=100.  
The smearing becomes apparently smaller when we reduce the time step 
from 10$^{-6}$ s to 10$^{-7}$ s.  
The improvement is not so drastic when we take 10$^{-8}$ s 
instead of 10$^{-7}$ s.  
The further reduction of numerical diffusions is obtained by improving 
the spatial resolution.  
When we increase the number of radial grid points to $N_r$=1000 from $N_r$=100, 
the smearing of the step becomes even smaller as we can see in the lower panel.  
We note that 
the time step is simultaneously reduced by a factor of 10 in this computation 
so that the Courant number remains to be 0.3.  

The smearing of sharp edges in the distribution function as observed above 
is rather common to radiation transfer codes \citep{sto92c,tur01,swe09}.  
The performance of new code is not so bad as it seems, 
being comparable to those obtained by other popular transfer codes.  
For example, the reduction of numerical diffusion with the spatial and 
temporal resolutions in this test is similar to what was shown 
in \S 4.3.3 with Fig. 16 of \citet{swe09}.  
No oscillation (or overshooting) around the edge in our results is found 
as expected for the implicit time-differencing we adopt in our code 
and is consistent with the results reported in \S 6.1 with Fig. 7 
of \citet{sto92c}.  

Although the simple advection of the step-like distribution is done 
as a standard test here, we stress that it is 
too stringent from a view point of core-collapse simulations, since 
we do not obtain such a sharp edge in the neutrino distribution function 
as we will see in \S \ref{1Dconfig}, for example.  
It is true that the smearing of the forward-peaked distribution 
in the optically thin region at large radii in the core is a concern.  
This is a problem more closely connected with the numbers 
of angular grid points in the momentum space as we will discuss in \S \ref{stationary} 
rather than that of the advection scheme, though.  
We note also that it is 
the neutrino transfer and its influences to the hydrodynamics of material 
below the stalled shock wave ($\sim$200 km) 
that we would like to address with the new code.  
Then the forward peak of the neutrino angular distribution is not so appreciable 
and the numerical smearing will be less problematic. 
The neutrino luminosities and spectra at much larger radii 
are certainly important particularly from an observational point of view 
and will be addressed quantitatively by our new code on 
supercomputers of the next generation.  
The current advection scheme is admittedly diffusive and is adopted 
just as a first step in the 3D neutrino transfer, 
which itself has become possible only recently and is in its infancy.  
Exploration of a more sophisticated scheme \citep[][for example]{sto92d} 
is hence a future issue.  

As a final test, we present 3D computations of 
a searchlight beam \citep{sto92c}.  
We note again that this test is also too severe 
for the code intended for supernova simulations 
in spite of its popularity as a benchmark test for radiative transfer codes.  
Indeed, there is no hot spot in the supernova core unlike in the sun 
which would require high angular resolutions.  
The main purpose of this test is to 
examine whether the code can give a correct propagation velocity of neutrinos, 
that is, the light speed in the genuinely 3D setting.  
We inject the neutrino beam 
with $\mu_{\nu}=-0.90412$ and $\phi_{\nu}=6.1771$ radian 
at a certain point on the boundary.  
The numbers of mesh points employed for this test are 
$N_r=50$, $N_{\theta}=48$, $N_{\phi}=48$ for the spatial grid 
and $N_{\theta_{\nu}}=12$, $N_{\phi_{\nu}}=12$, $N_{\varepsilon}=4$ 
for the momentum space.  

\begin{figure}
\epsscale{0.6}
\plotone{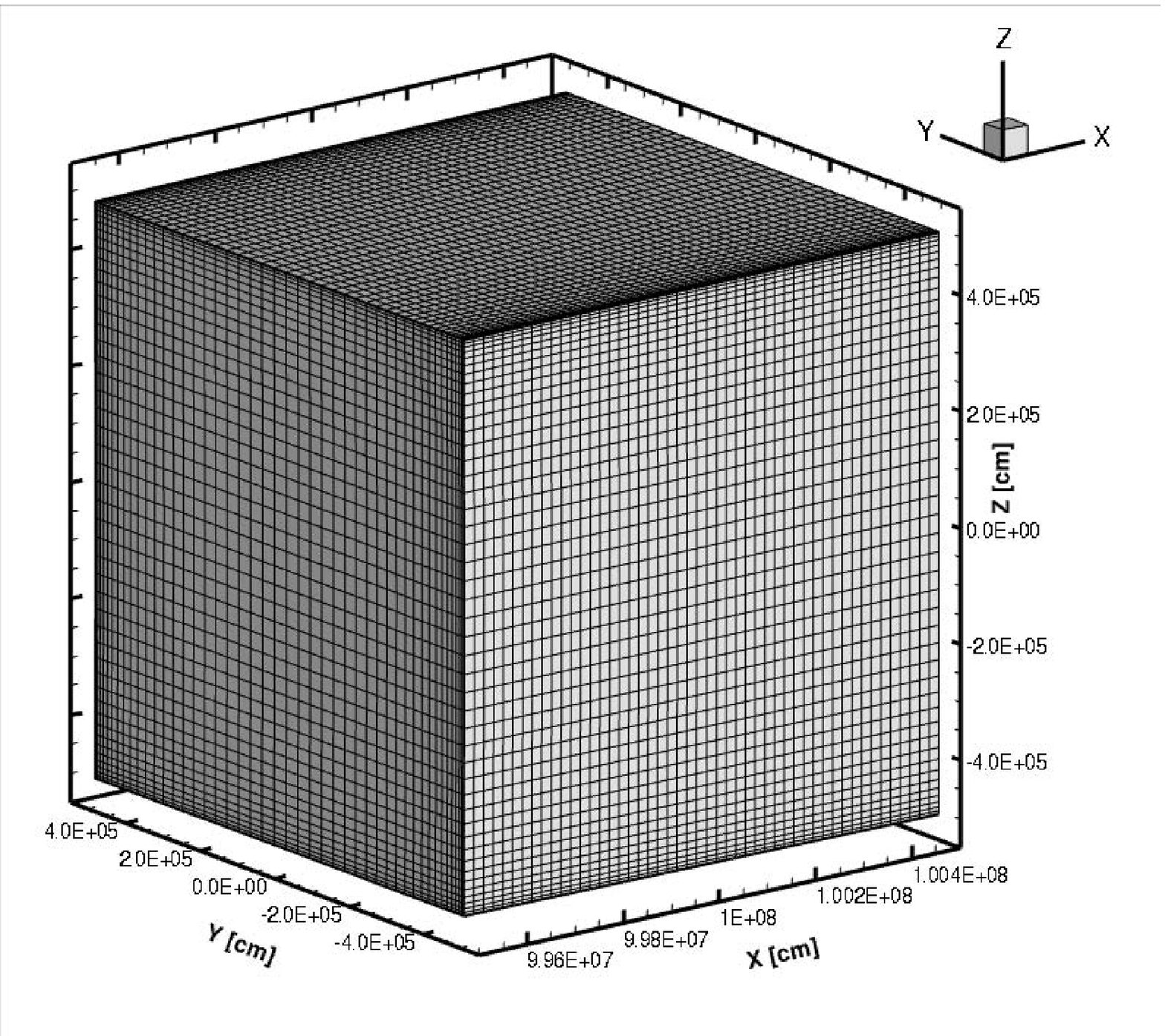}
\plotone{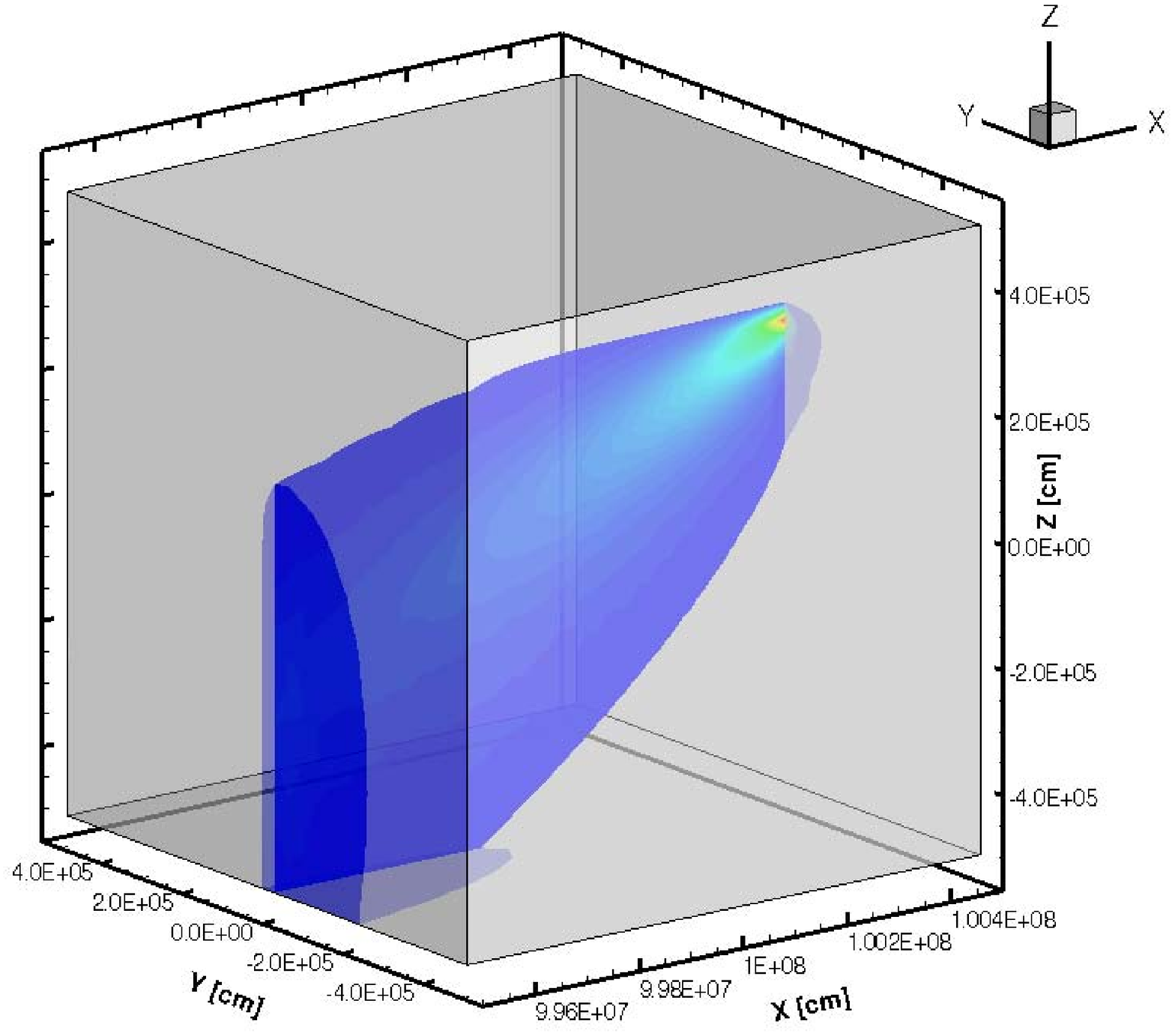}
\caption{Searchlight beam test in the 3D box.  
The neutrino densities are shown by color code in the lower panel.  
The source of neutrinos to one direction is set 
at the right boundary of the box.  
The beam propagates along the direction 
with a gradual spread due to the numerical diffusion.  
The numerical grid of the 3D box is shown in the upper panel.  }
\label{fig:3d-beam}
\end{figure}

In Figure \ref{fig:3d-beam}, we show the resultant neutrino densities 
in a 3D box with a side of $10^{6}$~cm at the time of $\sim$1$\times$10$^{-2}$ s 
when they become almost steady.  
We can see that the beam propagates along the designated direction, 
while the beam becomes broader as it propagates because of numerical diffusions.  
It is to be noted that the diffusion is mainly due to the small numbers 
of angular grid points in the momentum space rather than due to 
the relatively coarse spatial resolution.
(see the employed grid in the 3D box presented in the upper panel).  
This is different from the situation for the 1D advection discussed earlier.  
We also find that the beam shape is deformed as it propagates 
owing to the non-uniform angular grid in the momentum space along the beam.  
These results indicate that 
it is desirable in principle to deploy an angular mesh in the momentum space 
that is as fine as possible and 
covers the whole solid angle uniformly \citep[See][for example]{ott08}.  




\subsection{Collision Term}\label{collision}

In order to examine the collision term 
for the neutrino reactions in supernova cores, 
we investigate test cases with realistic profiles 
using the neutrino reaction rates described in \S \ref{reaction}.  
We concentrate here on basic tests under spherical symmetry.  
We will explore further the neutrino transfer 
inside the supernova core in \S \ref{applications}.  

As a typical situation, 
we utilize supernova profiles from the spherical calculation \citep{sum05}.  
We take the profiles of density, temperature and electron fraction 
at 100 ms after the core bounce in the collapse of the 15M$_{\odot}$ star 
(See also \S \ref{1Dconfig}).  
We adopt the radial grid points ($N_r=200$) for $r=0 \sim 1.4 \times 10^{3}$ km 
from the original profile.  
We cover the first octant of the sphere 
with the angular grid, though we treat a spherical profile.  
The neutrino energy and angular grids are determined by 
following the setting in the spherical simulations of supernovae 
by \cite{sum05}.  
The energy grid ($N_{\varepsilon}=14$) is placed logarithmically 
from 0.9 MeV to 300 MeV with a fine grid for high energy tails.  
%

\subsubsection{Stationary Case}\label{stationary}

We examine stationary cases through comparisons with 
the formal solutions of neutrino transfer.  
For this purpose, we take into account absorptions and emissions 
in the collision term and switch off scatterings.  
We check the neutrino propagation with the neutrino reactions 
in dense matter along a certain ray.  
This covers the intermediate regime between the opaque and 
transparent regimes.  
Therefore, this is a complementary check to the tests 
in the diffusion and free-streaming limits.  

We evaluate the neutrino distribution at a certain location 
by the formal solution \citep{mih99}.  
The neutrino distribution, $f(s)$, at the position, $s$, 
along the ray follows the transfer equation given by,
\begin{equation}
\label{eqn:formaleq}
-\frac{d}{ds}f(s) = - \chi(s) f(s) + \eta(s), 
\end{equation}
where $\chi(s)$ and $\eta(s)$ are the opacity and the emissivity, 
respectively.  
The path length, $s$, is measured backward 
(the opposite to the neutrino direction $\mbox{\boldmath $n$}$) 
from the position ($s=0$ at $r$ to obtain $f(\mbox{\boldmath $r$})$) 
to the outer boundary ($s=s_{bc}$ at $r=R_{out}$).  
The neutrino distribution at the position, $\mbox{\boldmath $r$}$, 
is obtained by the integral,
\begin{equation}
\label{eqn:formalsol}
f(s) = \int_{0}^{s_{bc}} e^{-\tau(s')}\eta(s') ds',
\end{equation}
along the ray designated by $\mu_{\nu}$ and $\phi_{\nu}$.  
The optical depth, $\tau(s)$, along the ray is given by, 
\begin{equation}
\label{eqn:formaltau}
\tau(s) = \int_{0}^{s} \chi(s') ds'.
\end{equation}
We assume in deriving Eq. (\ref{eqn:formalsol}) 
the incoming neutrino flux is zero at the outer boundary.  

We integrate numerically the above equations 
using the rates of emission and absorption on the grid points 
along the ray of neutrino propagation.  
At the same time, we perform computations of 
the neutrino transfer by solving the Boltzmann equation in 3D 
for a sufficiently long period to obtain a steady state solution.  
We fix the spherical profile of the supernova core 
at the post-bounce mentioned above 
by the spatial grid 
with $N_{r}=200$, $N_{\theta}=3$ and $N_{\phi}=3$.  
We set the neutrino angle grid 
with $N_{\theta_{\nu}}=6$, 12, 24 and $N_{\phi_{\nu}}=6$.  
We treat here the electron-type neutrino only.  

\begin{figure}
\epsscale{.80}
\plotone{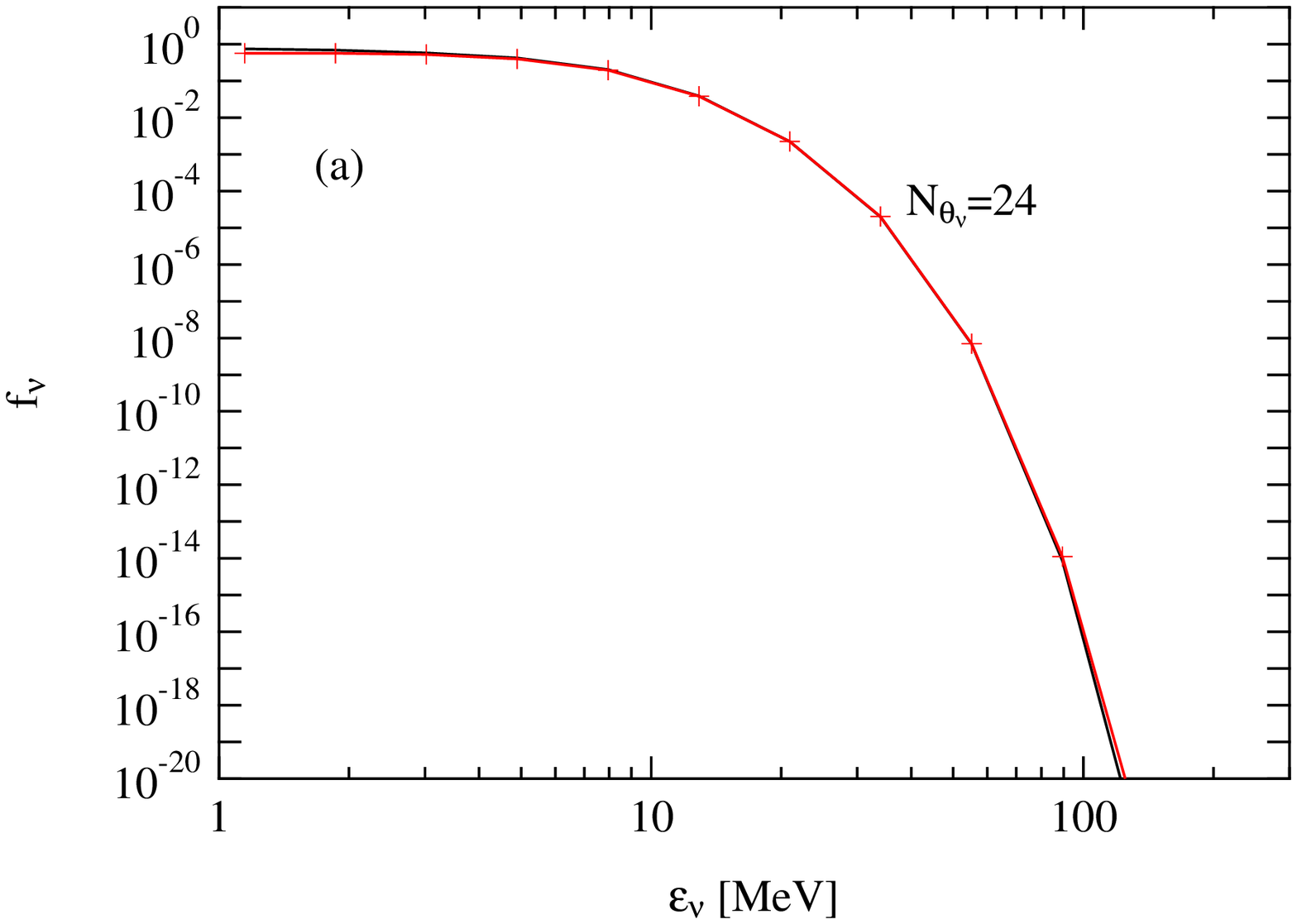}
\plotone{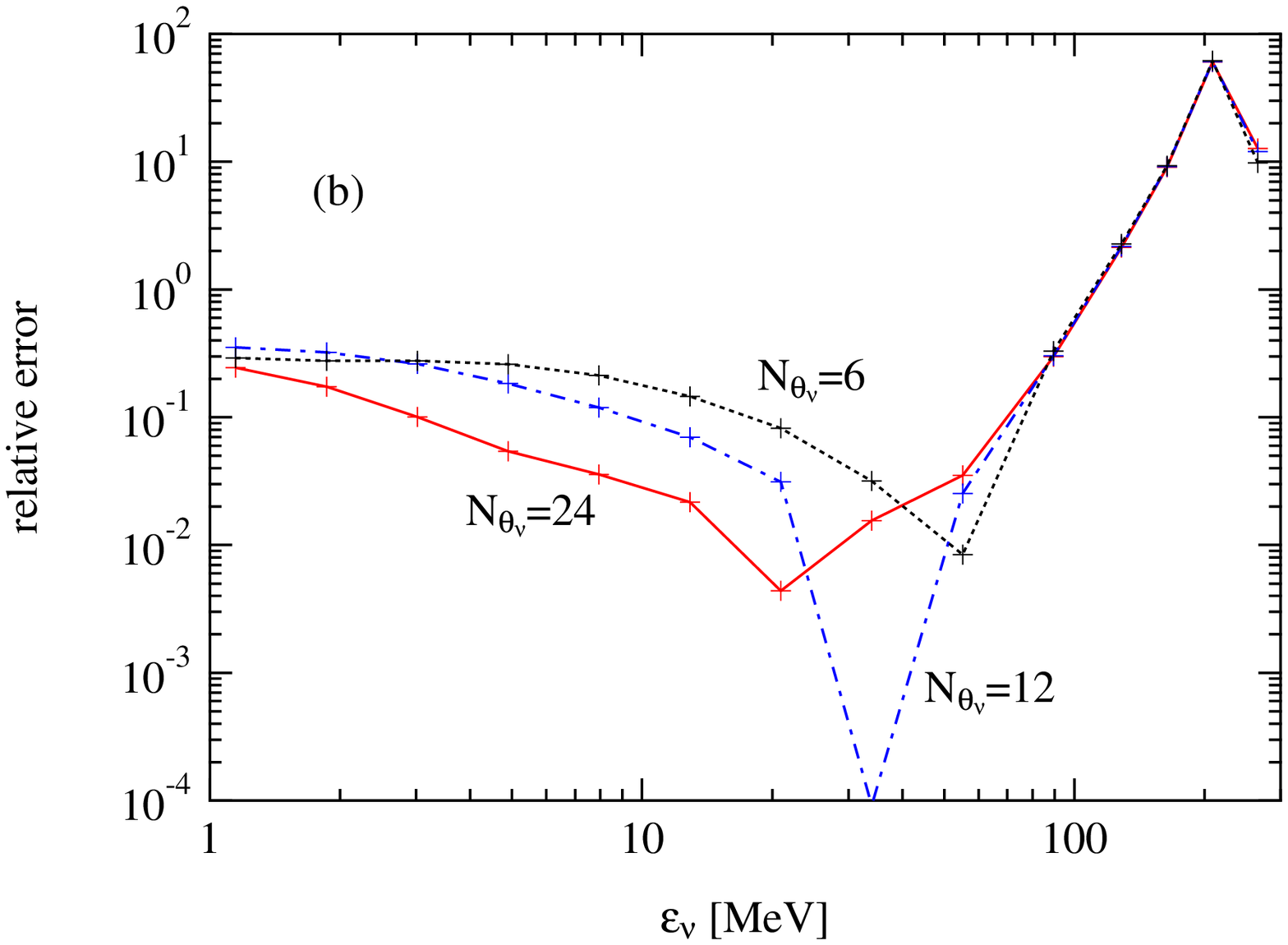}
\caption{(a) Energy spectra at the radial position of 98.4 km.  
The solid line and the solid line with symbol denote the spectra 
obtained by the formal solution and the computation 
($N_{\theta_{\nu}}=24$), respectively.  
(b) Relative errors of the spectra by the computation 
with respect to the formal solutions.  
The dotted, dot-dashed and solid lines denote the numerical results 
for $N_{\theta_{\nu}}=6$, 12 and 24, respectively.  }
\label{fig:formal}
\end{figure}

We compare the energy spectra obtained by the two methods 
in Fig. \ref{fig:formal} (a).  
We plot the spectra at the radial position of $r=$98.4 km 
for the neutrino direction with $\mu_{\nu}=0.99519$ (the most forward grid point) 
for $N_{\theta_{\nu}}=24$.  
The distributions accord well with each other for the wide range 
of neutrino energy, 
though there is a slight deviation at low and high energies.  
In Fig. \ref{fig:formal} (b), we show relative errors 
of the distributions by the computation with respect to the formal solutions.  
For the case of $N_{\theta_{\nu}}=24$, 
errors are less than 30$\%$ for neutrino energies, $\le100$ MeV.  
Errors become large for high energy tails above 200 MeV 
since tails of distributions become fairly small below 10$^{-20}$.  
As the energy goes below $\sim$3 MeV, errors increase beyond 10$\%$.  
This is because the matter becomes transparent for low energy neutrinos 
with small cross sections.  
A high angular resolution is necessary 
to follow the propagation of neutrinos 
in the transparent situation.  
For the medium range of energy, 
which is most important in the supernova study, 
the transition from the opaque central core to the transparent outer layers 
is well described by the computation.  

We next examine the angular resolution by changing $N_{\theta_{\nu}}$.  
In Fig. \ref{fig:formal} (b), 
we plot also the cases of $N_{\theta_{\nu}}=6$ and 12 
for comparison.  
The neutrino directions at the most forward grid points 
are $\mu_{\nu}=0.93247$ and 0.98156, respectively, in these cases.  
As $N_{\theta_{\nu}}$ increases, errors become small in principle, 
showing the improvement of the angular resolution.  
Although $N_{\theta_{\nu}}=24$ (or larger) is preferable 
for the precise evaluation of forward-peaked distributions, 
$N_{\theta_{\nu}}=6$ is sufficient to obtain errors less than 40$\%$. 
Note that $N_{\theta_{\nu}}=6$ is the minimally proper size 
for the supernova study as checked by the detailed tests \citep{yam99} 
and is adopted for the spherical calculations of core-collapse 
supernovae so far \citep[for example]{sum05}.  

The angular resolution to describe the peaked distribution 
is the intrinsic problem of the S$_{n}$ method in the field 
of radiation hydrodynamics. 
In the current problem of core-collapse supernovae, though, 
it is rather important to describe the phenomena around 
$\sim$200 km, 
where the shock wave is hovering, with the neutrino emission 
from the central core ($\sim50$ km).  
Therefore, 
the resolution for the angle factor of 
$\sim$0.25 
may be sufficient 
to describe the phenomena such as neutrino heating, 
at least for the first trial in 3D simulations.  
This is totally different from the situation in the solar physics, 
for example, where a small hot spot may be crucial for the radiation 
in outer layers at very large distance.  
Under the reasoning described above, 
we adopt $N_{\theta_{\nu}}=6$ for the following study 
of supernova cores (\S \ref{applications})
within the currently available computing resources.  

\subsubsection{Time Evolution toward Equilibrium}\label{equilibrium}

We examine the time evolution of the neutrino distribution 
toward the equilibrium determined by the condition of dense matter.  
We check the time scale to reach the equilibrium 
and the detailed balance under the chemical and thermal equilibrium 
through the neutrino reactions.  
We follow the evolution from an initial neutrino distribution 
with small values ($10^{-5}$ times the equilibrium value) 
to the equilibrium state.  
In the static background, 
the time evolution of the neutrino distribution at a certain energy grid point 
by the absorption and emission can be analytically expressed as 
$f=(f_{0} - f_{FD}) e^{-\frac{t}{\tau}} + f_{FD}$, 
where $f_{0}$ and $f_{FD}$ are 
the initial and equilibrium (Fermi-Dirac) values, respectively.  
The time scale, $\tau$, is given by $\tau=\frac{\lambda}{c}$ 
with the effective mean free path, $\lambda$.  
The effective mean free path is defined here by 
the inverse of the true absorption coefficient, $\kappa$.  
The true absorption coefficient is given by the sum of contributions 
\begin{equation}
\label{eqn:kappa-abs}
\kappa_{abs} = R_{abs}(\varepsilon, \Omega) [ 1 + e^{-\beta (\varepsilon - \mu_{\nu})} ] ~,
\end{equation}
for electron captures and neutrino absorptions and 
\begin{equation}
\label{eqn:kappa-pair}
\kappa_{pair} = 
 \int \frac{d \varepsilon' \varepsilon'^2}{(2 \pi)^3}~
\int d \Omega'~
R_{pair-emis}(\varepsilon, \Omega; \varepsilon', \Omega')
\left[
e^{\beta (\varepsilon + \varepsilon')}
\overline{f}(\varepsilon', \Omega')  \\
+ 
[ 1 - \overline{f}(\varepsilon', \Omega') ] 
\right],
\end{equation}
for the pair processes.  
Since we consider the iso-energetic scattering, the neutrino 
scattering contributes only to the realization of isotropy.  
The scattering coefficient, $\sigma_{scatt}$, is given by 
\begin{equation}
\label{eqn:kappa-scatt}
\sigma_{scatt} = 
\frac{\varepsilon^2}{(2 \pi)^3}~
\int d \Omega'~R_{scat}(\Omega; \Omega') ~, 
\end{equation}
which is larger than the absorption coefficient in this case.  
The effective mean free path due to the scattering process is short 
enough to realize the isotropic distribution within a short time period.  
We utilize the profile of the supernova core at 100 msec after the bounce 
as a background and choose a central grid point for the test.  
The neutrino reactions listed in \S \ref{reaction} are all included.  

\begin{figure}
\epsscale{.80}
\plotone{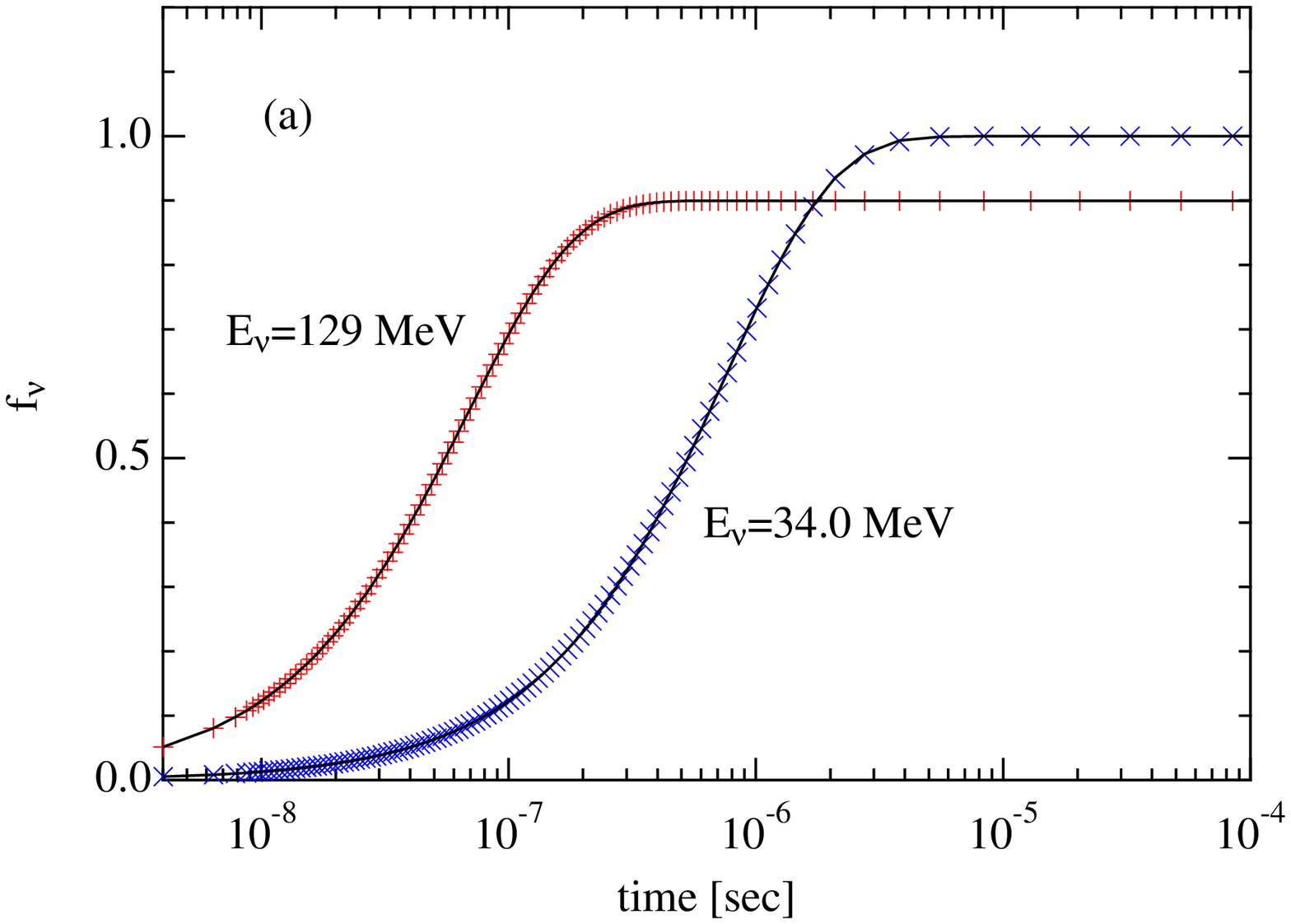}
\plotone{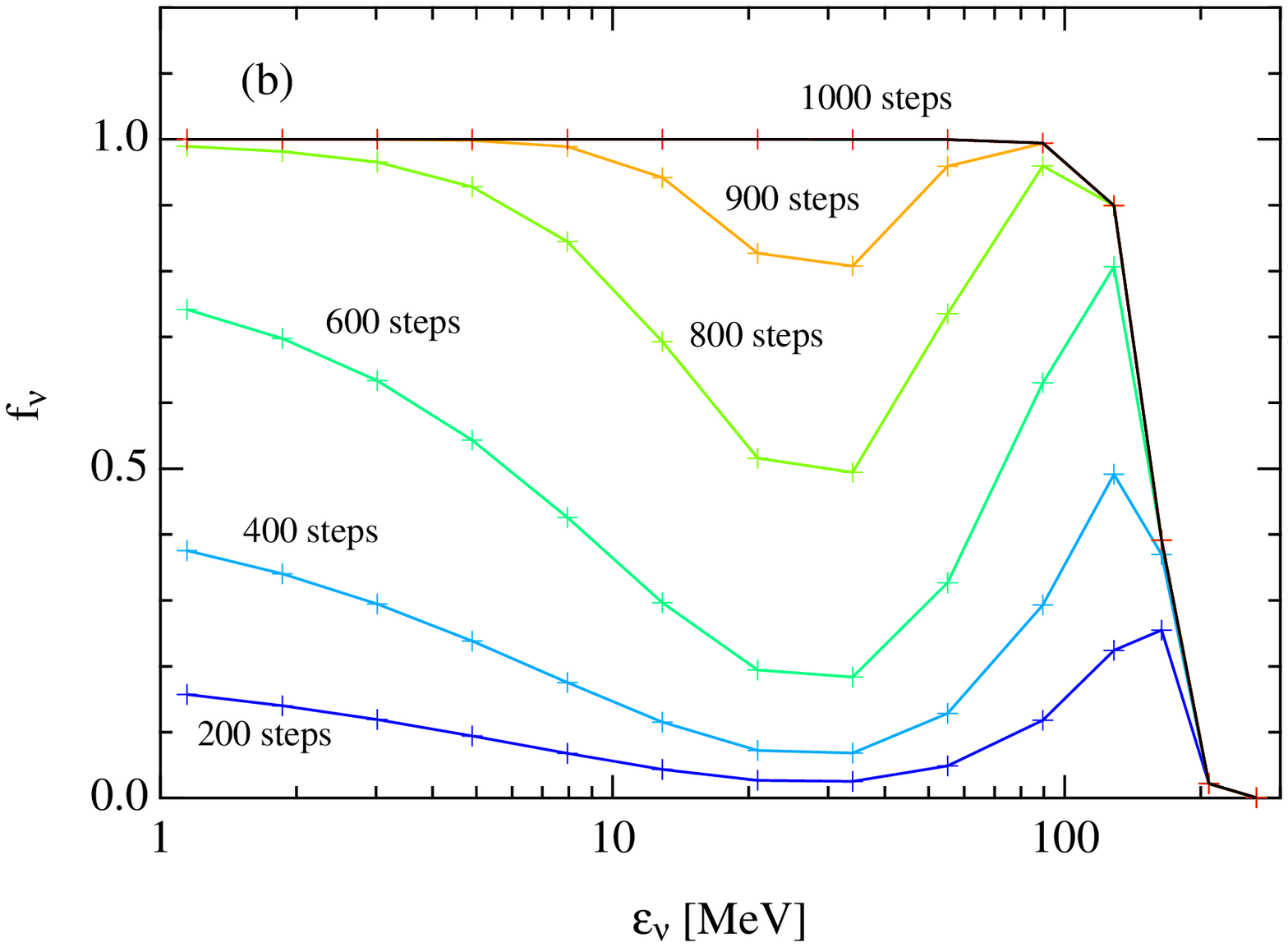}
\caption{(a) Time evolution of the neutrino populations at the two energy grid points 
toward the equilibrium value for the dense matter in the central core.  
(b) Evolution of the neutrino spectra at selected time steps toward the equilibrium.  
The lines correspond to the time steps, 
2.0$\times$10$^{-8}$, 5.4$\times$10$^{-8}$, 1.6$\times$10$^{-7}$, 
5.2$\times$10$^{-7}$, 1.3$\times$10$^{-6}$, 3.3$\times$10$^{-5}$ sec 
from bottom to top.  
The solid line expresses the Fermi-Dirac distribution.  }
\label{fig:equil}
\end{figure}

In Figure \ref{fig:equil}, we show the time evolution of the neutrino 
populations of $\nu_e$ for the energy grid points at $E_{\nu}$= 34.0 and 129 MeV 
at the center of the supernova core.  
The density, temperature and neutrino chemical potential are 
3.15$\times$10$^{14}$ g/cm$^{3}$, 13.4 MeV and 158 MeV, respectively.  
The time evolution in the computation is well in accord with 
the analytic solution.  
Time steps can be very long (over 1 sec) once the distribution 
reaches the chemical equilibrium owing to the implicit treatment.  
We show also the energy spectra in Fig. \ref{fig:equil}.  
The spectrum evolves toward the equilibrium and 
reaches the Fermi-Dirac distribution determined 
by the temperature and chemical potential.  
We note that the angle distribution becomes isotropic 
through scatterings during the evolution.  

We check also that the approximation for the pair process, 
which is taken for the linearization, is appropriate 
in the realistic profile of supernova core.  
In the central part of core, where the temperature is high, 
the neutrino distribution approaches soon the thermal equilibrium 
as shown above.  
The blocking factor in the reaction rate for the pair process 
can be hence expressed by the thermal distribution or 
the distribution at the previous step.  
The effective mean free paths due to the pair processes, 
Eqs. (\ref{eqn:nureactionpair1}) and (\ref{eqn:nureactionpair2}), 
in the new code are compared with those from the spherical calculation.  
They accord very well with each other once the thermal equilibrium 
is maintained after a short period (See \S \ref{1Dconfig}).  
This time scale is much shorter than the hydrodynamical time scale ($\sim$ms), 
therefore, the approximation can be safely used in the dynamical 
simulations of core-collapse supernovae.  

\section{Applications}\label{applications}

We investigate here the performance of our code 
for realistic profiles taken from supernova cores.  
We first employ two spherically symmetric core profiles, 
one during the collapse and the other for the post-bounce stage.  
In 2D and 3D cases, we deform by hand an originally spherically 
symmetric core profile rather arbitrarily 
and investigate the neutrino transfer in the non-spherical settings.  
We demonstrate that the new code can describe such features as 
fluxes and Eddington tensors in a qualitatively correct way both 
in 2D and 3D.  
We also use the formal solutions for more quantitative assessments.  

\subsection{1D Configurations}\label{1Dconfig}

We examine first the neutrino transfer under spherical symmetry 
through comparisons with the numerical results 
by the neutrino-radiation hydrodynamics in general relativity \citep{sum05}.  
Adopting the profiles of supernova cores, 
we follow the time evolution of the neutrino distributions 
from small initial values until they reach a steady state.  
We treat the three neutrino species, 
$\nu_e$, $\bar{\nu}_e$ and $\nu_{\mu}$ and 
implement the neutrino reactions listed in \S \ref{reaction}.  

\begin{figure}
\epsscale{1.0}
\plottwo{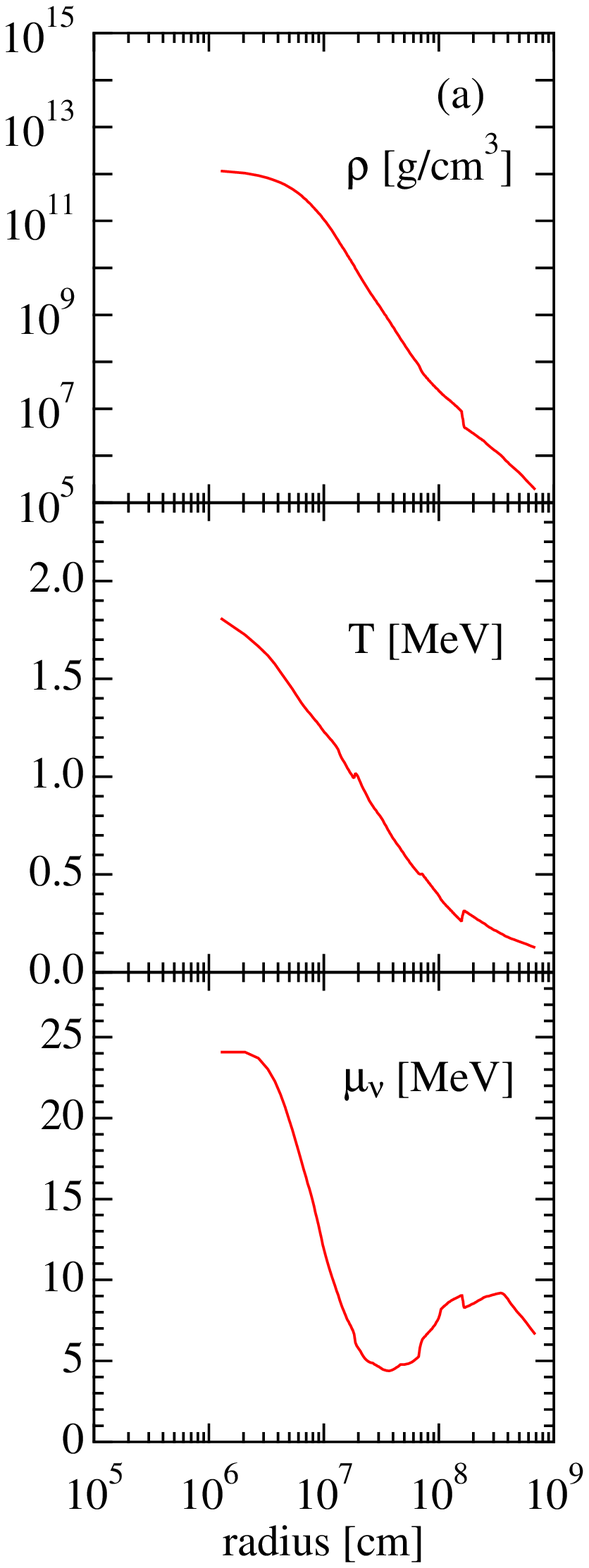}{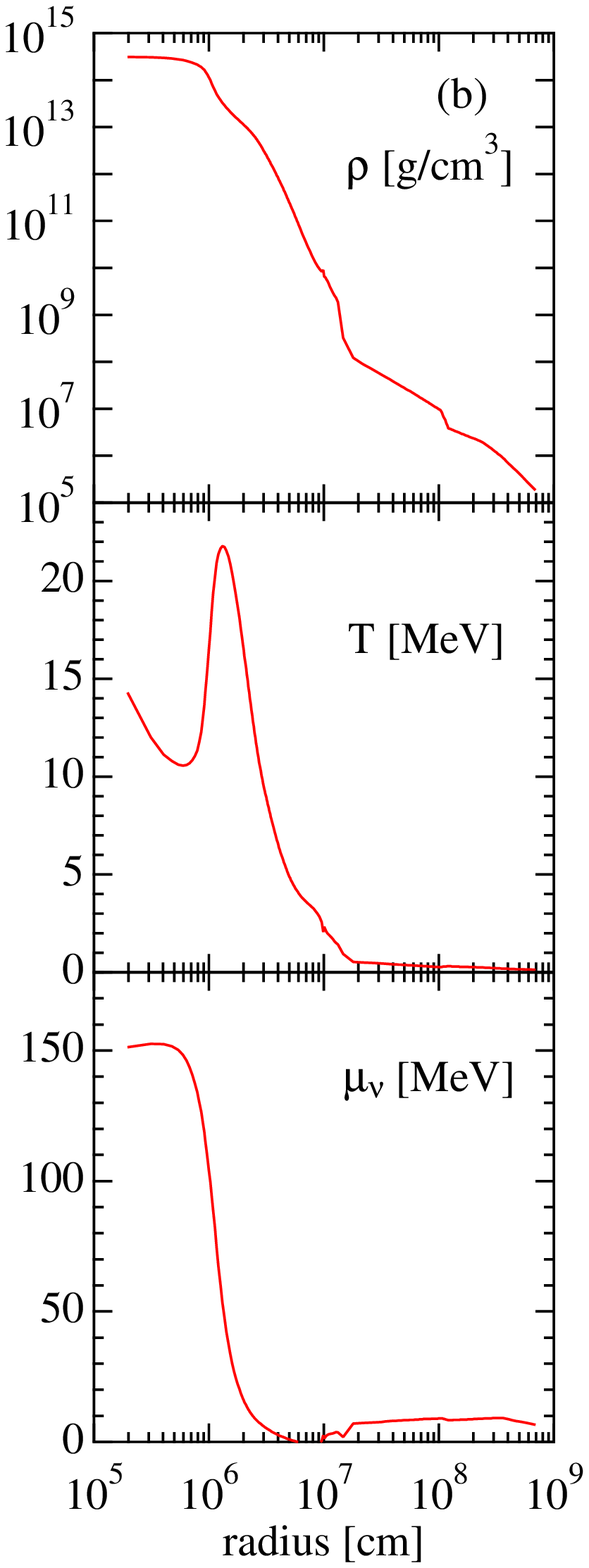}
\caption{Radial profiles of the density, temperature 
and neutrino chemical potential in the supernova core 
(a) when the central density is 10$^{12}$ g/cm$^{3}$ during the collapse 
and (b) at 100 ms after the bounce in the spherical simulation 
by \citet{sum05}.  }
\label{fig:1d-profile}
\end{figure}

Figure \ref{fig:1d-profile} shows the adopted profiles 
of the supernova core in the gravitational collapse of 
the 15 M$_{\odot}$ star \citep{woo95} from \citet{sum05}.  
The snapshots are taken at the timings 
when the central density is 10$^{12}$ g/cm$^{3}$ during the collapse 
and at 100 ms after the core bounce.  
The former is an example of the situation during the collapse, 
where the neutrinos are trapped by the reactions with nuclei.  
The latter is a typical situation of the stalled shock wave 
after the bounce, where the neutrino heating takes place 
by the neutrino flux from the central core.  
We set the spatial grid 
with $N_{r}=200$, $N_{\theta}=5$ and $N_{\phi}=5$ in the first octant 
and take the radial grid points from the original profiles 
as done in \S \ref{collision}.  
The neutrino angle and energy grids are set with 
$N_{\theta_{\nu}}=6$, $N_{\phi_{\nu}}=12$ and $N_{\varepsilon}=14$.  

\begin{figure}
\epsscale{1.0}
\plottwo{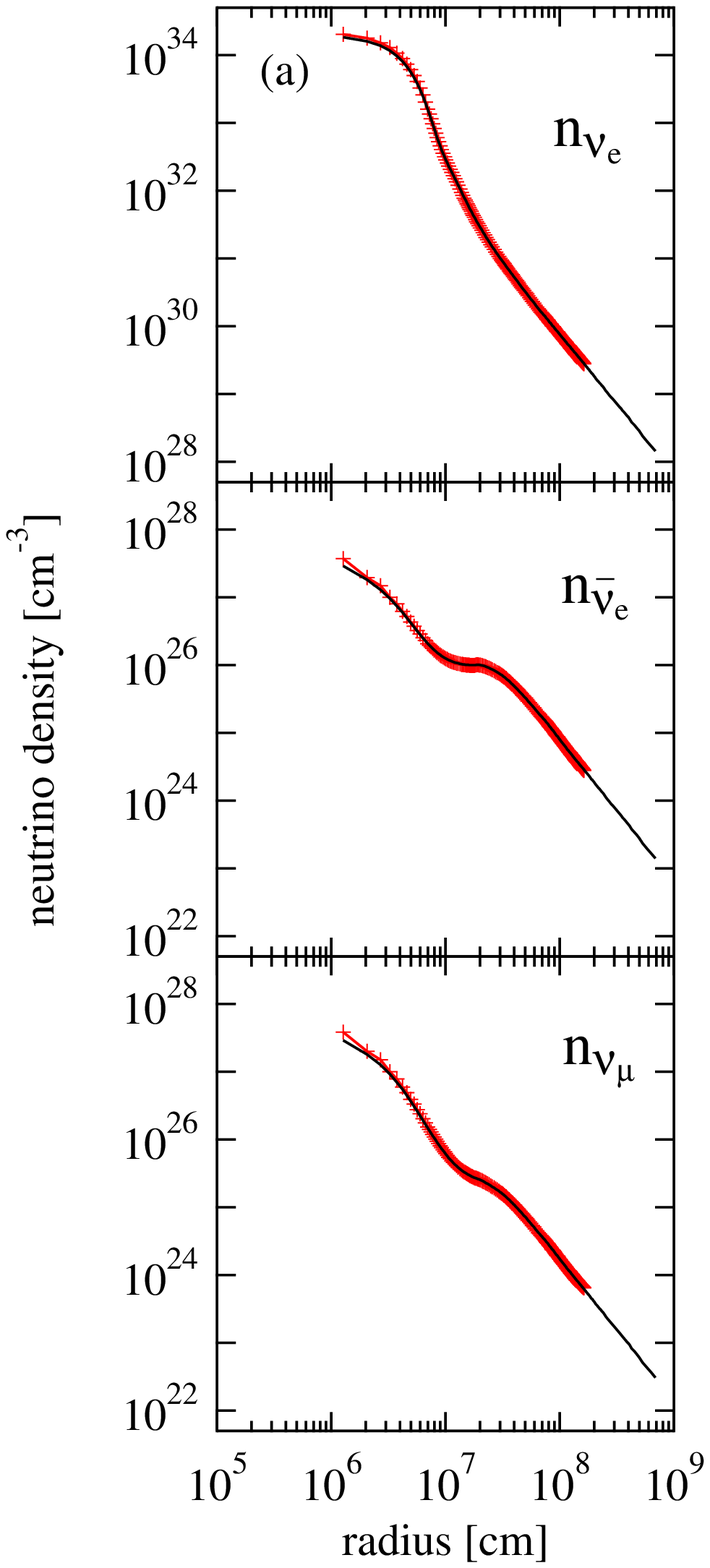}{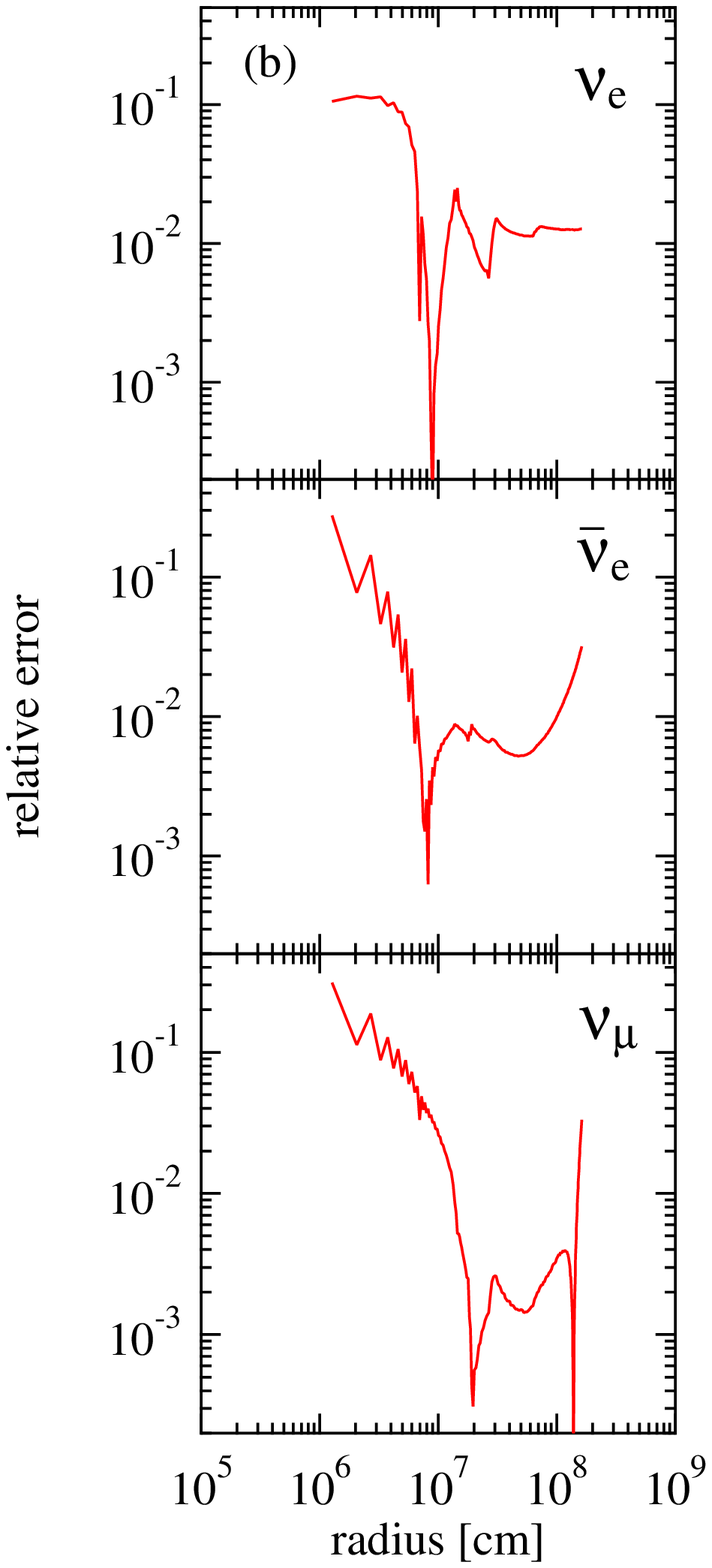}
\caption{(a) Radial distributions of neutrinos for the profile 
at the central density of 10$^{12}$ g/cm$^{3}$ during the collapse.  
The neutrino densities obtained by the 3D code are shown 
by cross symbols for three species.  
The densities from the spherical calculation 
(see the main text) are plotted by the solid lines.  
(b) Relative errors of the densities by the 3D code 
with respect to those by the 1D code.  }
\label{fig:1d-nudens.rhoc12}
\end{figure}

\begin{figure}
\epsscale{1.0}
\plottwo{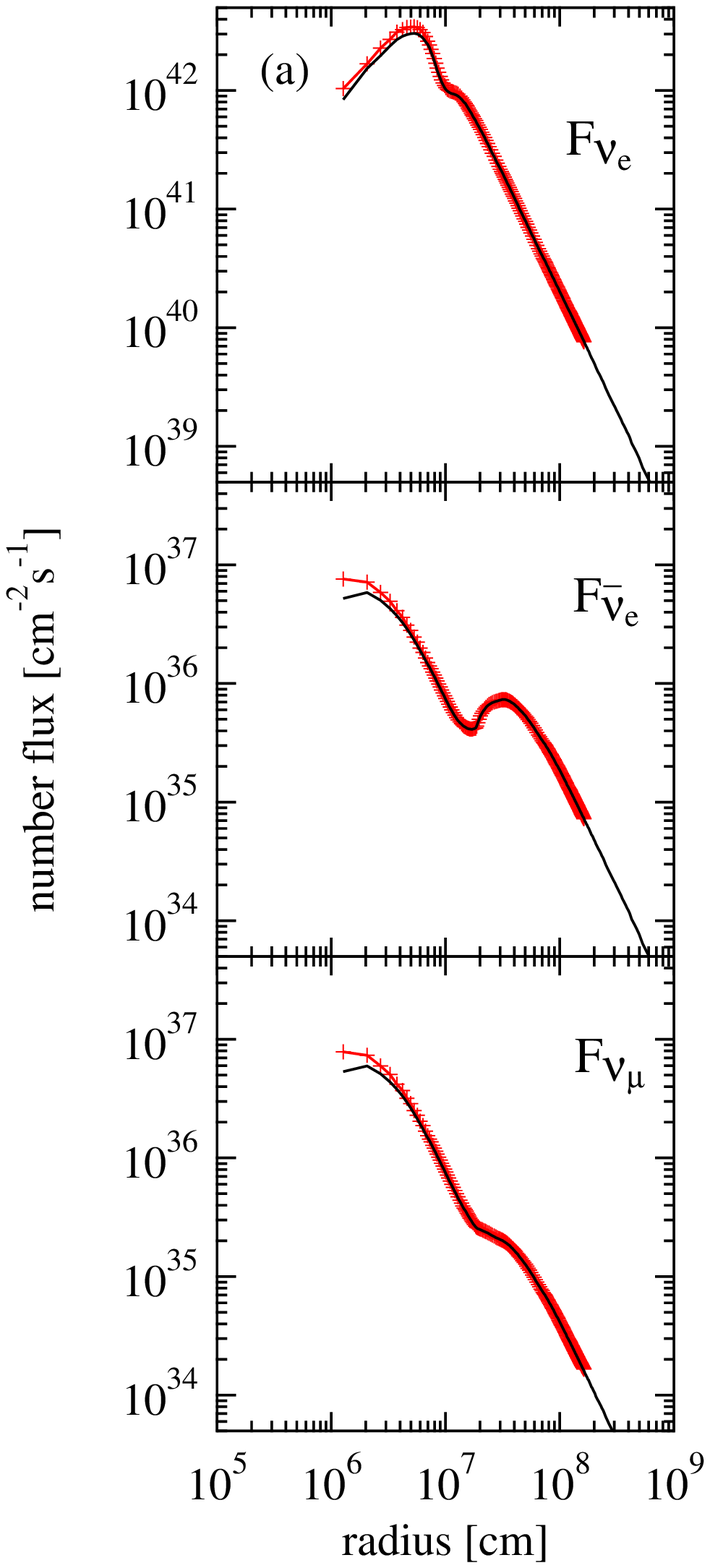}{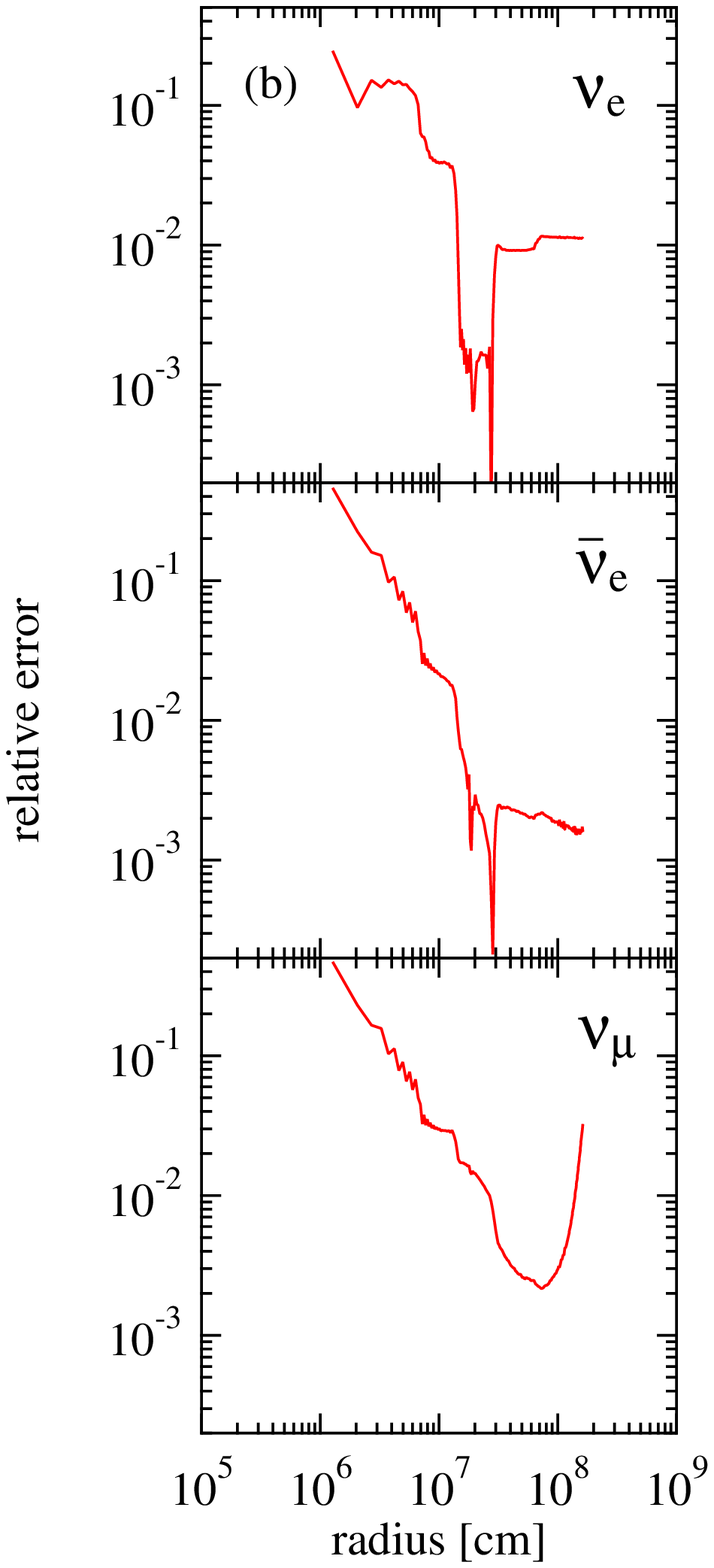}
\caption{(a) Neutrino number fluxes for the profile 
at the central density of 10$^{12}$ g/cm$^{3}$ during the collapse.  
The neutrino fluxes obtained by the 3D code are shown 
by cross symbols for three species.  
The neutrino fluxes from the spherical calculation 
(see the main text) are plotted by the solid lines.  
(b) Relative errors of the fluxes by the 3D code 
with respect to those by the 1D code.  }
\label{fig:1d-nuflux.rhoc12}
\end{figure}

\begin{figure}
\epsscale{1.0}
\plottwo{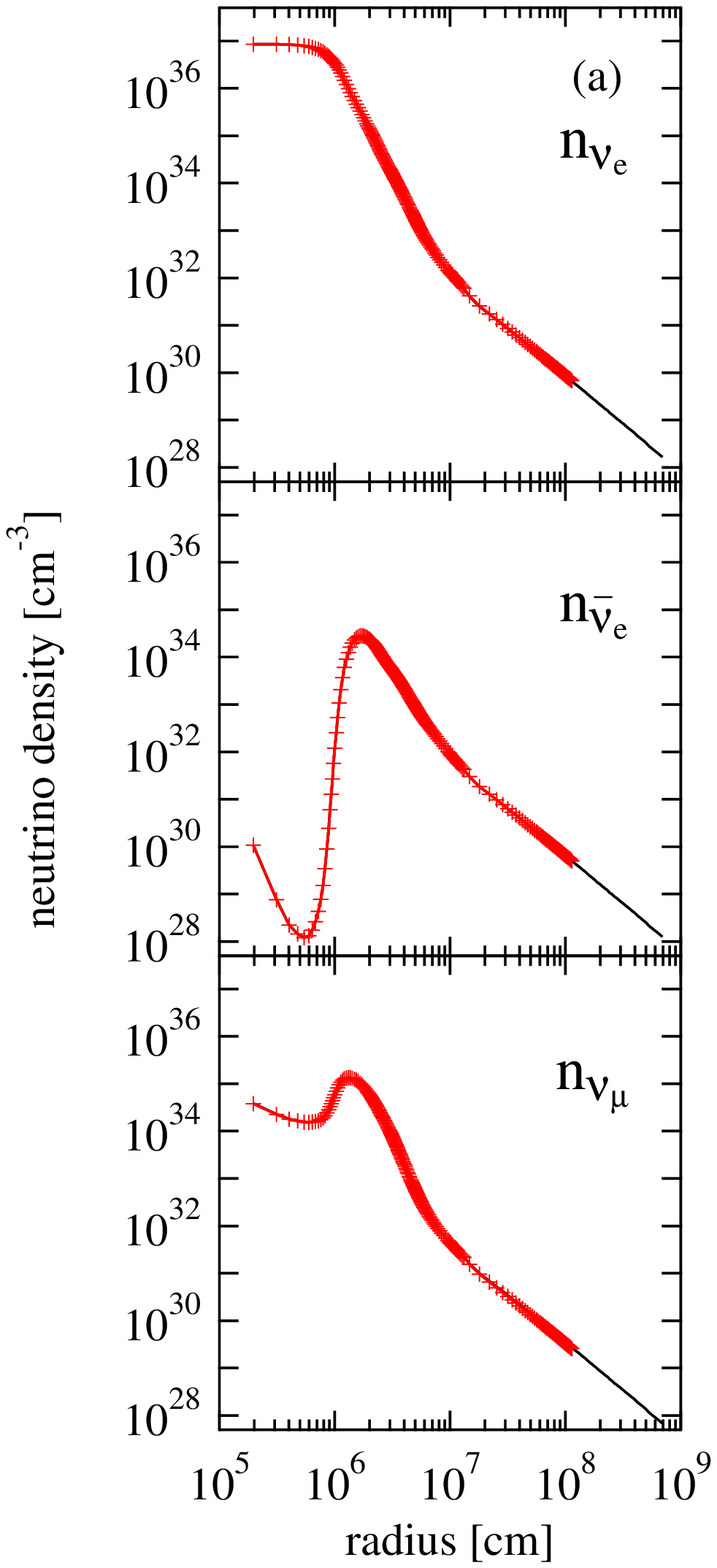}{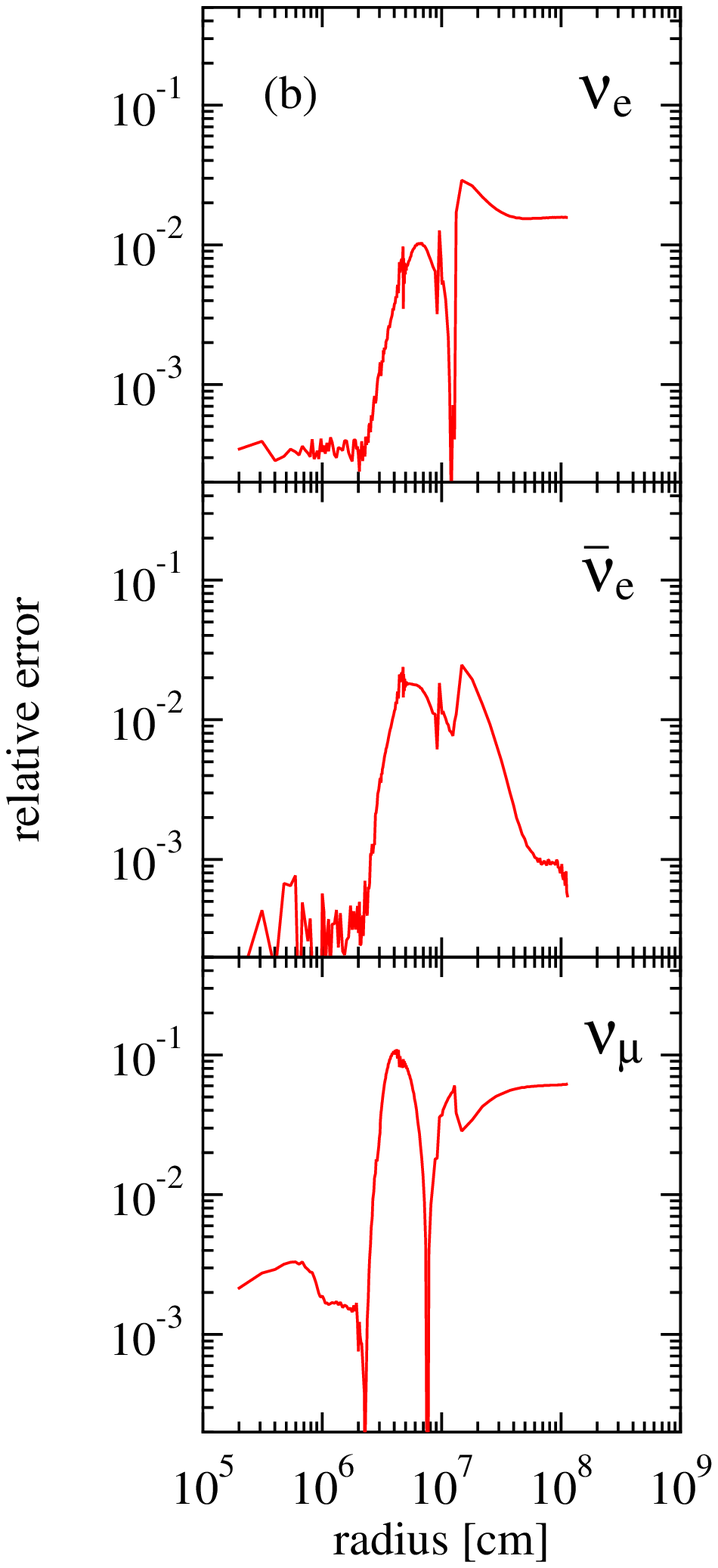}
\caption{Same as Fig. \ref{fig:1d-nudens.rhoc12}, 
but for the profile at 100 ms after the bounce.  }
\label{fig:1d-nudens.tpb100}
\end{figure}

\begin{figure}
\epsscale{1.0}
\plottwo{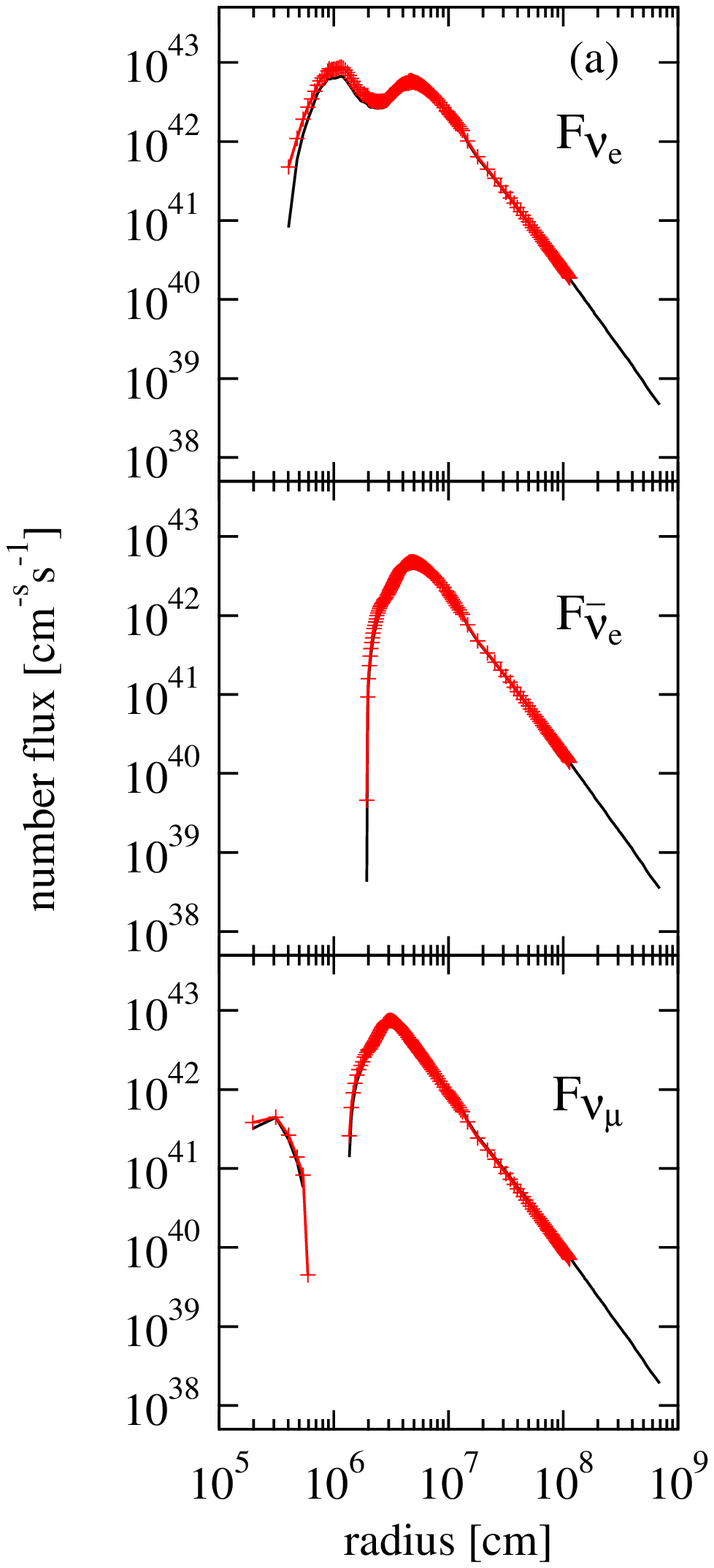}{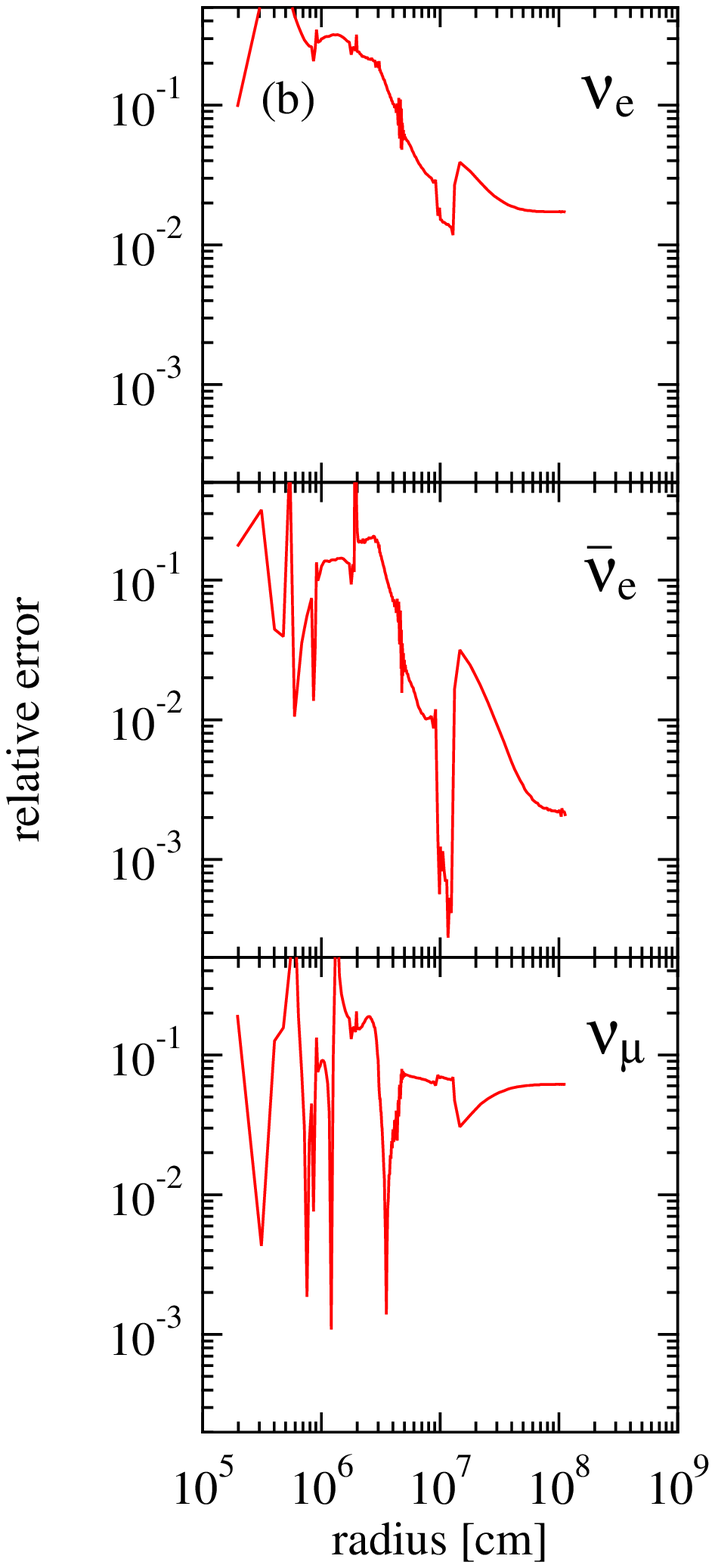}
\caption{Same as Fig. \ref{fig:1d-nuflux.rhoc12}, 
but for the profile at 100 ms after the bounce.  }
\label{fig:1d-nuflux.tpb100}
\end{figure}

We show the radial distributions of the number densities and fluxes 
for the three species of neutrinos 
in the left panel of 
Figs. \ref{fig:1d-nudens.rhoc12} and \ref{fig:1d-nuflux.rhoc12} 
for the profile during the collapse.  
We show the corresponding distributions 
for the profile after the bounce in the left panel of 
Figs. \ref{fig:1d-nudens.tpb100} and \ref{fig:1d-nuflux.tpb100}.  
The degenerate neutrinos ($\nu_e$) and thermal neutrinos 
($\bar{\nu}_e$ and $\nu_{\mu}$) at the central region are 
properly described with the tail of free-streaming fluxes 
in the outer region.  

In Figures \ref{fig:1d-nudens.rhoc12}, \ref{fig:1d-nuflux.rhoc12}, 
\ref{fig:1d-nudens.tpb100} and \ref{fig:1d-nuflux.tpb100}, 
we compare the current results (cross symbols) 
with the numerical results by the spherical code (solid lines) 
based on \citet{sum05}.  
Since we treat the steady state, 
we evaluate separately 
the neutrino distributions for the static background 
by solving the general relativistic Boltzmann equation 
under spherical symmetry (1D).  
Relative errors in the neutrino densities and fluxes 
between the two evaluations are shown in the right panel of each figure.  
In general, the numerical results accord very well with each other, 
while the errors of the density amount to a few tens of percent 
near the boundary.  
The errors of the fluxes are significant in the region 
where the errors of the density are appreciable.  
We found that the general relativistic treatment 
in the 1D code does not affect the comparison.  
This is because 
the neutrino distributions are determined locally 
by the thermodynamical condition at the central region 
and the neutrino fluxes at the outer layers are 
not affected by the general relativity due to large radii.  

\begin{figure}
\epsscale{1.0}
\plotone{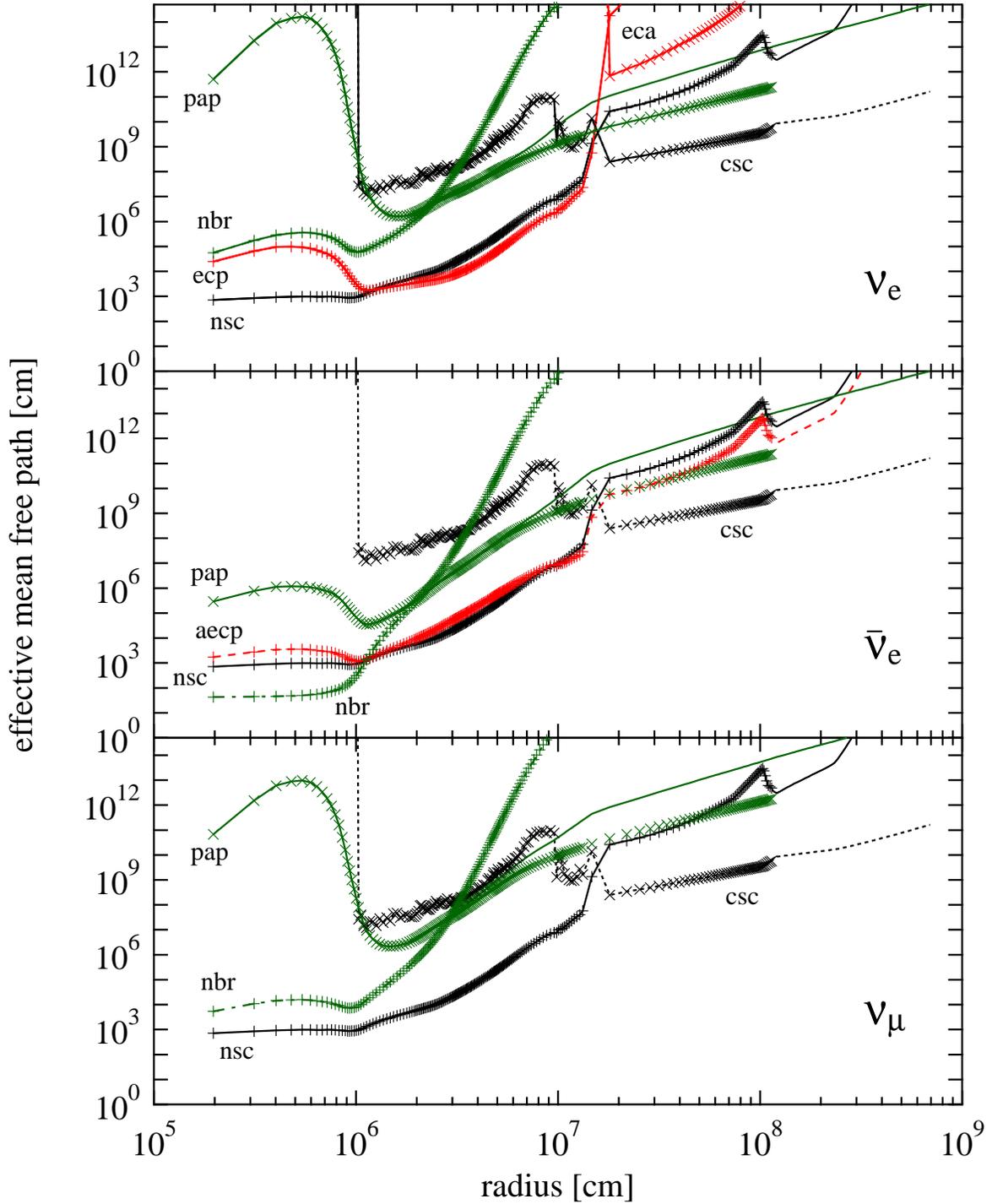}
\caption{Effective mean free paths 
for the three species of neutrinos 
with the energy of 34.0 MeV 
for the profile at 100 ms after the bounce.  
The results by the 3D code are shown by symbols as functions of radius.  
The mean free paths by the spherical calculation 
are shown by the lines.  
The name of neutrino reactions 
are indicated by the notation in \S \ref{reaction}.  }
\label{fig:1d-mfp}
\end{figure}

We show in Fig. \ref{fig:1d-mfp} 
the mean free paths of neutrino reactions in the profile 
after the bounce described above.  
The mean free path we discuss hereafter is the effective mean free path 
defined by the inverse of the absorption and scattering coefficients 
in Eqs. (\ref{eqn:kappa-abs}), (\ref{eqn:kappa-pair}) and (\ref{eqn:kappa-scatt}) 
for each process.  
We check the mean free paths 
by comparing with the numerical results obtained by the 1D code.  
As shown in  Fig. \ref{fig:1d-mfp}, 
the mean free paths by the 3D code accord with those in the 1D evaluation.  
The relative errors between the two evaluations are within $2 \times 10^{-3}$ 
except for the cases of the pair process and nucleon-nucleon bremsstrahlung 
to be discussed below.  
The mean free paths for the collapse phase have been 
also checked in the same way (not shown here in figure).  

\begin{figure}
\epsscale{1.0}
\plotone{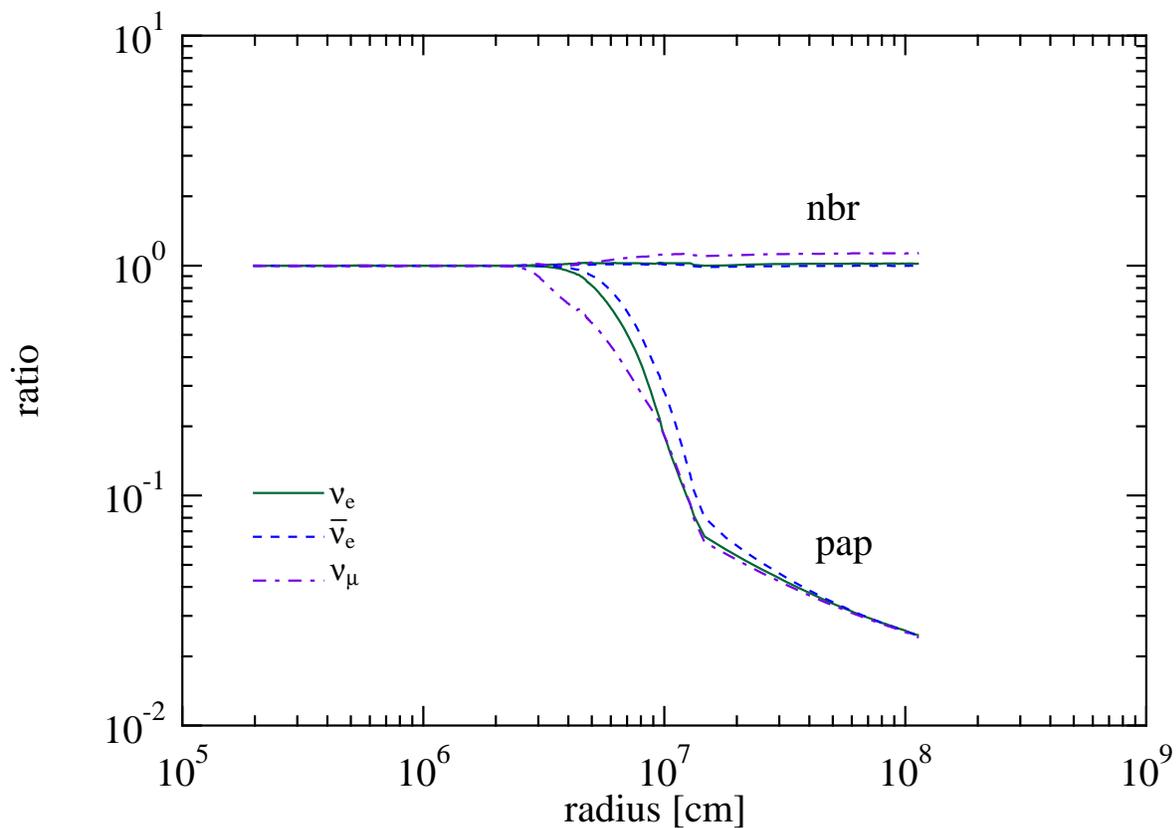}
\caption{Ratios of the effective mean free path 
calculated in the 3D code with respect to 
that used in the spherical calculation 
for the profile at 100 ms after the bounce.  
The ratios are shown for 
the pair process (pap) and 
the nucleon-nucleon bremsstrahlung (nbr) 
with the neutrino energy of 34.0 MeV.  
The solid, dashed and dot-dashed lines 
show the ratios 
for $\nu_e$, $\bar{\nu}_e$ and $\nu_{\mu}$, 
respectively.  }
\label{fig:1d-mfp.fac}
\end{figure}

We note that the mean free paths for the pair processes 
within the approximate expression accord very well 
with the full expression in the 1D code 
in the central region where these reactions are important.  
We show in Fig. \ref{fig:1d-mfp.fac} the ratios of 
the mean free path by the 3D calculation to 
that by the 1D code 
for the pair process and nucleon-nucleon bremsstrahlung.  
In the central region, the ratio is very close to 1 
and its deviation is within $4 \times 10^{-3}$.  
Deviations in the outer layers appear 
due to the approximation of the reaction rate and 
the angle-averaged distribution of couterpart-neutrinos.  
Deviations for the nucleon-nucleon bremsstrahlung 
are rather small since we adopt the same isotropic 
rate in the both codes.  
Large deviations in the pair-process are due to 
the approximation of taking isotropic reaction rates.  
We note, however, that the deviations in the outer layers 
are not crucial for the whole description 
since the material is in the transparent regime and 
contributions of these reactions are minor there.  

\begin{figure}
\epsscale{1.0}
\plotone{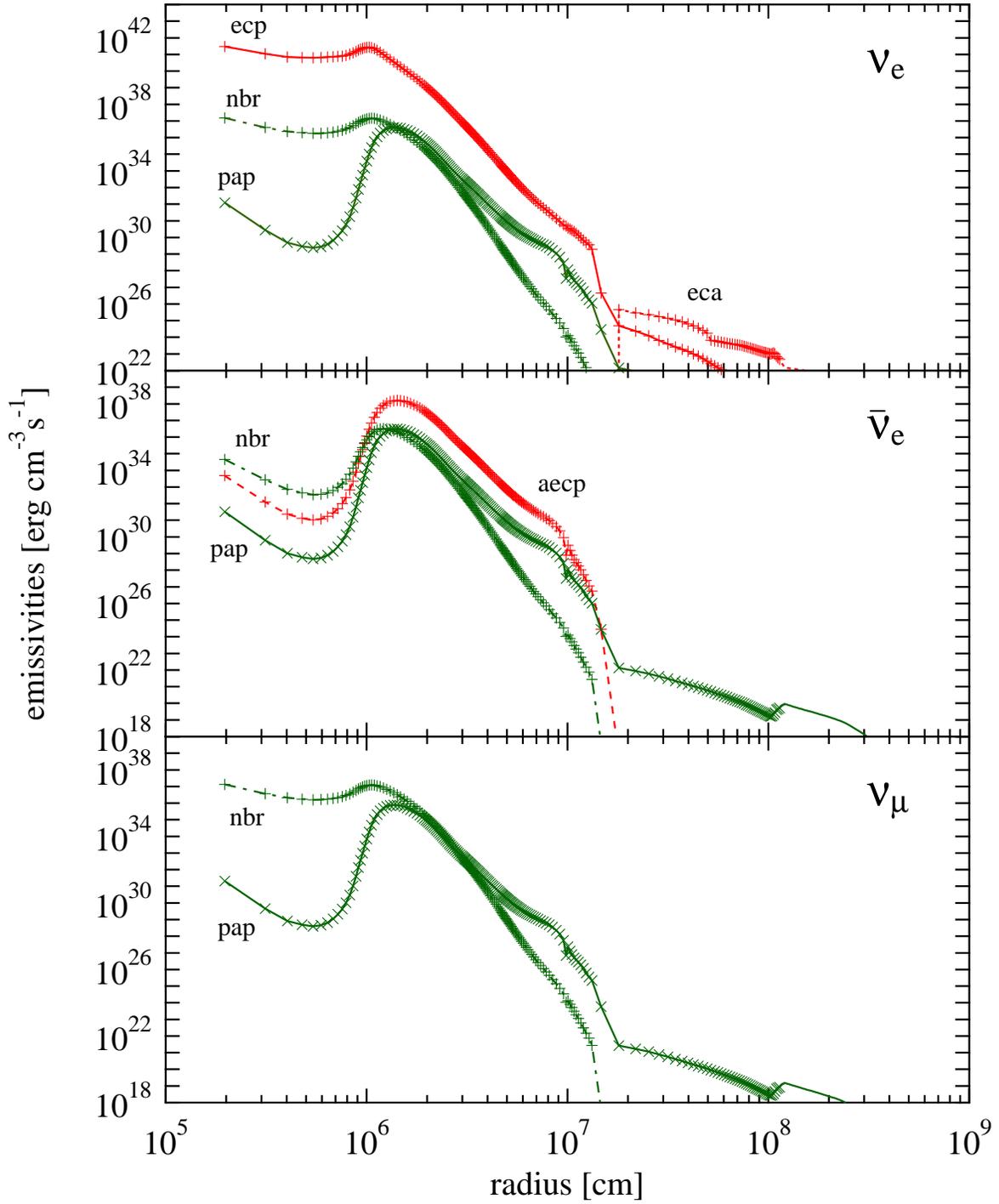}
\caption{Emissivities for the three species of neutrinos 
for the profile at 100 ms after the bounce.  
The results by the 3D code are shown by symbols as functions of radius.  
The emissivities obtained by the spherical calculation 
are shown by the lines.  
The name of neutrino reactions 
are indicated by the notation in \S \ref{reaction}.  }
\label{fig:1d-emis}
\end{figure}

In Figure \ref{fig:1d-emis}, we show the emissivities 
for the three neutrino species as a function of radius 
for the profile after the bounce.  
The emissivities are defined by 
\begin{equation}
\label{eqn:emis-ecp}
\epsilon_{emis} = 
\int \frac{d\varepsilon~\varepsilon^2}{(2 \pi)^3} 
\int d \Omega~ \varepsilon 
R_{emis}(\varepsilon, \Omega) ~, 
\end{equation}
and 
\begin{equation}
\label{eqn:emis-pap}
\epsilon_{emis-pair} = 
\int \frac{d\varepsilon~\varepsilon^2}{(2 \pi)^3} 
\int d \Omega 
\int \frac{d \varepsilon' \varepsilon'^2}{(2 \pi)^3} 
\int d \Omega' ~ \varepsilon 
R_{pair-emis}(\varepsilon, \Omega; \varepsilon', \Omega')
[ 1 - \overline{f}(\varepsilon', \Omega') ] ,
\end{equation}
for neutrino-emissions (ecp, aecp and eca) and 
pair-processes (pap and nbr), respectively.  
We compare the emissivities with those obtained by the 1D code.  
The emissivities in the two evaluations agree very well with each other 
within relative errors of $2 \times 10^{-3}$ in general.  
The largest errors arise 
in the transitional region between 20 km and 100 km 
but are within $\sim 1 \times 10^{-2}$.  
The agreement is good even for the pair processes 
despite the usage of the isotropic rates.  
This is because the isotropic term of the reaction rates 
is dominant in Eq. (\ref{eqn:emis-pap}) 
with isotropic distributions in the inner region 
or small distributions in the outer region.  
This situation is different 
from the case of the effective mean free path 
for the pair processes, 
where the angular distribution is important 
due to the exponential factor in Eq. (\ref{eqn:kappa-pair}).  

\begin{figure}
\epsscale{0.85}
\plotone{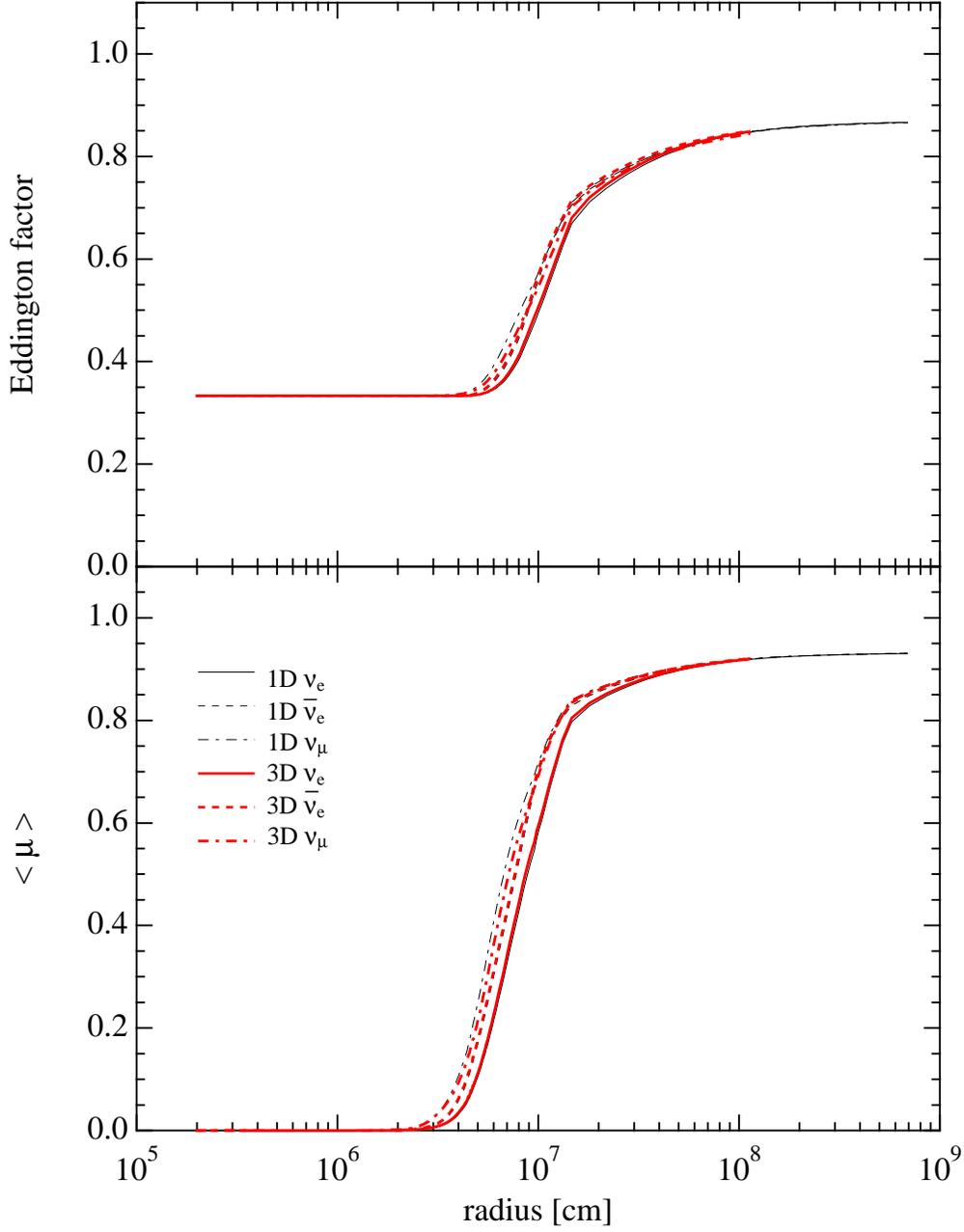}
\caption{Eddington factors and flux factors 
as functions of radius for the profile 
at 100 ms after the bounce.  
The solid, dashed and dot-dashed thick lines display 
the results for $\nu_e$, $\bar{\nu}_e$ and $\nu_{\mu}$, respectively.  
The solid, dashed and dot-dashed thin lines show 
the corresponding quantities in the spherical calculation.  }
\label{fig:1d-mom}
\end{figure}

\begin{figure}
\epsscale{0.9}
\plotone{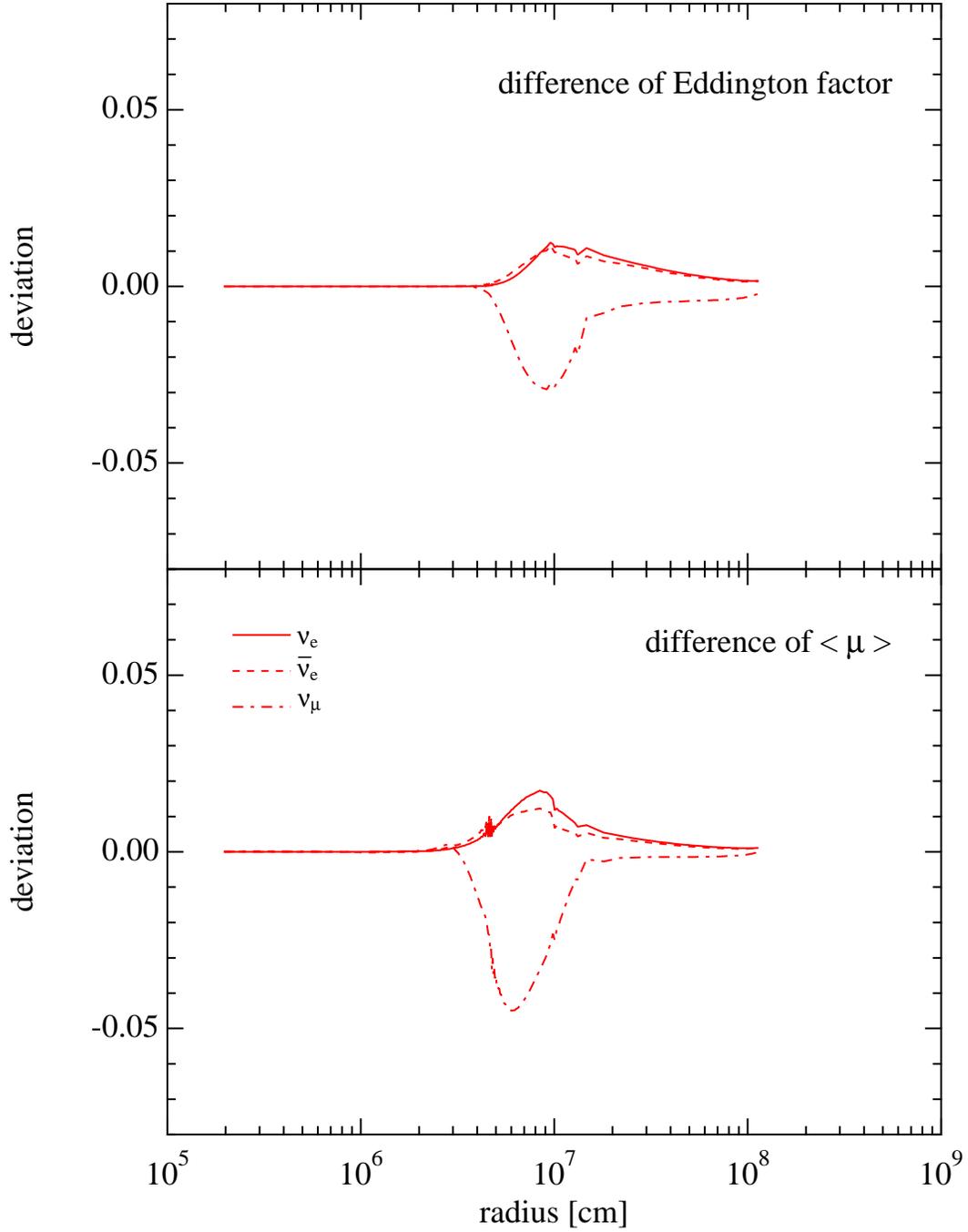}
\caption{Deviation of the Eddington factors 
and flux factors by the 3D code from those by the 1D code 
as functions of radius for the profile 
at 100 ms after the bounce.  
The solid, dashed and dot-dashed lines denote 
quantities for $\nu_e$, $\bar{\nu}_e$ and $\nu_{\mu}$, respectively.  }
\label{fig:1d-mom.err}
\end{figure}

We examine the energy and angle moments 
of the neutrino distributions obtained in the 3D code 
through the comparison with the 1D evaluation.  
We show in Fig. \ref{fig:1d-mom} 
the Eddington factors, $k^{rr}$, 
and the flux factors, $\langle \mu_{\nu} \rangle$, 
as functions of radius in the post-bounce profile.  
The definition of various moments of the neutrino distributions 
is summarized in \S \ref{moments}.  
We obtain the correct limits of quantities in the opaque and transparent regimes.  
The Eddington factor and the flux factor are 1/3 and zero, respectively, 
in the central core, where the distributions are isotropic, 
and they approach $\sim$1 (forward peaked) toward the outer layers.  
The moments by the 3D code 
accord generally well with those by the 1D evaluation.  
In Figure \ref{fig:1d-mom.err}, we show differences 
of the moments by the 3D code from the 1D evaluation.  
Deviations within $\sim$0.05 appear around the transitional region 
and near the boundary.  

\begin{figure}
\epsscale{0.9}
\plotone{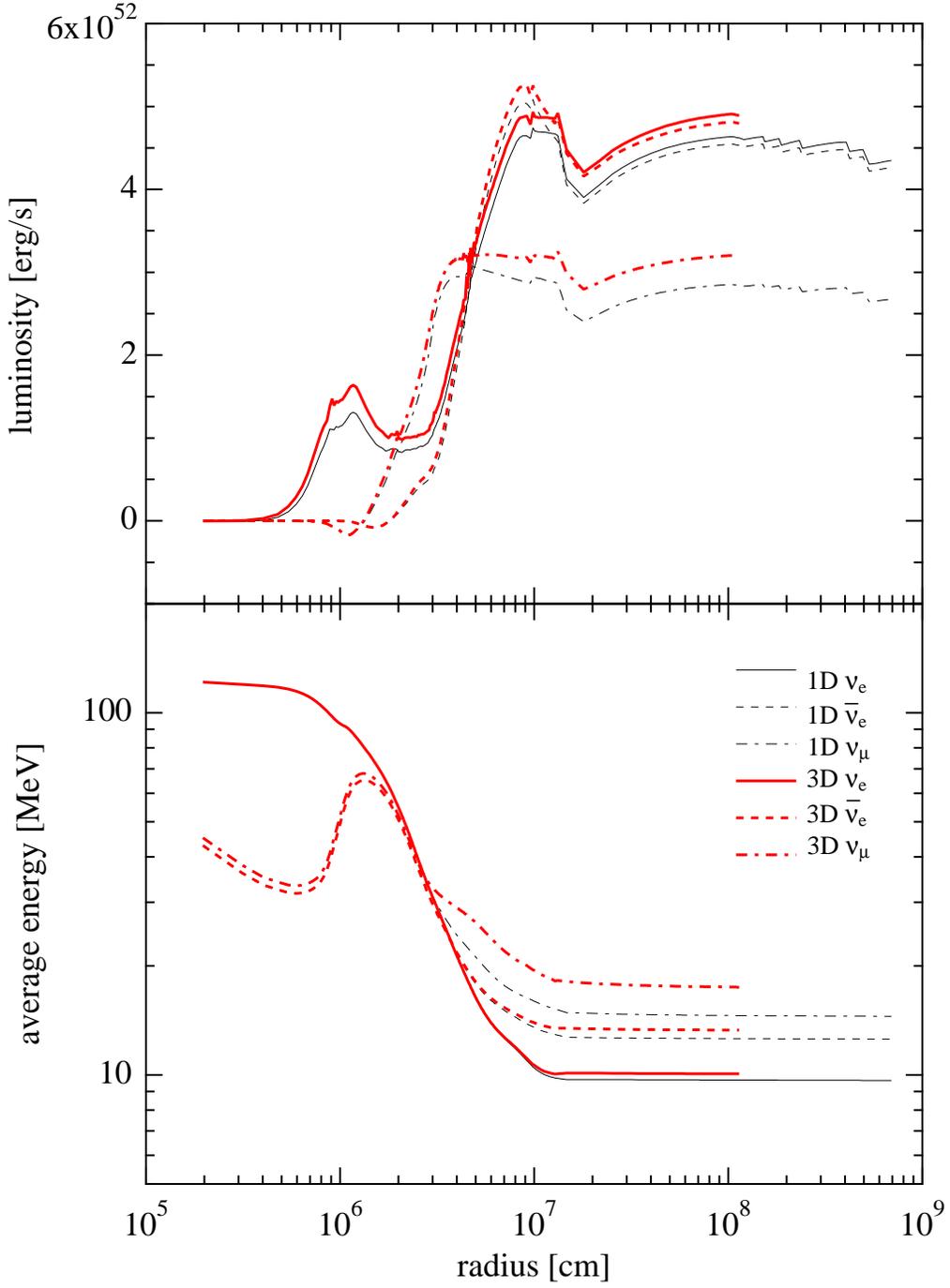}
\caption{Luminosities and average energies as functions of radius 
for the profile at 100 ms after the bounce.  
The solid, dashed and dot-dashed thick lines denote 
quantities for $\nu_e$, $\bar{\nu}_e$ and $\nu_{\mu}$, respectively.  
The corresponding quantities from the spherical calculation 
are shown by thin lines for comparison.  }
\label{fig:1d-enu}
\end{figure}

We show, in Fig. \ref{fig:1d-enu}, the luminosities and average energies 
of neutrinos for the post-bounce profile 
with the corresponding quantities from the 1D code for comparison.  
We plot here the average energy, $\langle \varepsilon \rangle$, and 
the luminosity, $L_{\nu} = 4 \pi r^2 F_{\nu}$, defined in \S \ref{moments}.  
The behavior of the luminosities and average energies 
accords generally well with the results by the 1D code.  

The numerical checks so far 
prove the new code with the microphysics 
of neutrinos in dense matter works properly with reliable accuracy 
in the spherical configurations.  
From the examination of the effective mean free paths 
and the emissivities, we judge that 
the approximate expressions adopted in the pair processes 
are efficient and sufficient for the numerical studies of 
supernova cores by the 3D code.  


\subsection{2D Configurations}\label{2Dconfig}

In order to demonstrate the ability of the new code 
in multi-dimensional realistic settings, we first study 2D case, 
utilizing artificially deformed profiles of supernova core 
based on the 1D core-collapse simulation \citep{sum05}.  
For the given background profile, we obtain a steady state solution 
of neutrino transfer by following the time evolution 
for a sufficiently long time.  
We compare the result with 
the formal solutions as discussed in \S \ref{stationary}.  

\begin{figure}
\epsscale{0.9}
\plotone{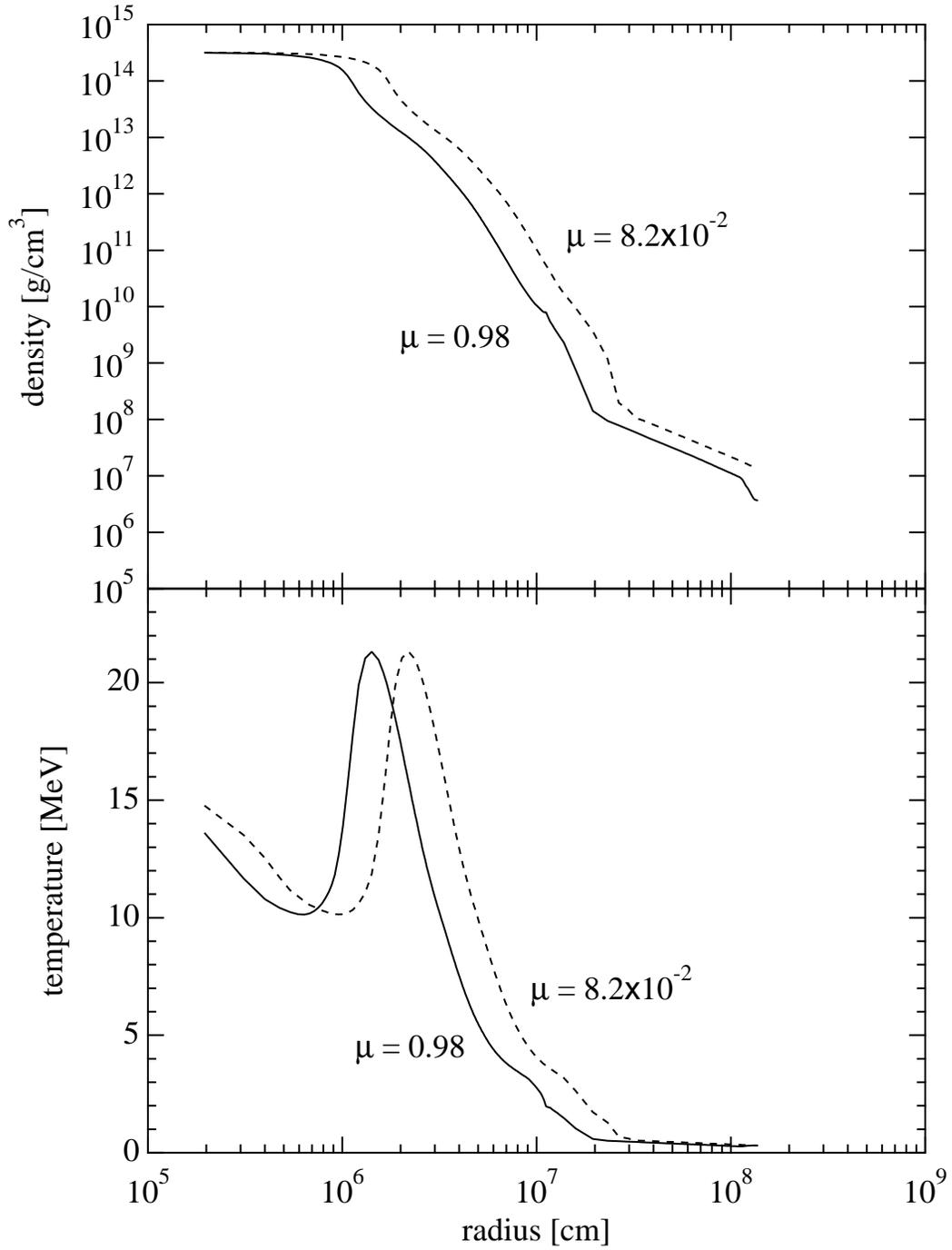}
\caption{Radial profiles of the density and temperature 
of the deformed supernova core for the two directions 
with the polar angles 
of $\mu=8.2\times10^{-2}$ (near the equator: dashed line) and 
$\mu=0.98$ (near the pole: solid line).  }
\label{fig:2d-profile}
\end{figure}
\begin{figure}
\epsscale{1.0}
\plotone{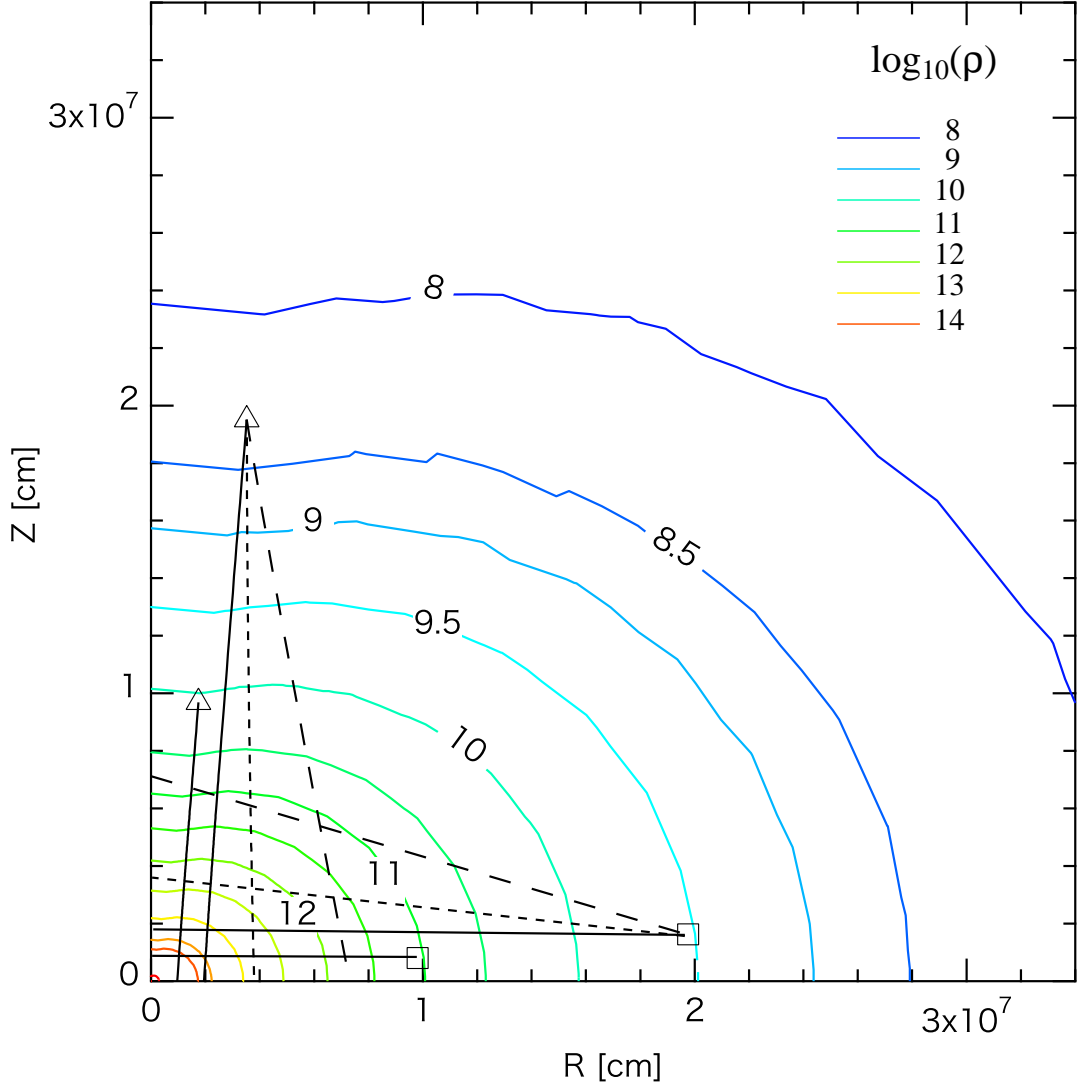}
\caption{Contour plot of the density of the deformed supernova core 
with the paths of neutrino propagation to evaluate the formal solution 
at the locations denoted by symbols.  
The square and triangle symbols denote the two locations 
along the polar directions with $\mu=8.2\times10^{-2}$ and $\mu=0.98$, 
respectively.  
The solid lines show the paths along the neutrino angle, 
$\mu_{\nu}=0.99519$ for the case of $N_{\theta_{\nu}}=24$.  
The paths along the neutrino angle, $\mu_{\nu}=0.93247$ and 0.98156, 
for the case of $N_{\theta_{\nu}}=6$ and 12 are also shown 
by long-dashed and dashed lines for the two locations, respectively.  
Note that interpolation is made to plot smoothly contours.  }
\label{fig:2d-contour}
\end{figure}

We utilize the same spherical profiles for the post-bounce stage 
as those studied in \S \ref{1Dconfig}.  
We modify the profiles of density, $\rho$($r$), temperature, T($r$), and 
electron fraction, Y$_e$($r$), along each radial direction, 
depending on its polar angle, $\theta$, by scaling the radius, $r$, as 
\begin{equation}
\label{eqn:2d-rscale}
\tilde{r}=r ( 1 - \epsilon \sin \theta ), 
\end{equation}
where $\epsilon$ is a parameter to specify the degree of deformation.
As a result, 
we obtain oblate profiles for positive $\epsilon$'s, which are crude 
approximations to rapidly rotating supernova cores.
We show in Figs. \ref{fig:2d-profile} and \ref{fig:2d-contour} 
some features of the profiles thus constructed for $\epsilon=0.4$.  

We first examine the neutrino transfer through comparisons 
with the formal solutions.  
For this purpose, we treat the electron-type neutrino alone, 
taking into account only the emissions and absorptions.  
We set the spatial grid with 
$N_{r}=200$, $N_{\theta}=9$ and $N_{\phi}=9$ in the first octant.  
We try four sets of angular meshes in the momentum space: 
($N_{\theta_{\nu}}$, $N_{\phi_{\nu}}$)=(6,6), (12,6), (24,6) and (12,12).  
The number of energy grid points is fixed to $N_{\varepsilon}=14$.  
The relatively small grid sizes are mainly due to the limitation 
of available computing resources.  

The formal solution is obtained by integrating the Boltzmann equation 
along the representative paths shown in Fig. \ref{fig:2d-contour}.  
These paths go through one of four grid points 
that are rather arbitrarily chosen and have a radius of 
$r$=98.4 or 198 km with an angle cosine of $\mu=8.2\times10^{-2}$ or $\mu=0.98$.  
The solid lines in the figure represent the paths 
with a polar angle cosine in the momentum space of $\mu_{\nu}=0.99519$, 
which corresponds to the most forward grid point for $N_{\theta_{\nu}}=24$.  
For comparison, 
the long-dashed and dashed lines show the paths with 
$\mu_{\nu}=0.93247$ and $\mu_{\nu}=0.98156$, 
which correspond to the most forward grid points 
for $N_{\theta_{\nu}}=6$ and 12, respectively.  

\begin{figure}
\epsscale{1.1}
\plottwo{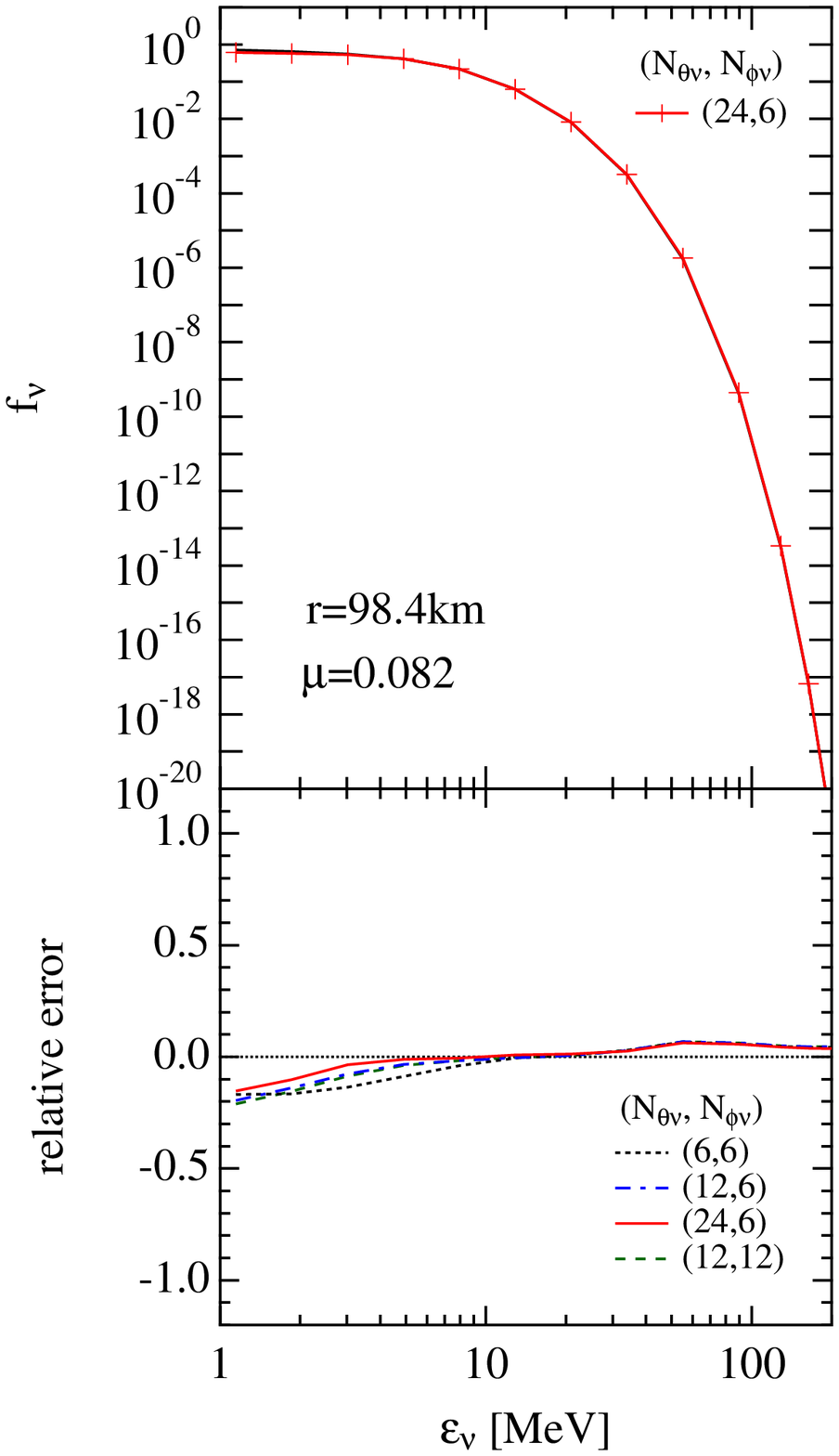}{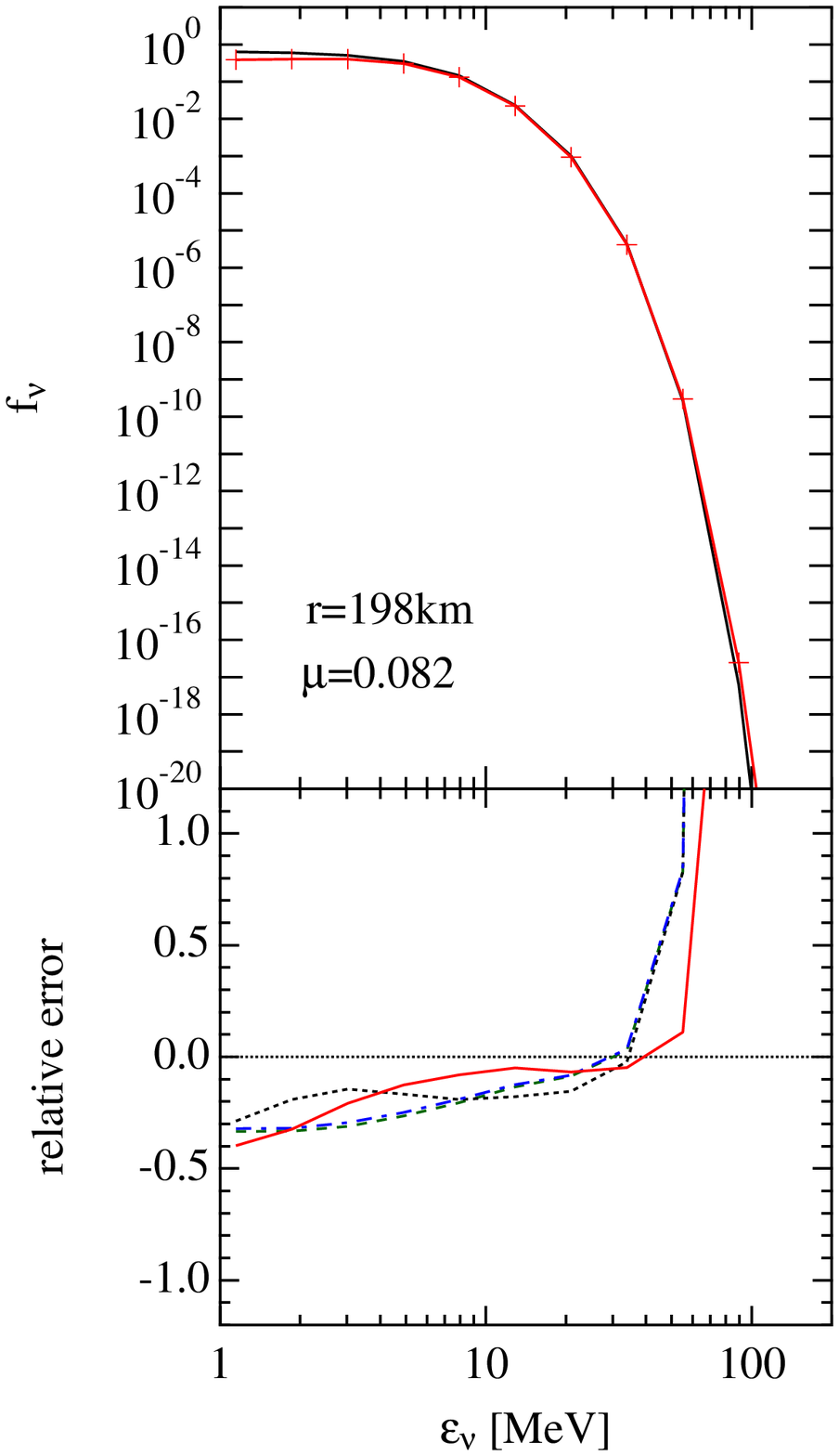}
\caption{Upper panels: energy spectra at the radial grid points 
98.4 km (left) and 198 km (right) along the polar direction $\mu=8.2\times10^{-2}$.  
The energy spectra evaluated by 
the 3D code (solid lines with cross symbols) 
with $N_{\theta_{\nu}}=24$ and $N_{\phi_{\nu}}=6$ 
are shown with the formal solutions (solid line).  
Lower panels: relative errors of the spectra by the 3D code using 
($N_{\theta_{\nu}}$, $N_{\phi_{\nu}}$)=(6,6), (12,6), (24,6) and (12,12) 
with respect to the formal solutions
are shown by dotted, dot-dashed, solid and dashed lines, respectively.  } 
\label{fig:2d-formal.ith2}
\end{figure}

\begin{figure}
\epsscale{1.1}
\plottwo{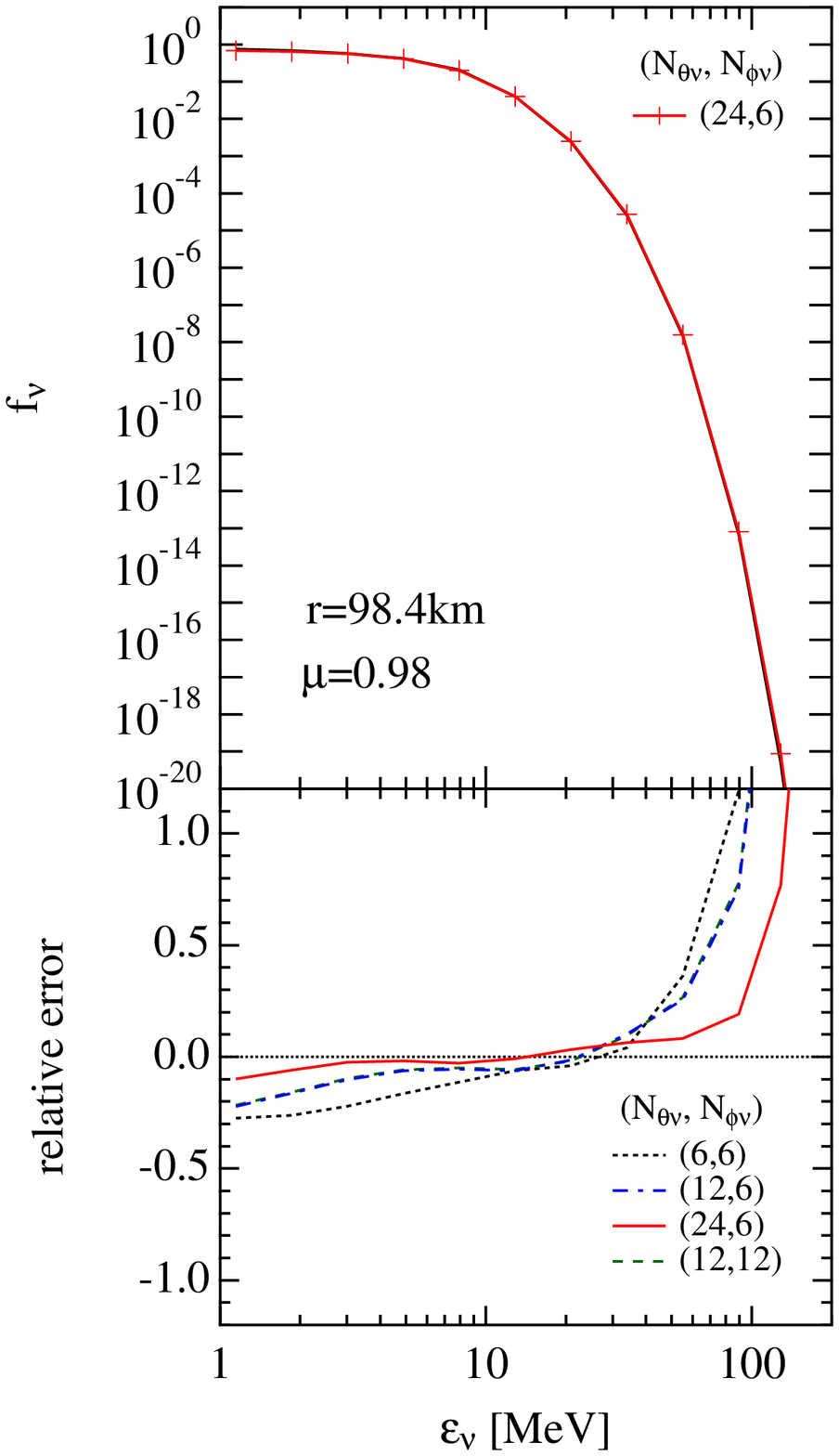}{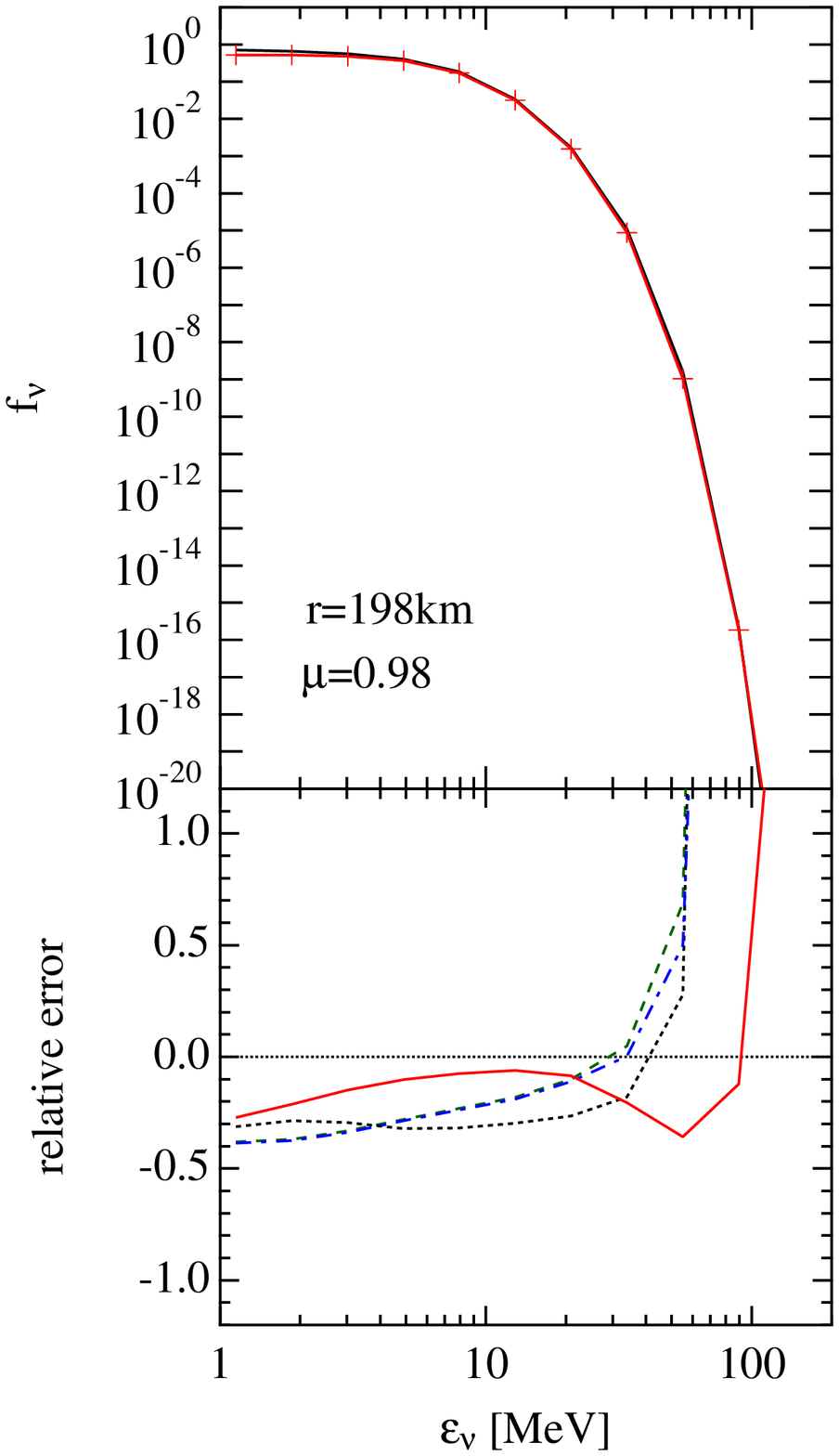}
\caption{Same as Fig. \ref{fig:2d-formal.ith2}, 
but for the polar direction $\mu=0.98$.  }
\label{fig:2d-formal.ith9}
\end{figure}

Figures \ref{fig:2d-formal.ith2} and \ref{fig:2d-formal.ith9} show 
the comparison of the numerical results by the new code 
with the formal solutions.  
In Figure \ref{fig:2d-formal.ith2}, we present the energy spectra 
for the points close to the equator (square symbols in Fig. \ref{fig:2d-contour}).  
The numerical results agree 
with the formal solutions within relative errors of 20 \% at r=98.4 km.  
In this case, the density at the grid point is rather high 
($\sim$10$^{11}$ g/cm$^{3}$) and the matter is already opaque to neutrinos there.  
Moreover, the path runs through the proto-neutron star.  
As a result, the neutrino spectrum at the grid point is mainly formed 
by the contributions from relatively near-zones and 
the agreement of the numerical result with the formal solution 
is excellent even with the rather coarse grid adopted in this computation.  
At r=198 km, on the other hand, the agreement is 
not so good as shown in the right panel of Fig. \ref{fig:2d-formal.ith2}.  
This is because the density at the grid point is much lower 
and the matter in the vicinity is transparent to neutrinos 
as well as because the path barely touches the periphery of 
the proto-neutron star.  
Then the neutrino distribution at the grid point is a superposition 
of the contributions from very far regions.  
In general, the larger the distance to the source is, 
the more anisotropic the neutrino distribution becomes and 
the more difficult it is to numerically reproduce it.  
It is important to see, however, that the relative error is still 
within $\sim$50 \% 
even at very low energies, where the opacities are the lowest.  

These features are common to the points near the pole 
(triangle symbols in Fig. \ref{fig:2d-contour}) 
as seen in Fig. \ref{fig:2d-formal.ith9}:  
the energy spectra obtained numerically agree with the formal solutions 
within $\sim$50 \% except for the high energy tail beyond $\sim$100 MeV, 
where the neutrino populations are very small, f$_{\nu}$ $\le10^{-10}$.  
The relative errors near the pole are somewhat 
larger in general than those near the equator.  
This is because of the oblate shape of the artificially deformed core 
in this computation.  
The matter densities at the two points near the pole are 
lower than those at the corresponding points near the equator.  
The matter is hence more transparent in the vicinity of the former, 
hence, the evaluation of the forward peak is further difficult.  
The paths near the pole, on the other hand, run into deeper 
inside the proto-neutron star.  
Although these effects are counteracting each other, 
the former turns out to be more important.  

The relative errors become smaller 
for $N_{\theta_{\nu}}=24$ than those for $N_{\theta_{\nu}}=6$, 12.  
The increase of $N_{\phi_{\nu}}$ to 12, on the other hand, does not 
improve the accuracy very much for $N_{\theta_{\nu}}=12$.  
This feature is common to all the paths considered here.  
Although we need further systematic studies with finer grids, 
the above facts suggest that 
it is more important to employ a sufficiently large $N_{\theta_{\nu}}$ 
in the momentum space.  

\begin{figure}
\epsscale{1.0}
\plotone{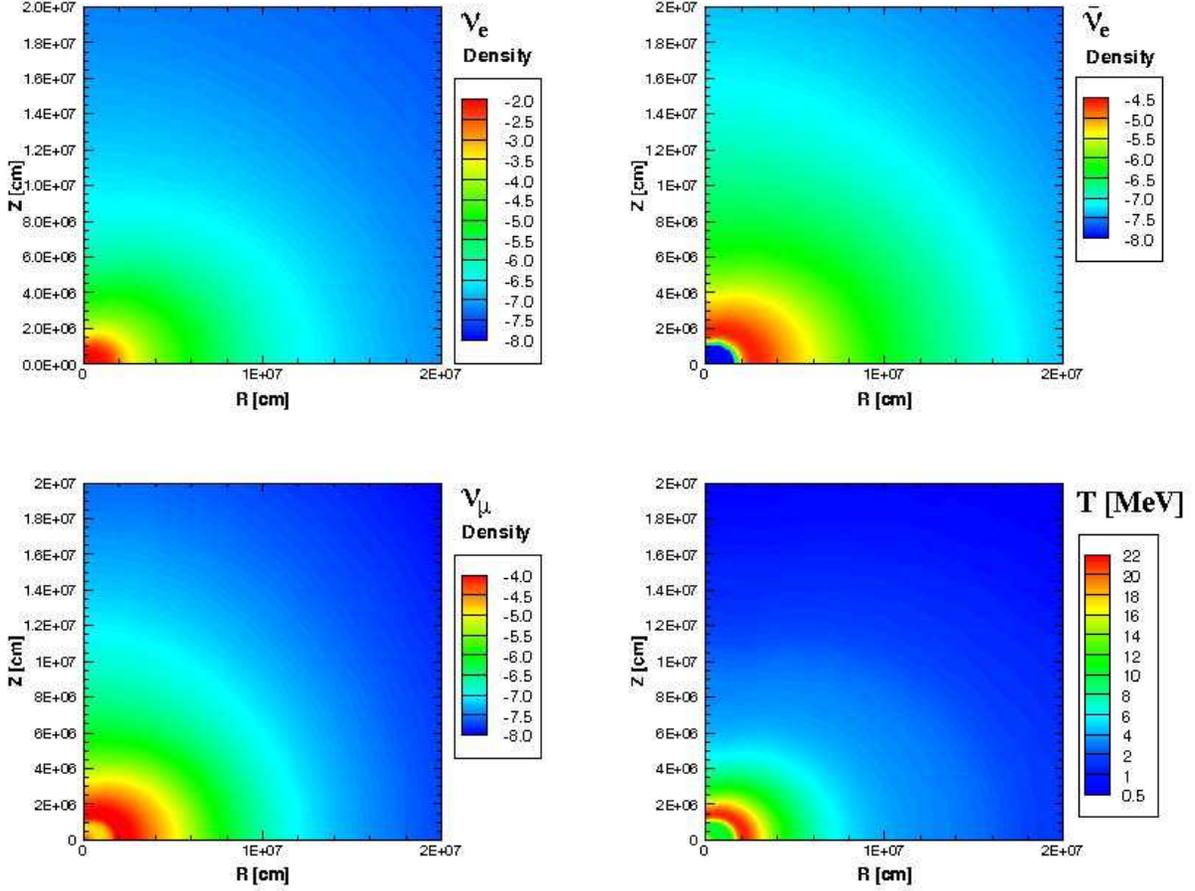}
\caption{Contour plots of the profiles 
in the deformed (axially symmetric) supernova core 
on the meridian slice with a constant $\phi$.  
The densities of 
electron-type neutrinos, electron-type anti-neutrinos and mu-type neutrinos 
are shown by color codes 
in left top, right top and left bottom panels, respectively.  
The neutrino densities are shown here in the unit of fm$^{-3}$=10$^{-39}$ cm$^{-3}$ 
and in log-scale.  
The contour plot of the temperature is also shown in right bottom panel.  }
\label{fig:2d-nudens}
\end{figure}

We proceed to the numerical results with the full set of neutrino reactions 
for the three neutrino species.  
We set the number of grid points to 
$N_{r}=200$, $N_{\theta}=9$, $N_{\phi}=9$, 
$N_{\theta_{\nu}}=6$ and $N_{\phi_{\nu}}=12$.  
We follow the evolution 
for a sufficiently long time period ($\sim$10 ms) to obtain a steady state.  
We show in Fig. \ref{fig:2d-nudens} the contour plots of the neutrino density 
on the meridian slice with a constant $\phi$=0.44 radian.  
The neutrino distributions apparently reflect the deformed profiles 
of density and temperature.  
The electron-type neutrinos are mostly populated in the central region, 
while the electron-type anti-neutrinos and mu-type 
neutrinos mainly exist off-center, where the temperature is high.  

\begin{figure}
\epsscale{1.1}
\plottwo{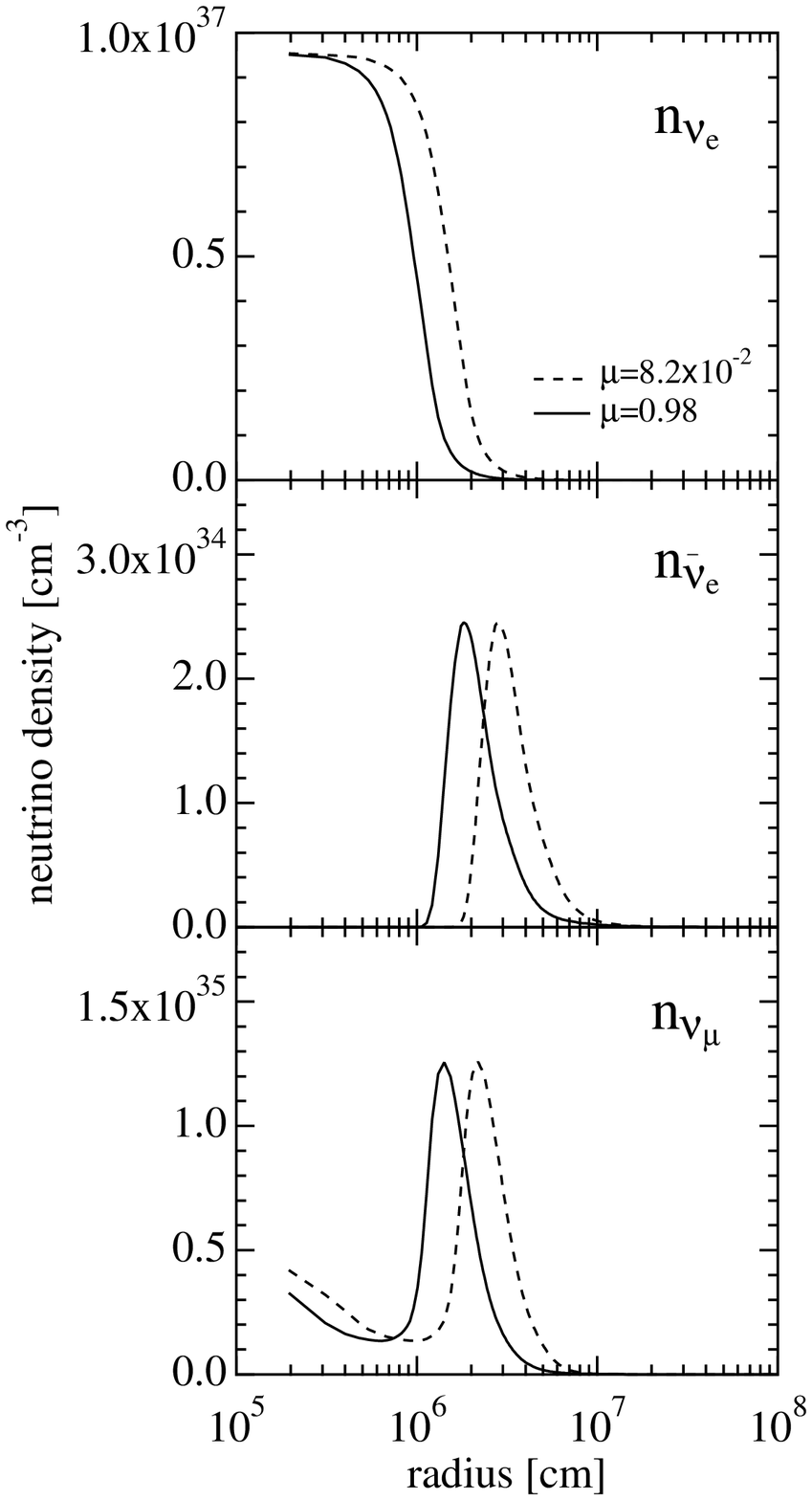}{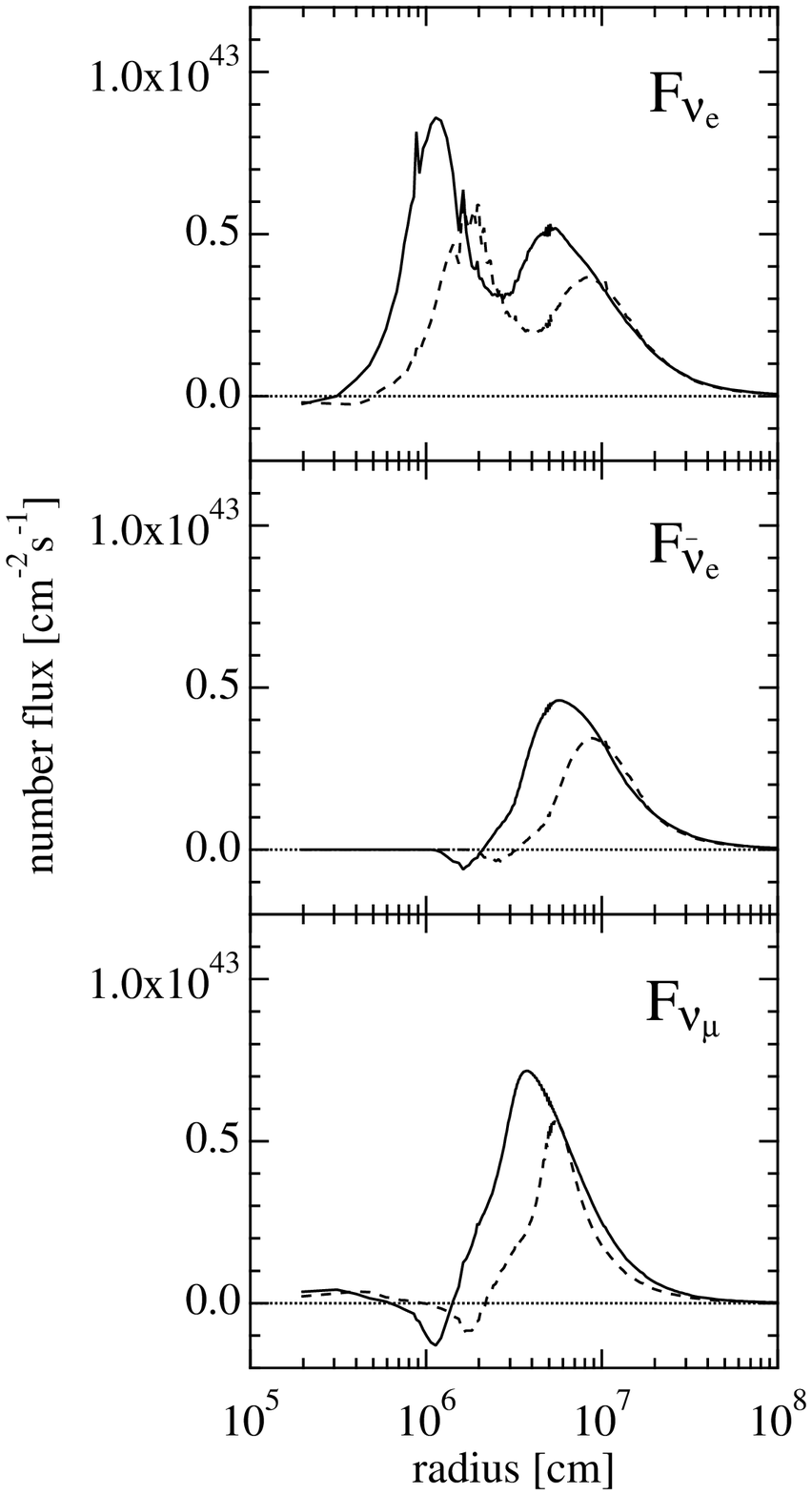}
\caption{Radial profiles of the densities and fluxes 
for the three neutrino species in the deformed supernova core 
along the two polar directions on the constant $\phi$-slice.  
The densities 
and fluxes 
are shown as a function of the radius in the two directions 
with $\mu=8.2\times10^{-2}$ (near the equator: dashed line) and 
$\mu=0.98$ (near the pole: solid line).  }
\label{fig:2d-nudens.radial}
\end{figure}

Figure \ref{fig:2d-nudens.radial} presents the radial profiles 
of the neutrino density and flux along the directions with 
$\mu=8.2\times10^{-2}$ (near the equator) and $\mu=0.98$ (near the pole).  
The peaks of the densities for electron-type neutrinos 
are located at the center with different widths, 
reflecting the deformation of the proto-neutron star.  
The peaks for electron-type anti-neutrinos and mu-type neutrinos, 
on the other hand, are located off-center at different radial positions 
corresponding to the temperature peaks.  
These neutrinos are mainly produced by the pair-processes 
in non-degenerate and positron-abundant environment.  
This is the reason why they are abundant off-center.  
We find that their populations agree very well 
with the local equilibrium distributions.  
The radial neutrino fluxes also reflect the deformed matter distributions.  
In fact, the radial fluxes near the pole are 
larger than those near the equator at the same radius, 
since the density gradient is steeper near the pole (See Fig. \ref{fig:2d-nudens}).  

\begin{figure}
\epsscale{0.9}
\plotone{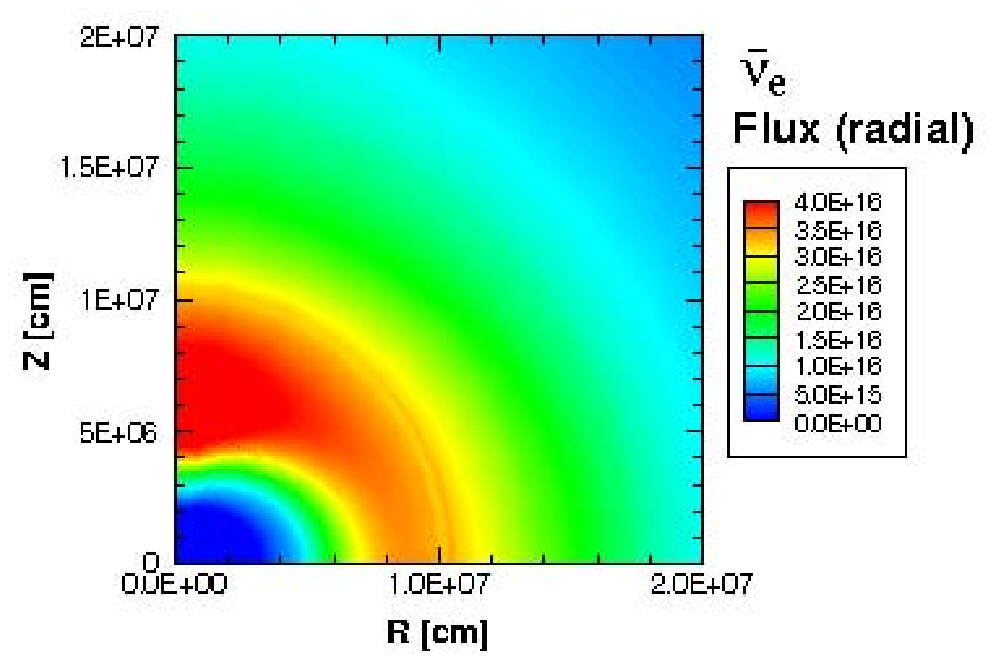}
\plotone{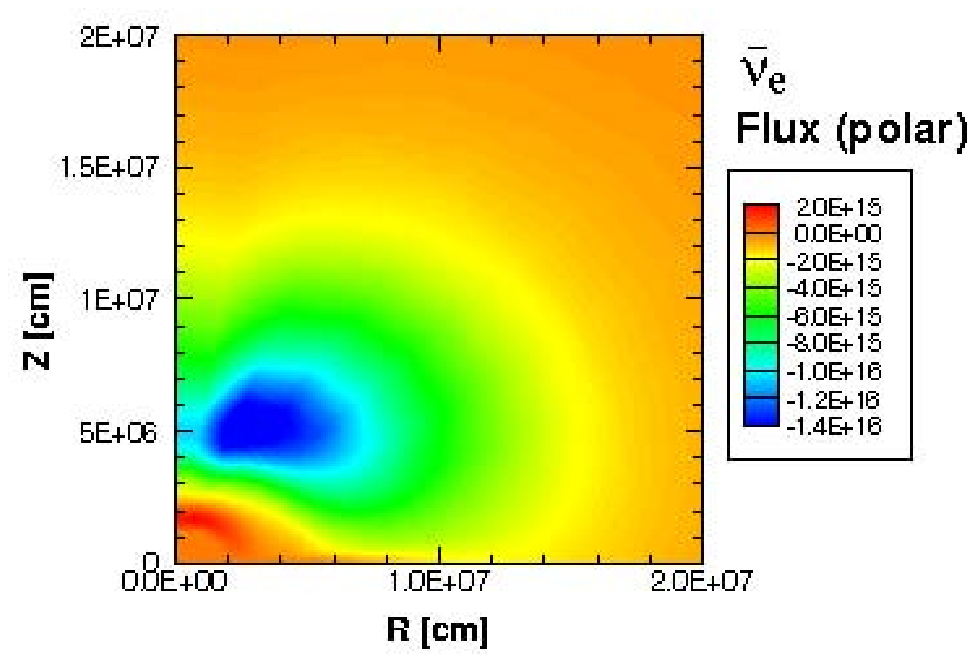}
\caption{Contour plots of the neutrino fluxes for electron-type 
anti-neutrinos in the deformed (axially symmetric) supernova core 
on the constant $\phi$-slice.  
The radial and polar components of flux 
are shown by color codes in top and bottom panels, respectively.  
The fluxes are shown in the unit of fm$^{-2}$s$^{-1}$.  }
\label{fig:2d-nuflux}
\end{figure}

We show not only the radial neutrino flux but also the polar flux 
of electron-type anti-neutrinos in Fig. \ref{fig:2d-nuflux} 
to elucidate multi-dimensional transfer.  
The radial flux is enhanced near the pole as just mentioned.  
The polar flux, on the other hand, is significant in the middle region 
between the pole and the equatorial plane.  
Because of the deformation, the density gradient 
is directed to the pole in general and is greatest 
in the middle region, since the axial symmetry and equatorial symmetry 
imposed in this computation force the density gradients 
radially directed near the pole and equator.  
We stress that this is a feature that can be captured properly only 
by the multi-dimensional transfer computations such as 
done in the current study and not by the ray-by-ray approximation.  
We note that the polar flux is significant 
even beyond 100 km, where an approximation 
by the polar gradient of the neutrino pressure \citep{mul10} 
may break down.  

\begin{figure}
\epsscale{0.8}
\plotone{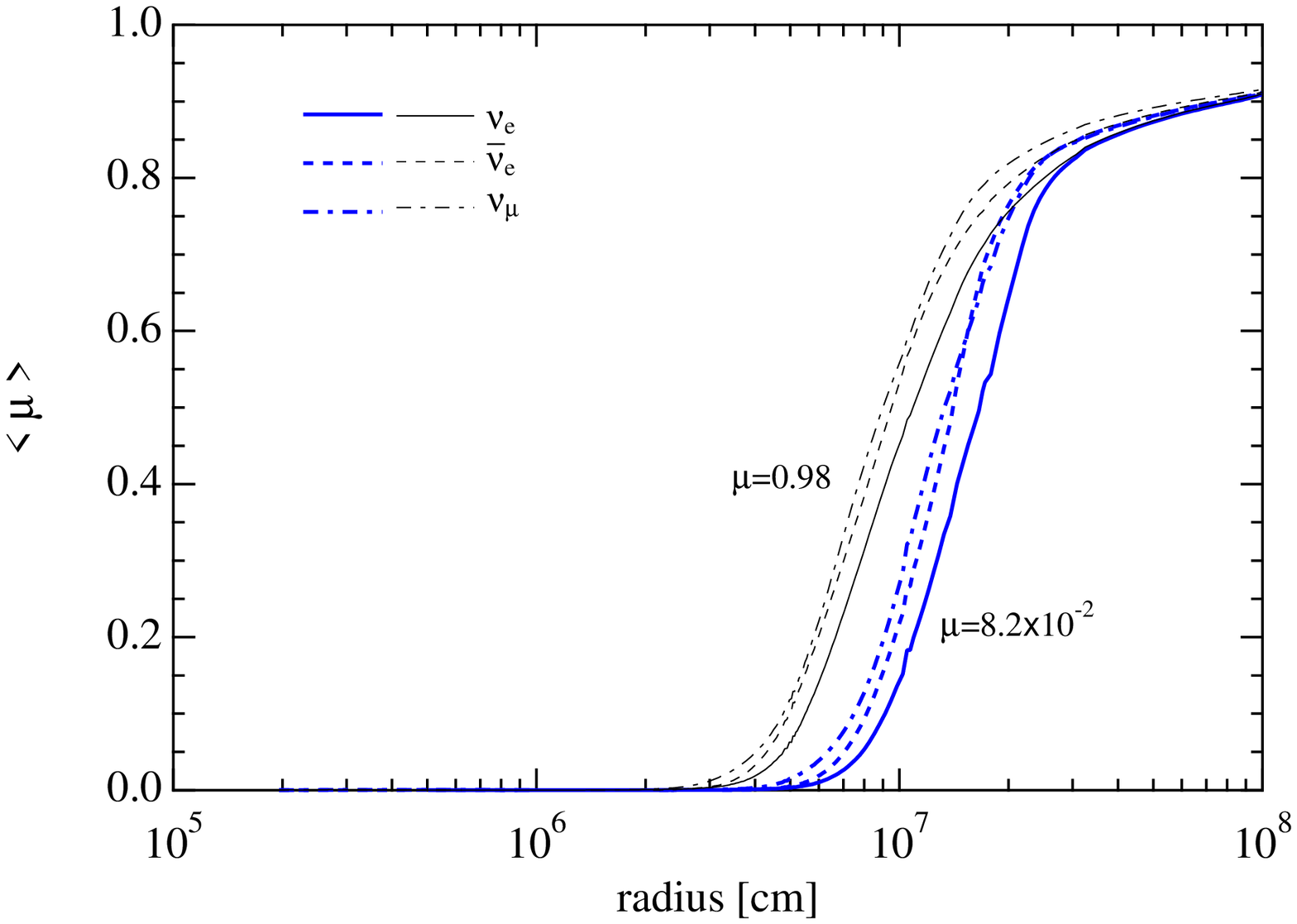}
\plotone{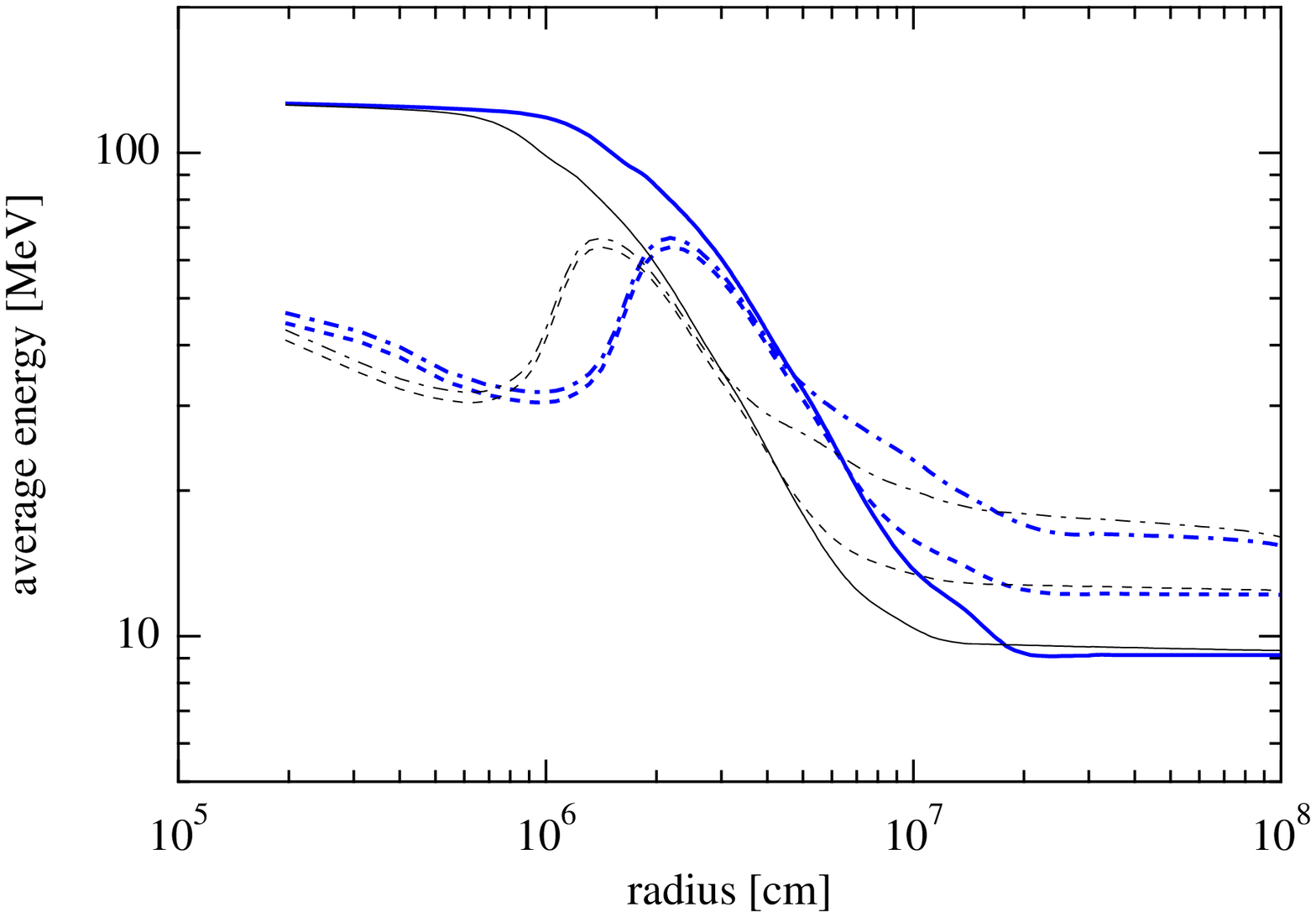}
\caption{Radial profiles of the flux factors (top) and 
average energies (bottom) in the deformed supernova core 
along the two polar directions with $\mu=8.2\times10^{-2}$ (thick lines) 
and $\mu=0.98$ (thin lines).  
The solid, dashed and dot-dashed lines denote 
quantities for $\nu_e$, $\bar{\nu}_e$ and $\nu_{\mu}$, respectively. }
\label{fig:2d-nueave}
\end{figure}

\begin{figure}
\epsscale{0.7}
\plotone{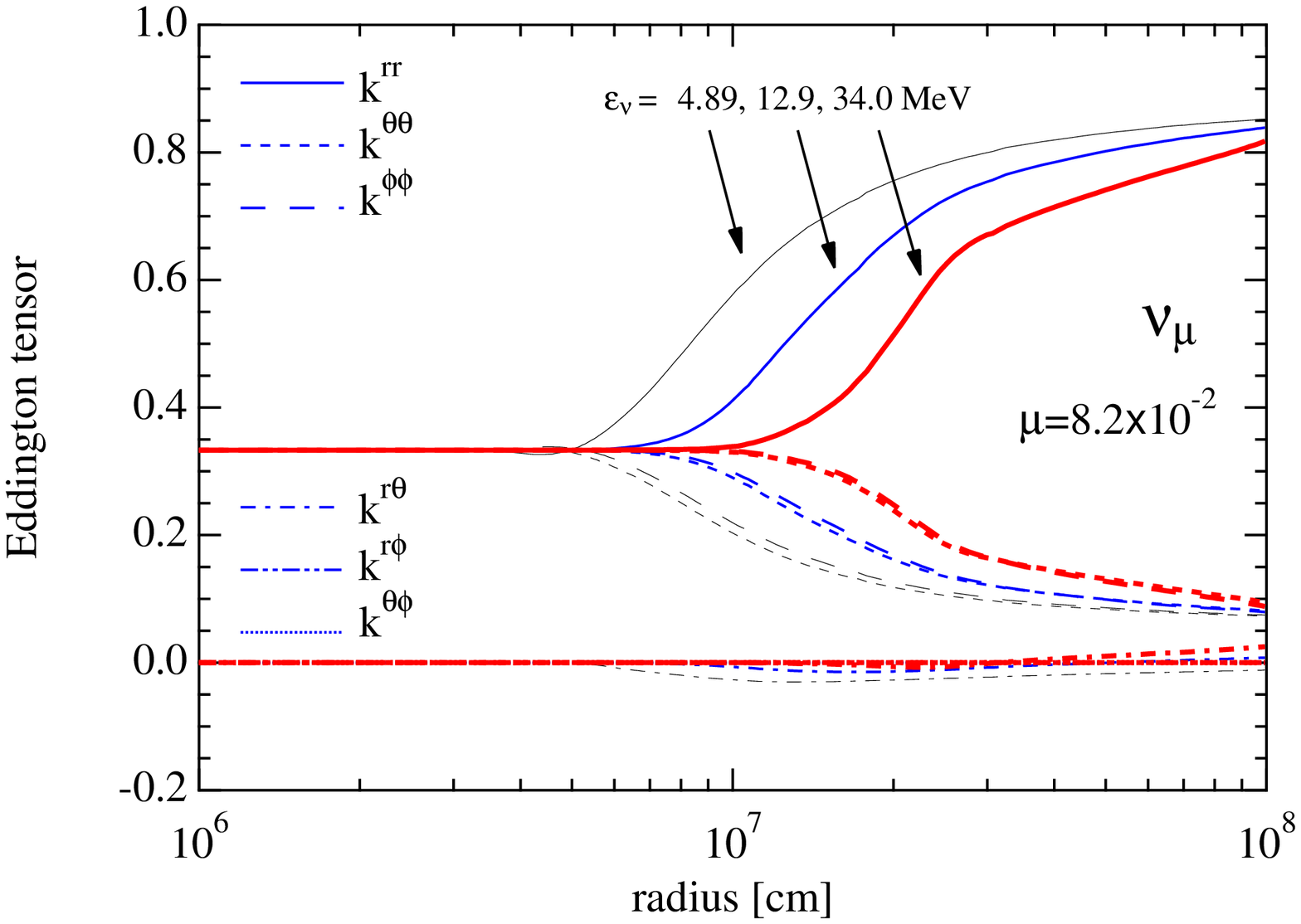}
\plotone{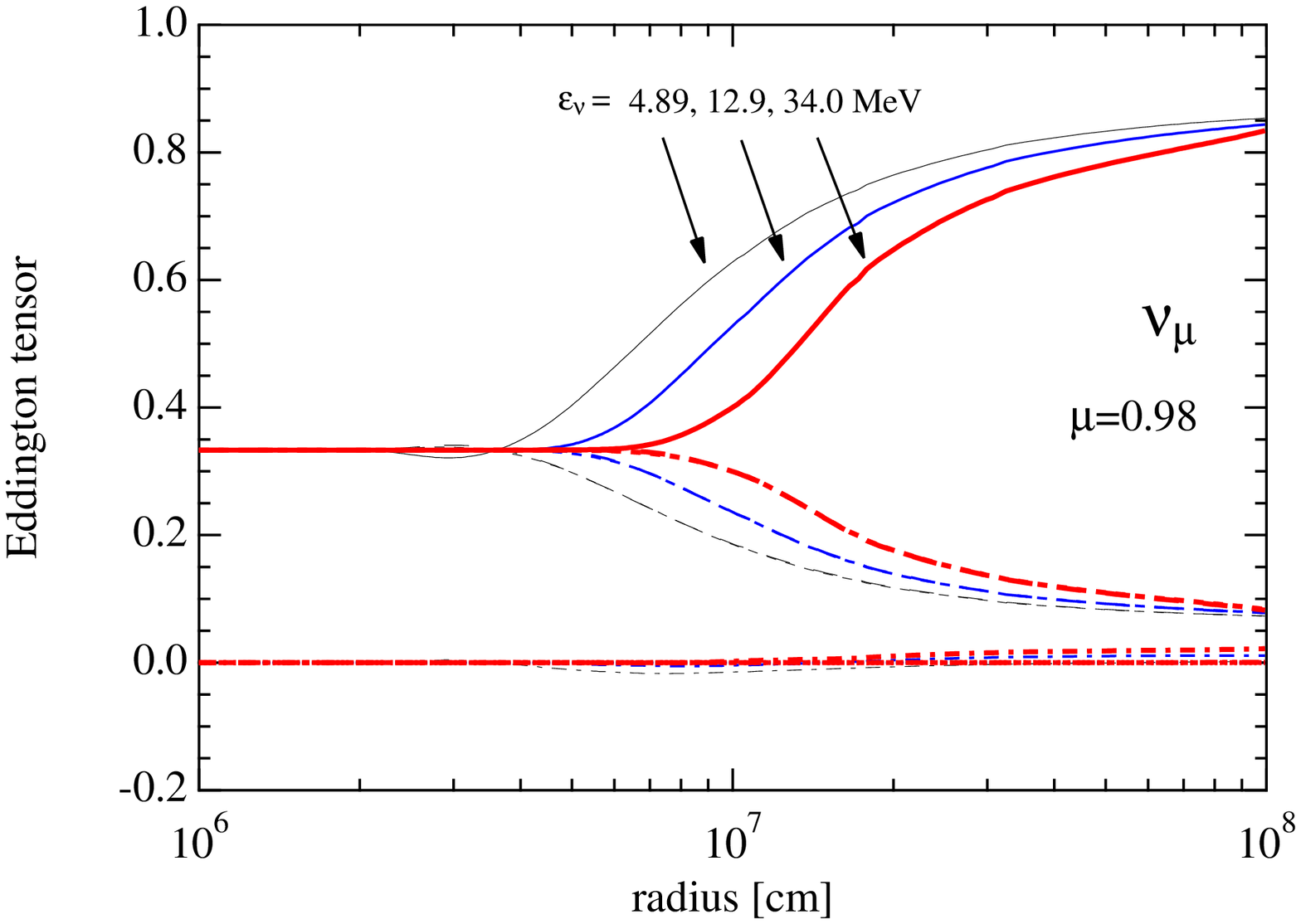}
\caption{Elements of the Eddington tensors in the deformed supernova core 
as a function of radius along the two polar directions 
with $\mu=8.2\times10^{-2}$ (near the equator: top) 
and $\mu=0.98$ (near the pole: bottom).  
The diagonal elements 
($k^{rr}$, $k^{\theta\theta}$, $k^{\phi\phi}$) 
are shown by solid, dashed and long-dashed lines, respectively.  
The non-diagonal elements 
($k^{r\theta}$, $k^{r\phi}$, $k^{\theta\phi}$) 
are shown by dot-dashed, dot-dot-dashed and dotted lines, respectively.  
The thin, normal and thick lines denote the elements 
for the neutrino energies, 4.89, 12.9 and 34.0 MeV, respectively.  }
\label{fig:2d-eddington}
\end{figure}

We examine the energy and angle moments of neutrino distributions 
in Figs. \ref{fig:2d-nueave} and \ref{fig:2d-eddington}.  
The flux factors, $\langle \mu_{\nu} \rangle$, 
are shown in the upper panel of Fig. \ref{fig:2d-nueave} 
as functions of radius along the two directions discussed above.  
The flux factors change 
from 0 at center to 1 at large radii for both directions.  
The transitional zone corresponds to the region 
with the optical depth $\sim$1 and has a different radial location 
for each direction.  
The average energies, $\langle \varepsilon \rangle$, 
are shown in the lower panel of 
Fig. \ref{fig:2d-nueave} as functions of radius.  
The radial dependence of energies for the three species of neutrinos 
reflects the thermodynamical states (degenerate or not) in the deformed core.  
The profile of the average energy for electron-type neutrinos follows 
roughly the density profile.  
The average energy declines rapidly as the density 
(and hence the degeneracy parameter given by chemical potential 
divided by temperature) decreases with radius.  
The profiles for other types of neutrinos 
(electron-type anti-neutrinos and mu-type neutrinos), on the other hand, 
reflect the temperature profile, having a peak around 10--20 km.  
The radial profiles near the equator are shifted outwards 
from those near the pole just as the other quantities seen above.  

As another measure of the transition from the opaque to transparent regimes, 
we show the Eddington tensor 
as a function of radius along the same two directions in Fig. \ref{fig:2d-eddington}.  
The diagonal elements of the Eddington tensor 
($rr$-, $\theta \theta$- and $\phi \phi$-components) 
are shown for mu-type neutrinos with three different energies.  
The diagonal elements are 1/3 
in the central region, where the matter is opaque and 
the neutrino distributions are isotropic.  
The $rr$-component increases with radius 
as the neutrino distributions become more forward-peaked, 
whereas $\theta \theta$- and $\phi \phi$-components decrease.  
All off-diagonal elements are found to be nearly zero 
at the central region and have small values also at large radii.  
The transition of the diagonal elements 
from 1/3 to 1 or 0 occurs around the neutrino sphere, 
which has larger radii for higher neutrino energies.  

We remark that the energy dependence is not so strong 
for electron-type neutrinos and anti-neutrinos 
and the radial profiles of the Eddington tensor 
are more close to each other for the three neutrino energies 
(not shown in the figure).  
This is because 
the isotropy of the neutrino distributions is maintained up to $\sim$100 km 
through the charged current process.  
This behavior is consistent with the analysis 
by \citet{ott08}.  
Although 
the density profile is rather more extended in our model 
compared with theirs, 
they indeed found the delayed decoupling 
near the equator in their rotating model 
as seen in the right panel of their Fig. 6.  

The radial distributions of the Eddington tensor also depends 
on the polar angle (see top and bottom panels of Fig. \ref{fig:2d-eddington}).  
Because of smaller density scale heights near the pole in the deformed core, 
the transition from the opaque to transparent regimes 
occurs at deeper and narrower locations near the pole 
as seen in Fig. \ref{fig:2d-profile}.  
These are just as expected intuitively for the oblate core and demonstrate that 
the new code works appropriately at least qualitatively for 
the 2D configurations considered here.  

\subsection{3D Configurations}\label{3Dconfig}

We demonstrate here the performance of the new code in 3D realistic settings.  
We study the 3D neutrino transfer utilizing deformed profiles 
in a similar manner to those studied in \S \ref{2Dconfig}.  
We modify the deformed profiles further 
by adding the dependence on azimuthal angle, $\phi$, in the scaling as 
\begin{equation}
\label{eqn:3d-rscale}
\tilde{r}=r ( 1 - \epsilon \sin \phi \sin \theta ).  
\end{equation}
The resulting 3D profiles are deformed maximally with polar dependence 
in the yz-plane ($\phi$=$\pi$/2), 
whereas they have no polar dependence in the zx-plane ($\phi$=0).  
These profiles are simple examples inspired by the asymmetric 
shape of supernova cores in the 3D standing accretion shock instability 
(SASI) \citep{blo07b,iwa08}.  

\begin{figure}
\epsscale{1.0}
\begin{center}
\plottwo{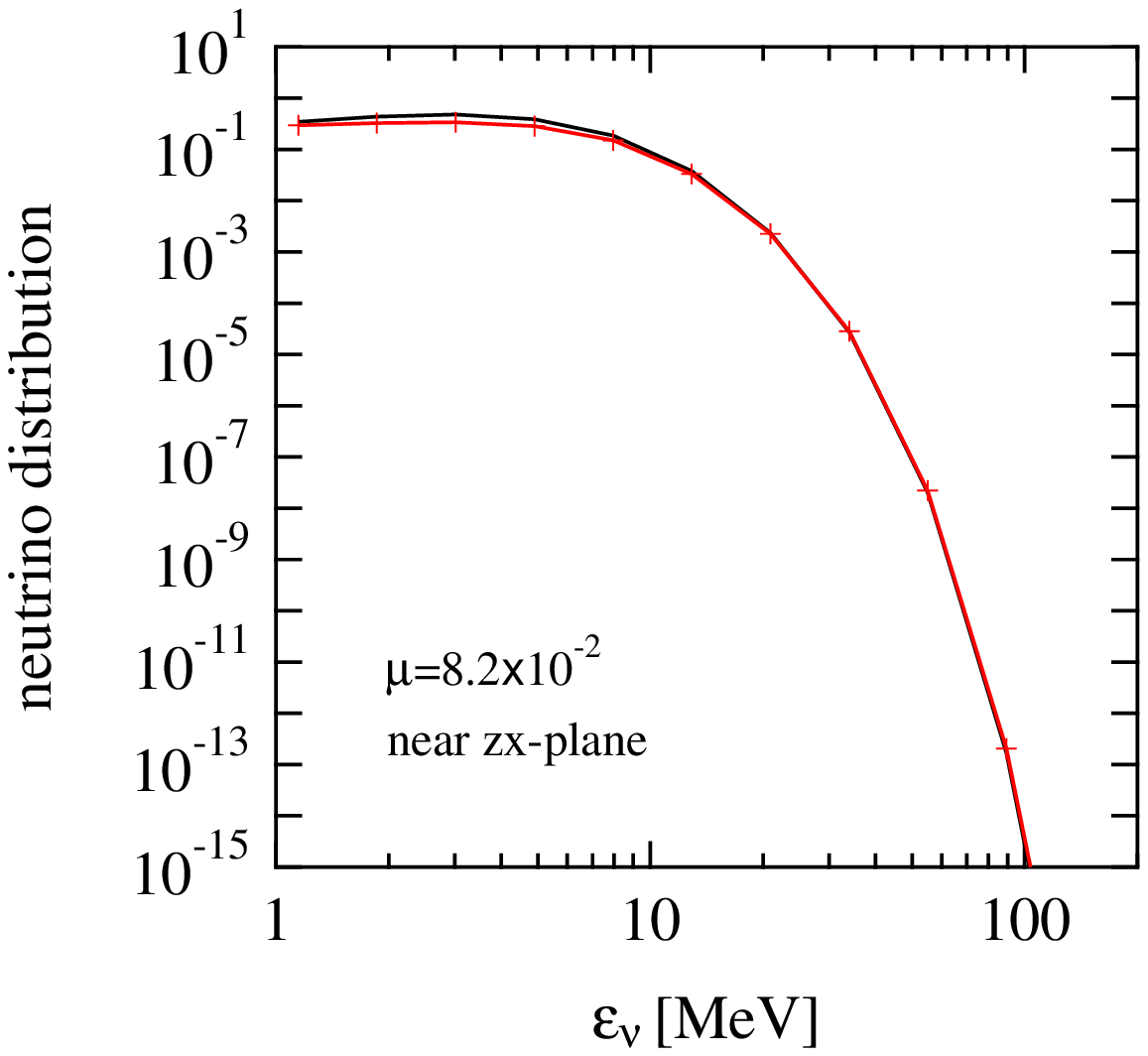}{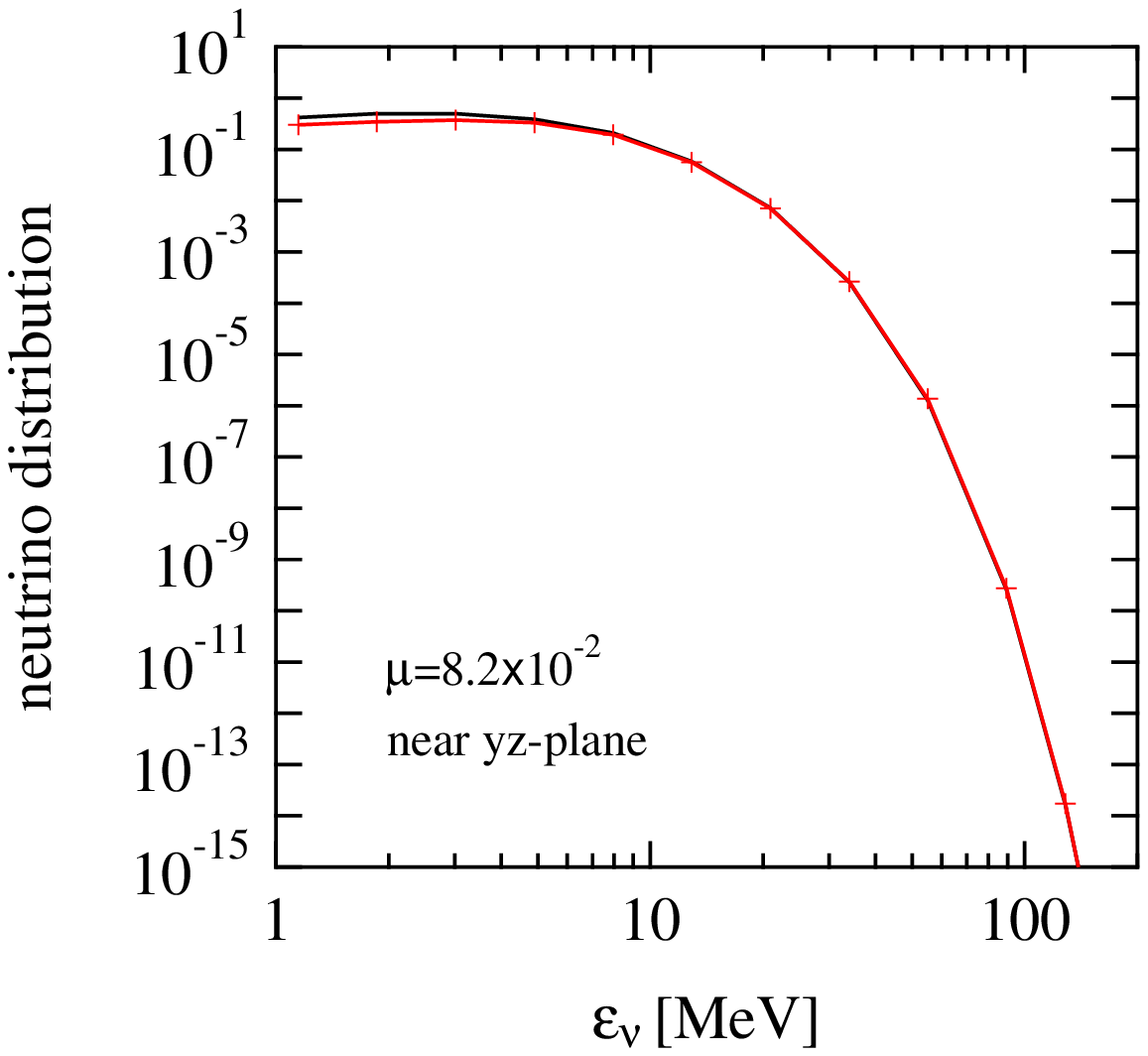}
\plottwo{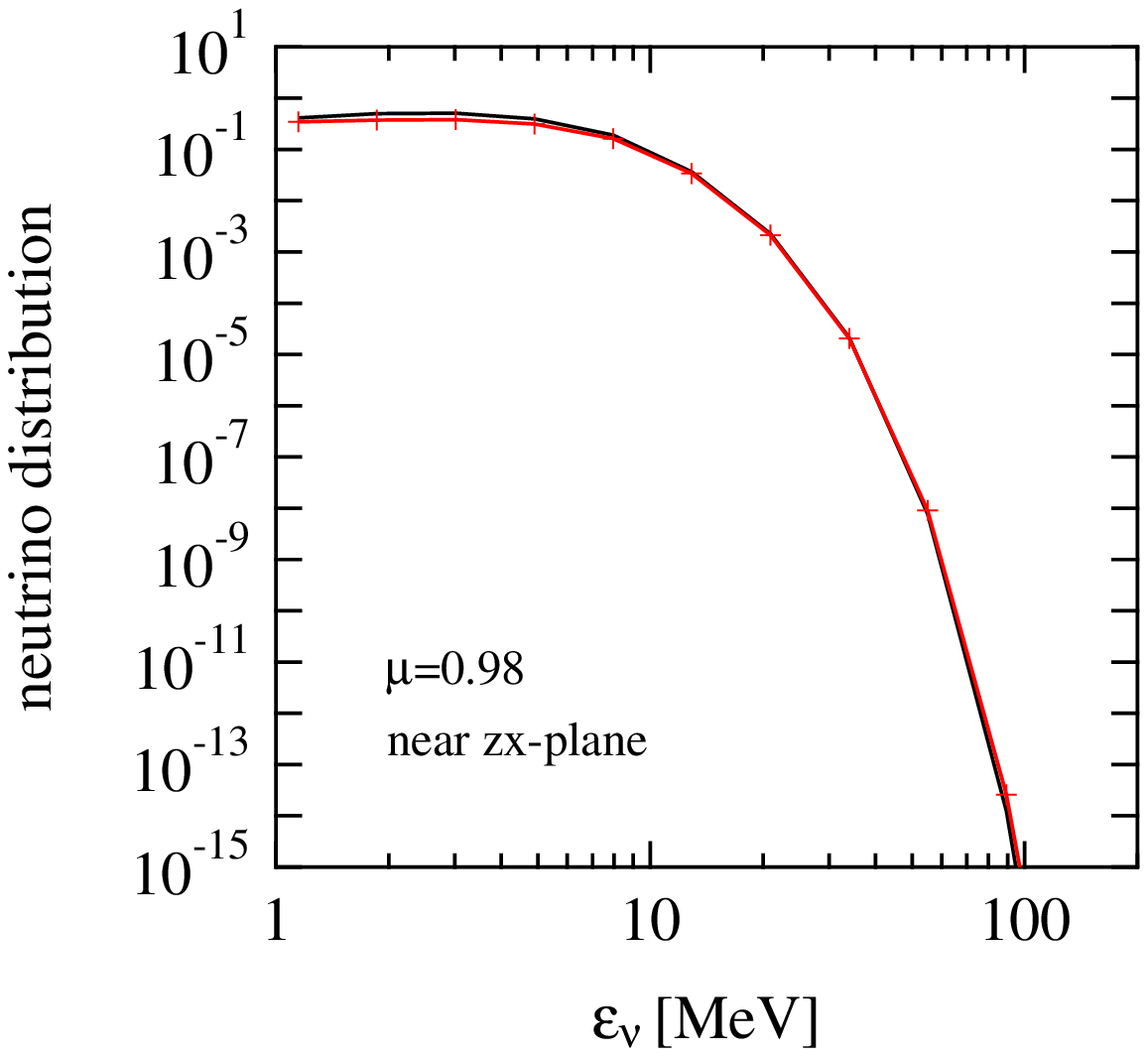}{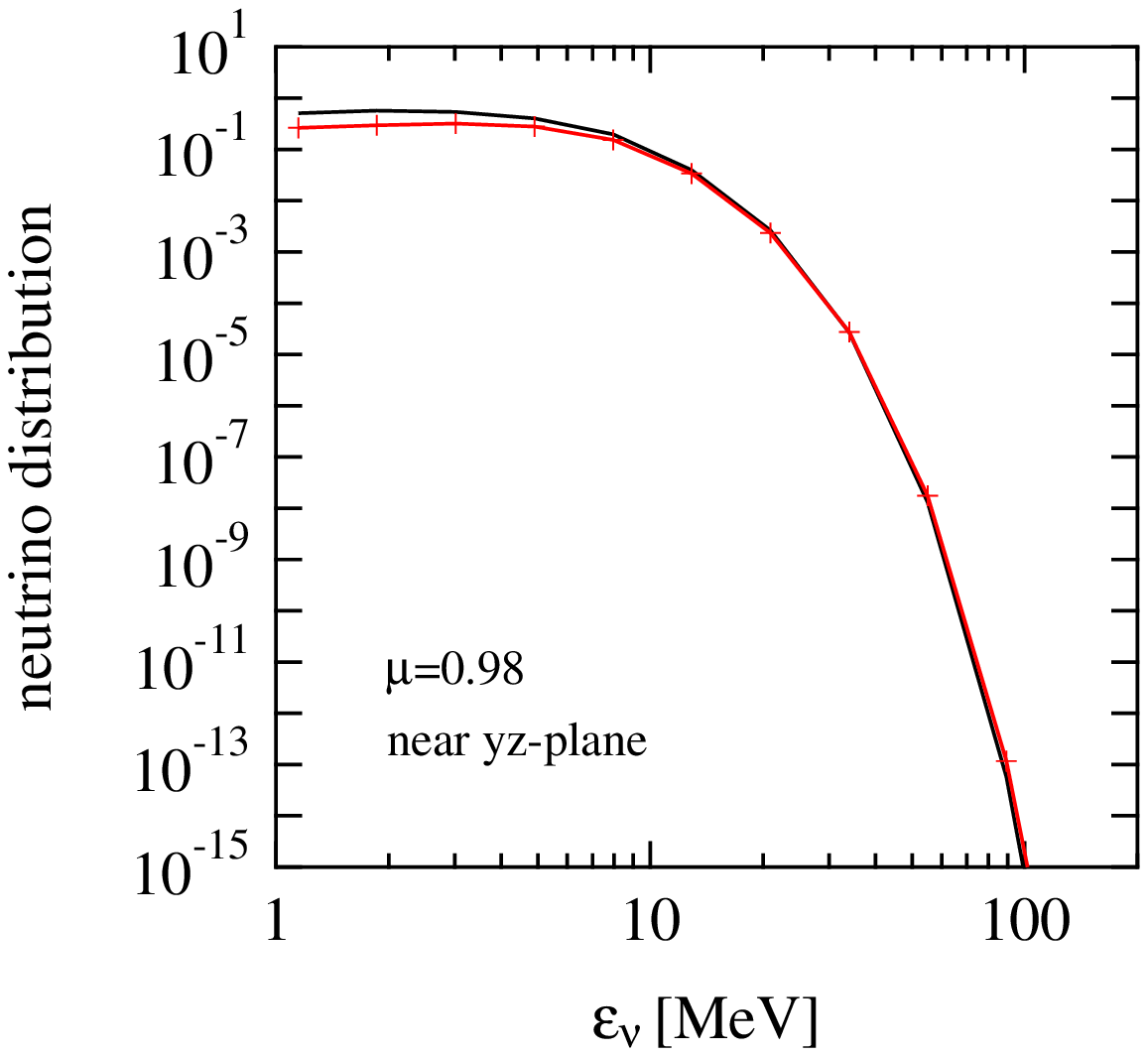}
\end{center}
\caption{Energy spectra 
by the formal (black) and steady state (red with cross symbols) solutions 
at the four locations with the radius of 98.4 km 
in the 3D deformed supernova core.  
Left and right panels show the spectra on the meridian slices 
near the zx- and yz- planes, respectively.  
The top and bottom panels correspond to the directions 
near the equator and pole, respectively.  
See the main text for the details of the locations.  }
\label{fig:3d-formal}
\end{figure}

We first check the neutrino transfer through comparisons 
with the formal solutions in the same way as in 2D.  
We treat the electron-type neutrino with the emissions and absorptions.  
We set the spatial grid with 
$N_{r}=200$, $N_{\theta}=9$ and $N_{\phi}=9$ 
and the angle grid with 
$N_{\theta_{\nu}}=6$ and $N_{\phi_{\nu}}=6$.  
Figure \ref{fig:3d-formal} shows the comparison between the steady 
and formal solutions.  
Four panels show the energy spectra at four locations 
with $r$=98.4 km in different directions.  
The left panels correspond to the spectra 
for two polar directions, 
$\mu=8.2\times10^{-2}$ (top) and 0.98 (bottom) 
on the meridian slice with $\phi$=0.262 radian (near the zx-plane).  
The right panels correspond to the spectra 
for $\mu=8.2\times10^{-2}$ (top) and 0.98 (bottom) 
on the slice with $\phi$=1.309 radian (near the yz-plane).  
The agreement between the two solutions is generally good 
for the wide range of energy, 
having relative errors within $\sim$20 \% at energies around 20 MeV.  
Larger errors at low and high energies arise 
due to the same reason as described in \S \ref{stationary}.  
The energy spectra in the right panels differ each other 
due to the deformation with polar dependence.  
The energy spectrum near the equator (right top) 
is harder than that near the pole (right bottom) 
due to different densities and temperatures at the two locations.  
The two energy spectra in the left panels 
are similar each other due to a nearly spherical 
geometry of the background.  

\begin{figure}
\epsscale{1.0}
\plotone{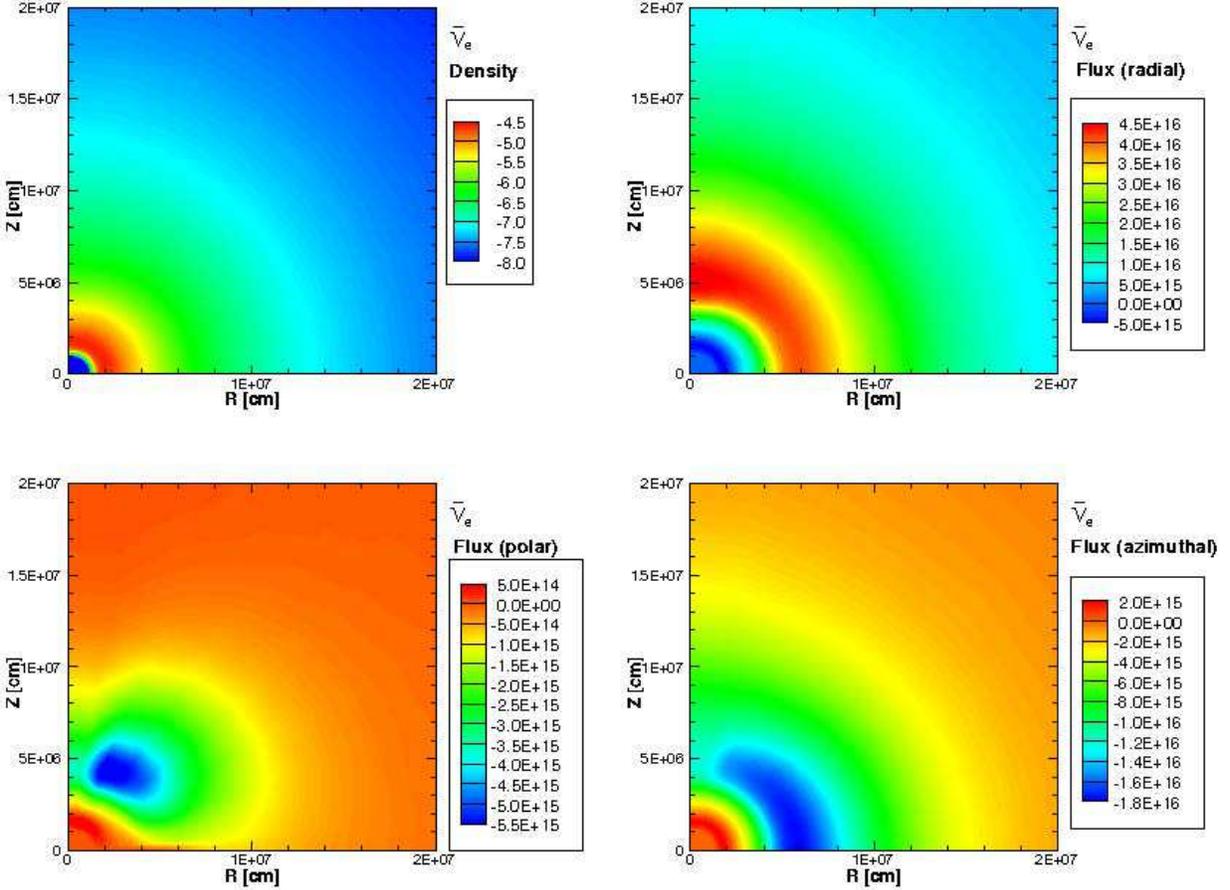}
\caption{Contour plots of the neutrino distributions 
on the meridian slice with $\phi=0.436$ radian (near zx-plane) 
for electron-type anti-neutrinos in the 3D deformed supernova core.  
The density, radial, polar and azimuthal components of the flux 
are displayed by color codes 
in left top, right top, left bottom and right bottom panels, respectively.  
The densities are shown in the unit of fm$^{-3}$=10$^{-39}$ cm$^{-3}$ 
and in log-scale.  
The fluxes are shown in the unit of fm$^{-2}$s$^{-1}$.  }
\label{fig:3d-profile.iph3}
\end{figure}

\begin{figure}
\epsscale{1.0}
\plotone{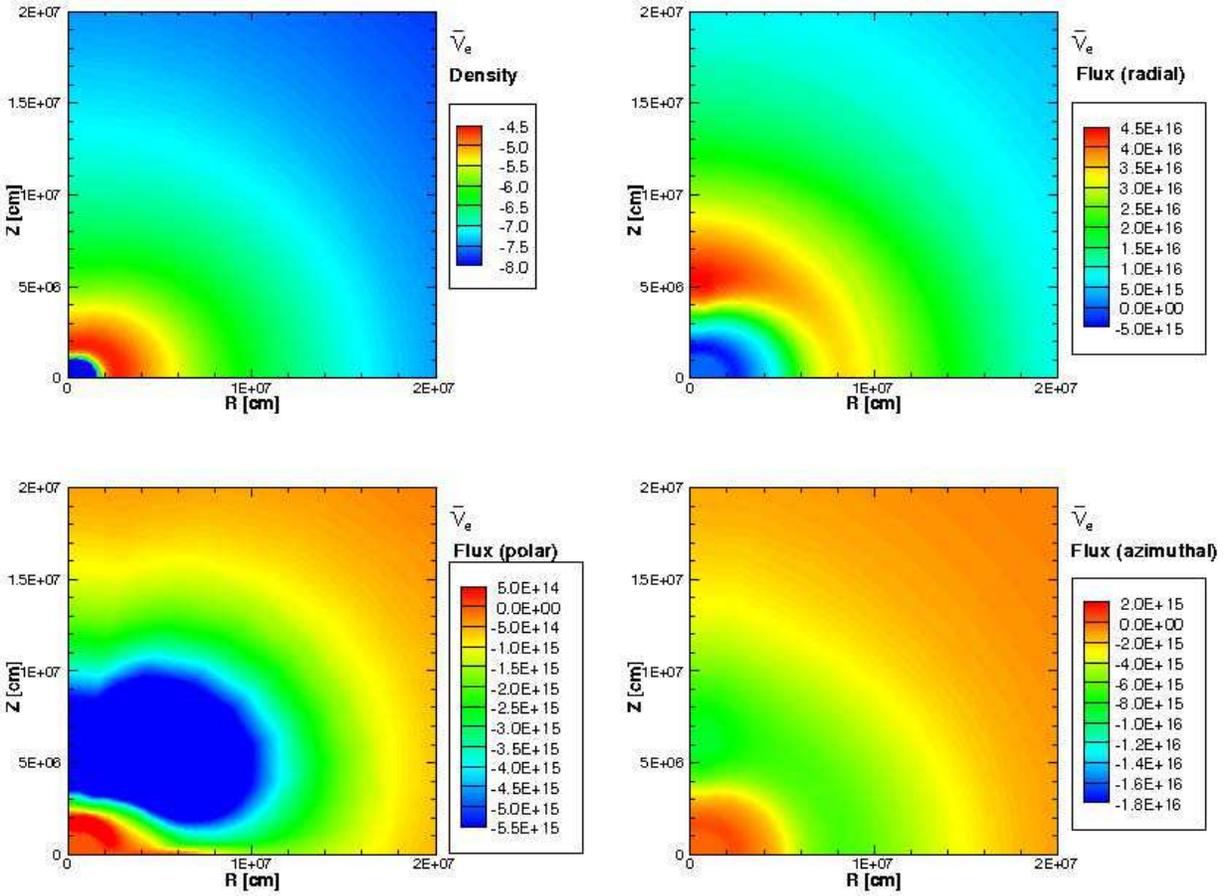}
\caption{Same as Fig. \ref{fig:3d-profile.iph3}, 
but for the meridian slice with $\phi=1.309$ radian (near yz-plane).  }
\label{fig:3d-profile.iph8}
\end{figure}

Next, we examine the numerical results with the full set of neutrino reactions.  
We set the angle grid with $N_{\theta_{\nu}}=6$ and $N_{\phi_{\nu}}=12$.  
We display 
in Figs. \ref{fig:3d-profile.iph3} and \ref{fig:3d-profile.iph8} 
the density and fluxes of electron-type anti-neutrinos 
on the slices with $\phi$=0.436 and 1.309 radian, respectively, 
by color contour maps.  
Figure \ref{fig:3d-profile.iph8} demonstrates the deformations 
of density and flux profiles near the yz-plane, 
having the background with strong polar-dependence.  
The deformed profile of neutrino density follows exactly 
the shape of the deformed profile of temperature.  
The radial flux is enhanced around the pole and 
the polar flux is appreciable in the middle region 
as discussed in \S \ref{2Dconfig}.  
The azimuthal flux is not significant 
due to a small gradient in the azimuthal direction 
near the yz-plane.  
In contrast, the density and radial flux near the zx-plane 
in Fig. \ref{fig:3d-profile.iph3} 
have less deformed shapes than those near the yz-plane.  
While the polar flux has a certain contribution in the middle, 
the azimuthal flux has a significant magnitude 
due to the azimuthal dependence of the background 
near the zx-plane.  

We stress that our code properly describes 
the polar and azimuthal transfer, 
reflecting the deformation of the 3D supernova core.  
This is the advantage of the 3D Boltzmann solver in 3D profiles.  
The non-radial transfer cannot be described 
correctly in the ray-by-ray approach.  
We found that the polar and azimuthal fluxes 
are appreciable as compared with the radial flux in a wide region 
as seen in Figs. \ref{fig:3d-profile.iph3} and \ref{fig:3d-profile.iph8}. 
The non-radial fluxes are significant even at $r \ge$100 km
and spread beyond the diffusion regime.  
In this case, the reliability of the flux limited diffusion 
approximation is doubtful.  
Our code is, therefore, a unique tool to study 
the neutrino transfer in 3D configurations 
and to examine the approximations used 
in the state-of-the-art supernova simulations.  

\begin{figure}
\epsscale{1.0}
\begin{center}
\plottwo{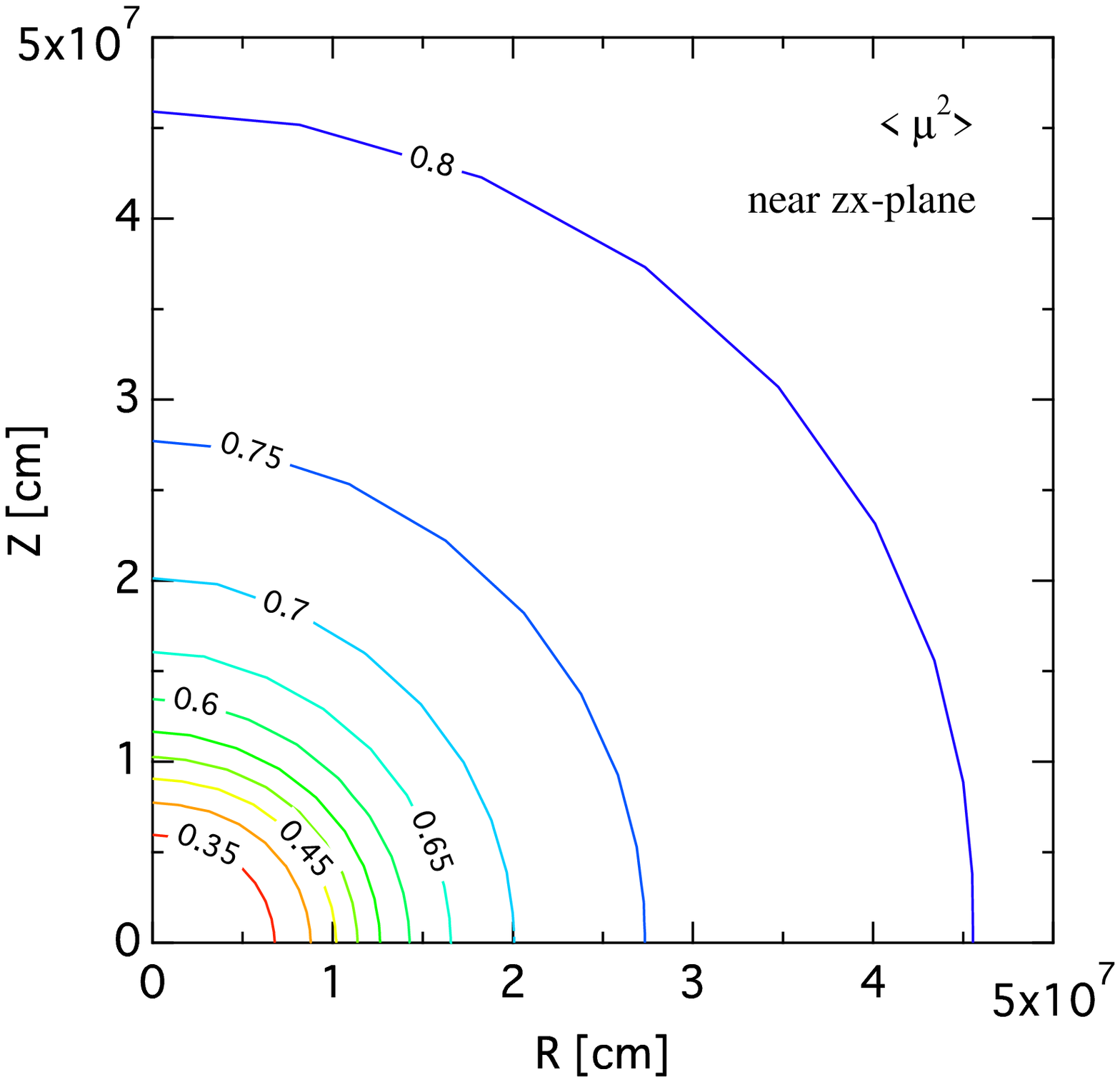}{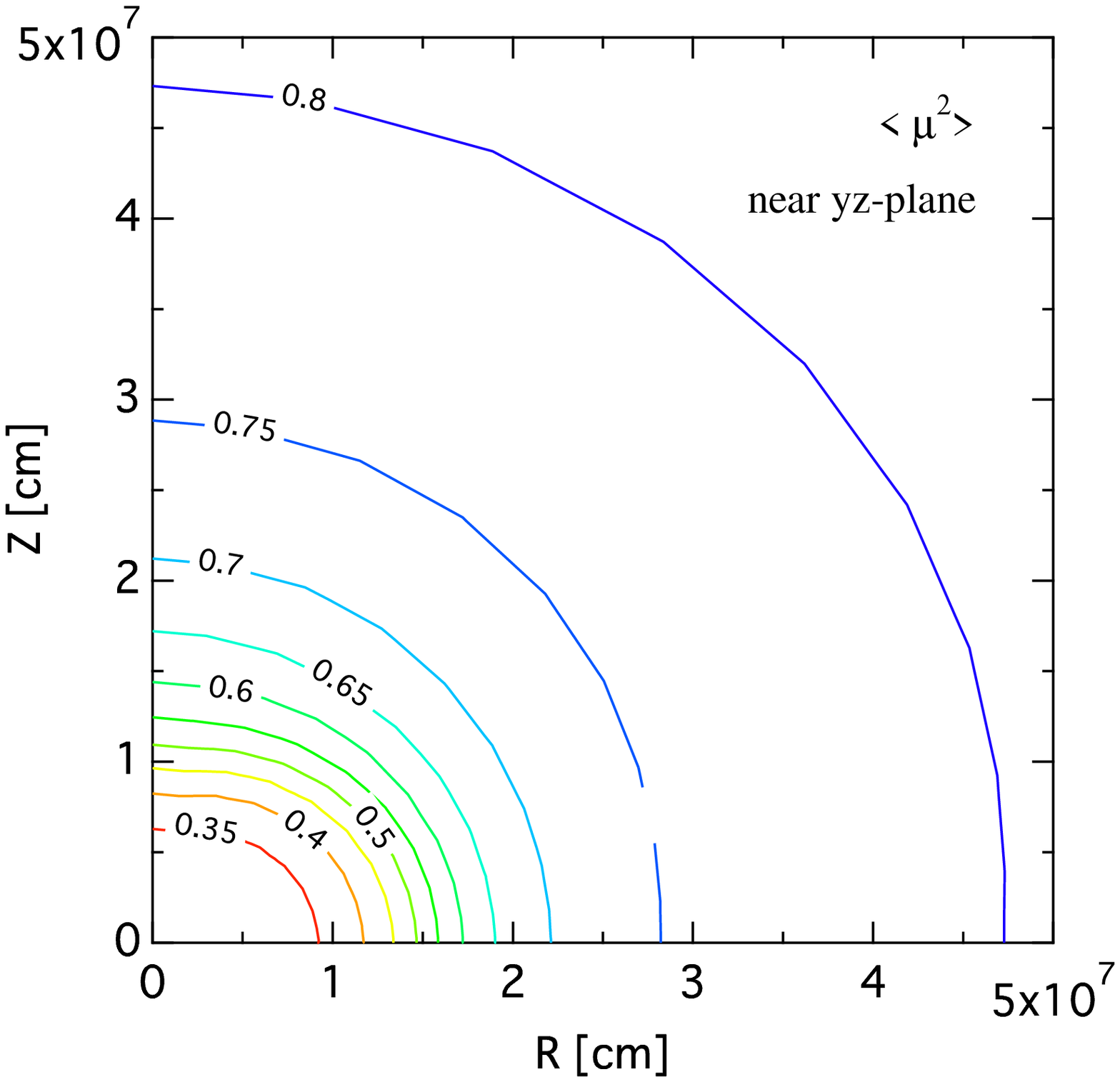}
\plottwo{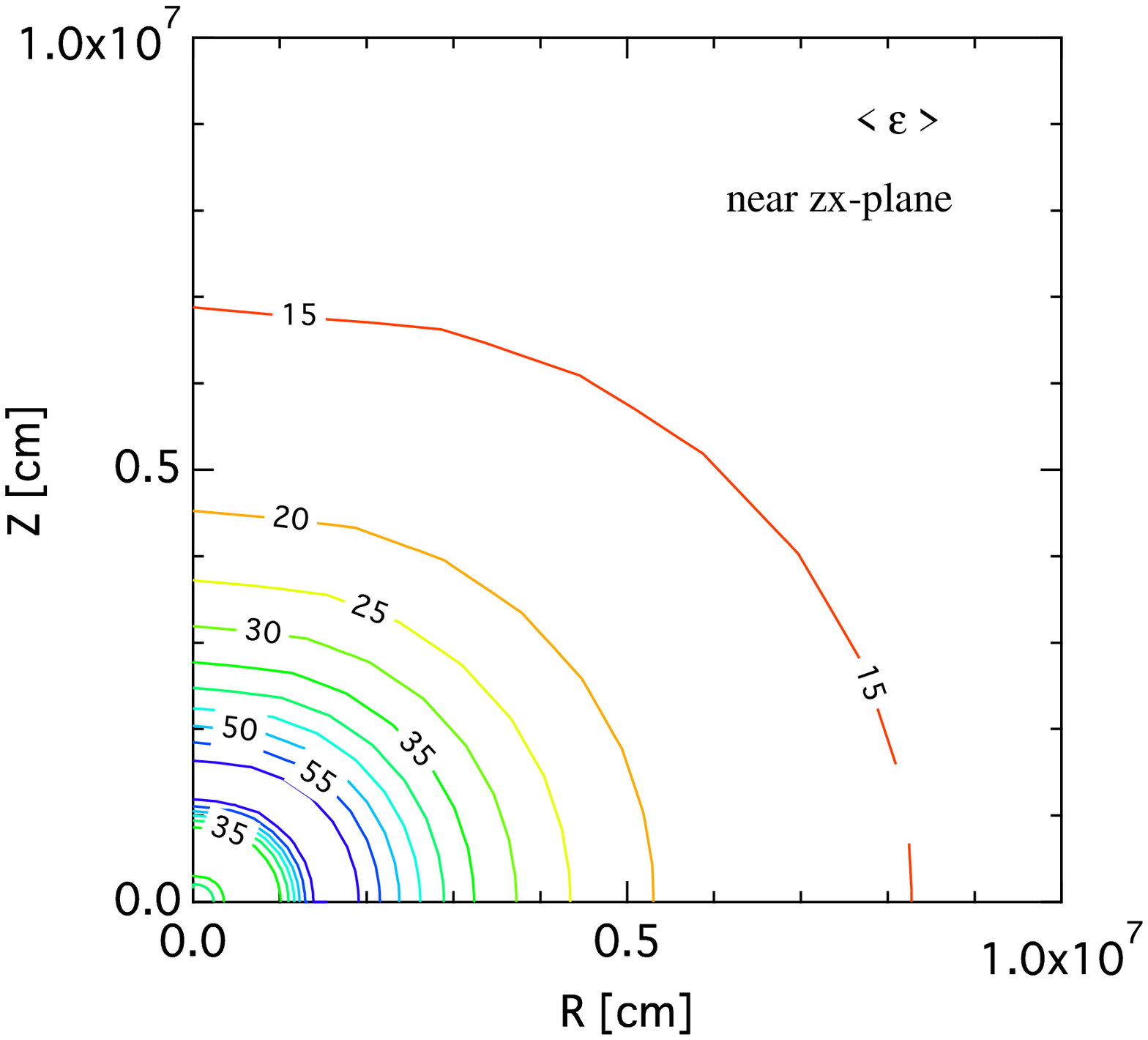}{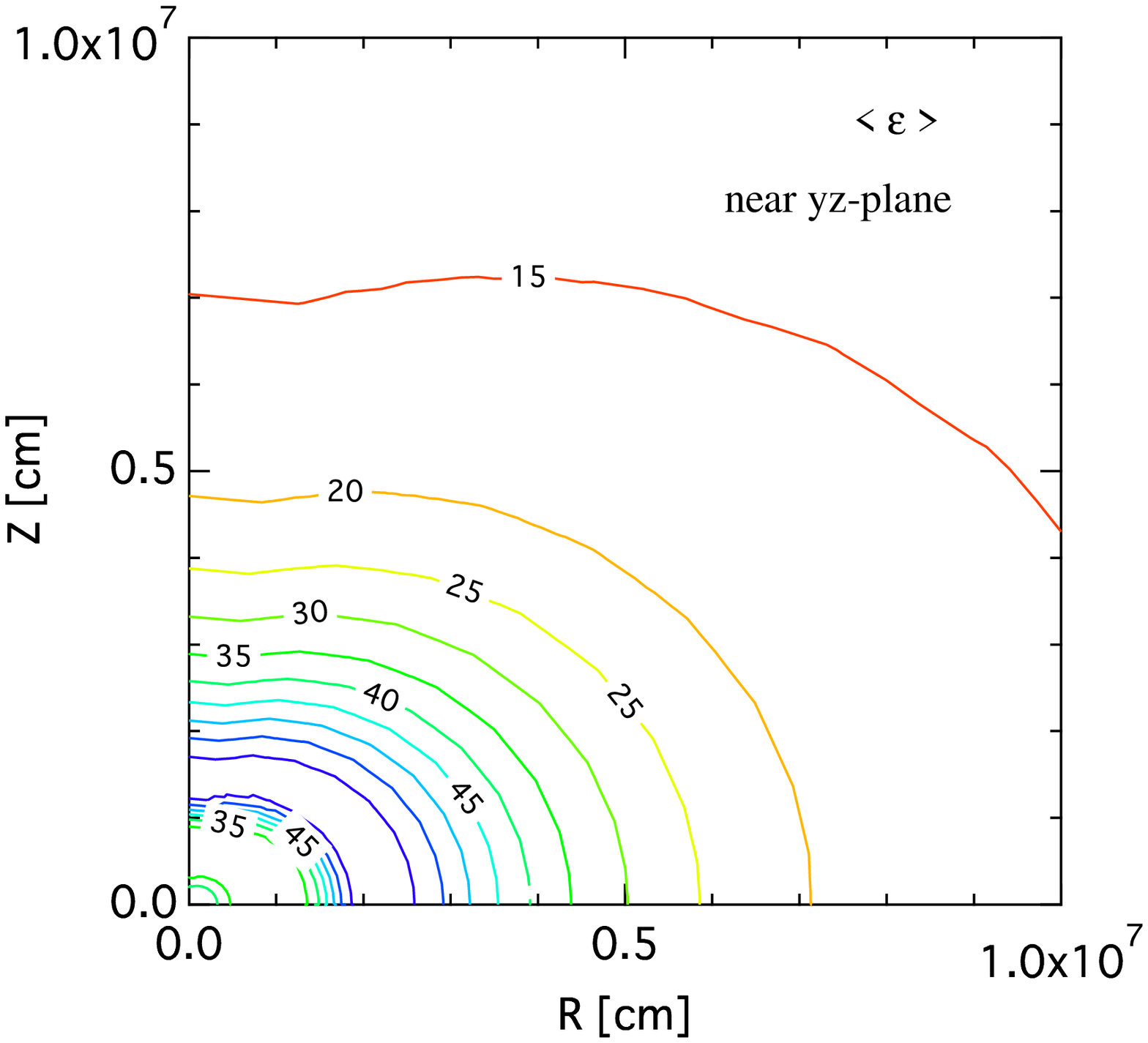}
\end{center}
\caption{Contour plots of 
the angle moments, $\langle \mu^2 \rangle$, (top) and 
the average energies, $\langle \varepsilon \rangle$, (bottom) 
for electron-type anti-neutrinos 
for the slices 
with $\phi=0.436$ (left) and $\phi=1.309$ (right) radian.  
Note that the covered scales are different between top and bottom panels.  
}
\label{fig:3d-eavemu}
\end{figure}

We examine the energy and angle moments of the neutrino distributions 
in Figure \ref{fig:3d-eavemu}.  
Contour plots show the angle moments, $\langle \mu^2 \rangle$, and 
the average energies, $\langle \varepsilon \rangle$, 
for electron-type anti-neutrinos on the slices with 
$\phi=0.436$ (near the zx-plane: left) and 
$\phi=1.309$ (near the yz-plane: right) radian.  
Deformation is strong near the yz-plane 
and weak near the zx-plane due to the polar and azimuthal dependence.  
The angle moments have an isotropic value of 1/3 at the central region 
as shown by the deformed shape of contour lines.  
They approach $\sim$0.8 at $\sim$500 km 
independent of the polar direction 
as the angle distributions approach a forward-peaked shape.  
The contour lines for the average energy have also a deformed 
shape, reflecting the deformed supernova core.  
The average energies at large radii ($\sim$1000 km, not shown in the figure) 
do not show any significant asymmetry.  

\begin{figure}
\epsscale{0.7}
\plotone{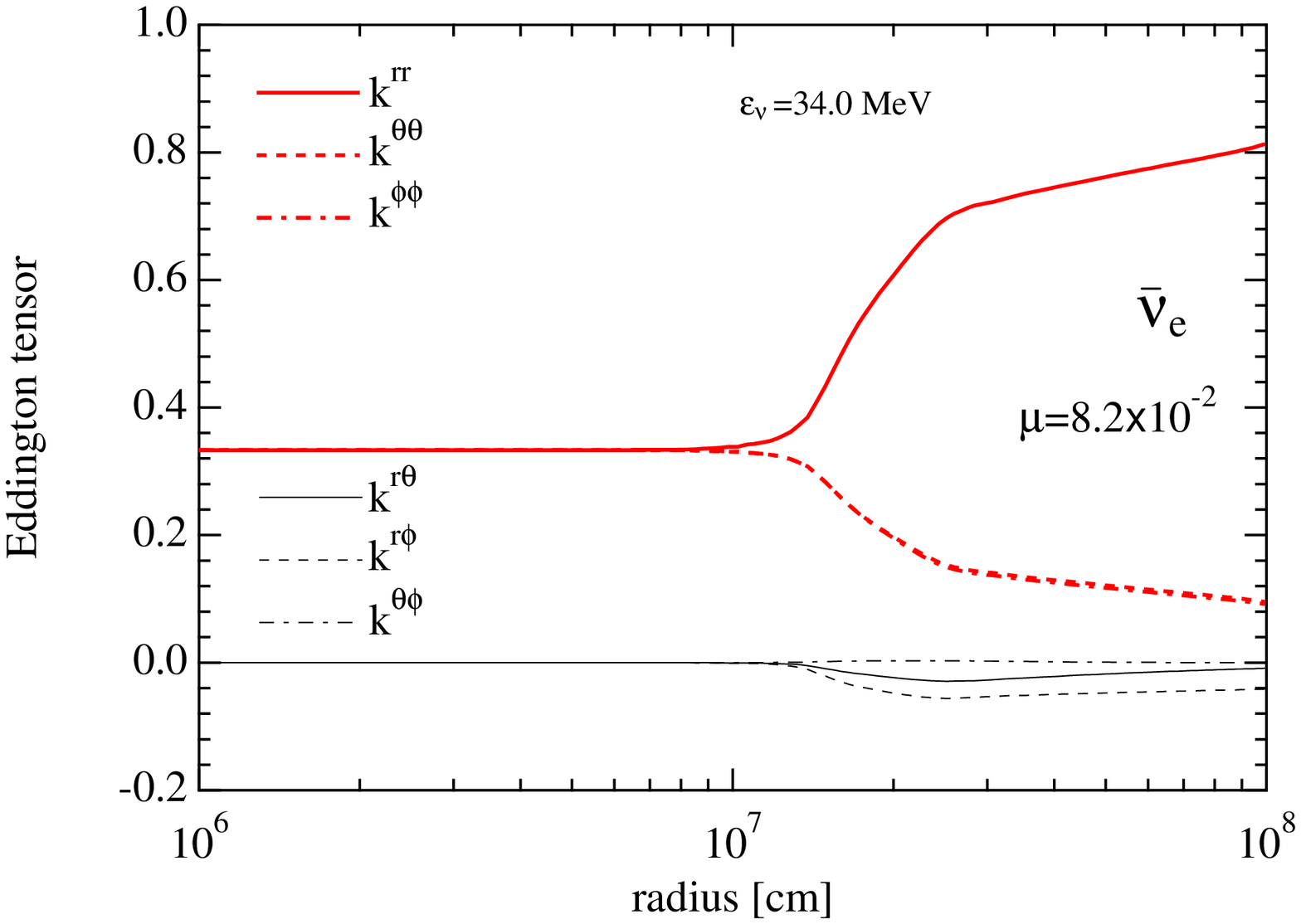}
\plotone{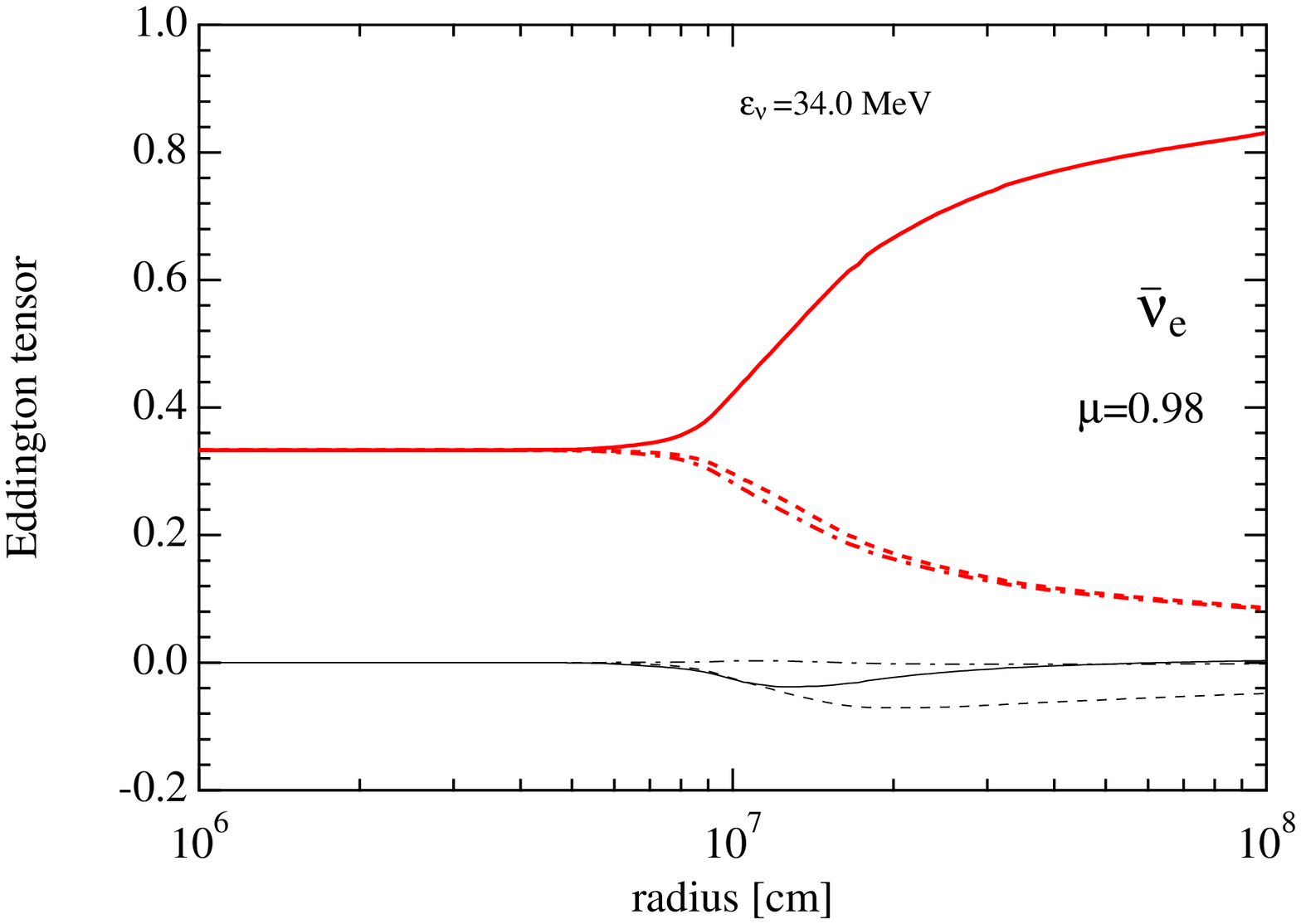}
\caption{Radial profiles of the elements of the Eddington tensor 
along the two polar directions 
with $\mu=8.2\times10^{-2}$ (top) and $\mu=0.98$ (bottom) 
on the slice with $\phi=1.309$ radian (near yz-plane).  
The elements are shown 
for electron-type anti-neutrinos 
with the neutrino energy of 34.0 MeV.  
The diagonal elements (thick) are shown by 
solid, dashed and dot-dashed lines 
for $k^{rr}$, $k^{\theta\theta}$, $k^{\phi\phi}$, respectively.  
The non-diagonal elements (thin) are shown by 
solid, dashed and dot-dashed lines 
for $k^{r\theta}$, $k^{r\phi}$, $k^{\theta\phi}$, respectively.  
}
\label{fig:3d-eddington}
\end{figure}

Figure \ref{fig:3d-eddington} shows the elements of 
the Eddington tensor as functions of radius along the two polar directions, 
$\mu=8.2\times10^{-2}$ (top) and $\mu=0.98$ (bottom), 
on the slice with $\phi=1.309$ radian (near the yz-plane).  
Six elements of the tensor are shown for electron-type anti-neutrinos 
at energy grid point of 34.0 MeV.  
The transitions from the isotropic values (1/3 for diagonal 
and 0 for non-diagonal elements) occur further locations 
in the direction near the equator than those near the pole.  
This dependence on the polar direction is small 
near the zx-plane and the transitions occur similar locations 
(not shown here).  
The magnitude of non-diagonal elements is appreciable 
at outer layers having the azimuthal dependence in the 3D supernova core.  
The transition from the isotropic regime to the free-streaming regime 
is described well by the new code as in the 1D and 2D cases.  
Our code is also useful to examine the Eddington 
tensors for general 3D profiles and to provide valuable 
information for the approaches using variable Eddington factors.  

\section{Summary and Discussion}\label{summary}

We have developed the numerical code to calculate the neutrino 
transfer with multi-energy and multi-angle in three dimensions 
for the study of core-collapse supernovae.  
The numerical code solves the Boltzmann equations 
for the neutrino distributions by the discrete-ordinate (S$_n$) method.  
The time step is advanced by the fully implicit differencing.  
The neutrino distribution is a function of time, 
three spatial coordinates, neutrino energy and two angles.  
We solve, therefore, the time evolution of the distributions 
in six phase-space dimensions.  
An essential set of neutrino reactions 
(emission, absorption, scattering and pair processes) is implemented 
in the collision term of the Boltzmann equation.  
The Boltzmann equations are formulated in the inertial frame 
as the basis of our developments.  
In the current study for the static background, 
we drop the velocity dependent terms in the collision term.  
The collision term of the pair-processes is linearized 
by assuming the neutrino distributions for the counterpart 
of the neutrino pair.  
The set of equations for the discretized form 
of the Boltzmann equations is formulated as 
a linear equation for the vector of neutrino distributions.  
The large sparse matrix with block diagonal matrices 
is solved by an iterative method.  

We have performed the numerical simulations 
by the 3D Boltzmann solver 
to validate the numerical code and 
to apply it to the realistic profiles of supernova core.  
The numerical code of the 3D Boltzmann solver 
describes correctly the neutrino transfer in the diffusion and 
free streaming limits.  
The collision term by the neutrino reactions 
has been checked by the formal solutions as well as 
the analytic evolution toward the equilibrium.  
The pair processes are well described in an approximate expression 
for the linearization of the collision term.  
We have examined the neutrino transfer in the profiles 
of the supernova cores before and after the bounce 
using the numerical results from the previous simulations 
under spherical symmetry.  
The 3D calculations using the spherical configuration show 
the good agreement in neutrino distributions 
with the spherical calculations.  

We have demonstrated that 
the ability of the 3D numerical code to solve the neutrino transfer 
for artificially deformed supernova cores in 2D and 3D.  
The new code describes properly the multi-dimensional feature 
such as lateral and azimuthal fluxes, which cannot be handled 
in approximate methods.  
The moments of energy and angle of neutrino distributions in 3D 
can be revealed by the new code and will be helpful to gauge 
simplified transfer approaches.  

Toward the full calculation of core-collapse supernovae 
by the neutrino-radiation hydrodynamics, 
there are necessary developments on top of the current 3D code.  
First of all, we need to link the 3D Boltzmann solver 
with the numerical code for the 3D hydrodynamics.  
This development for the neutrino-radiation hydrodynamics 
is now under way.  
In order to describe the neutrino transfer in hydrodynamics, 
it is necessary to take into account the relativistic effects 
($v/c$ terms) 
such as the Doppler shifts of neutrino energy and 
the aberration of neutrino angles.  
We plan to evaluate the collision term by considering 
the Lorentz transformation of the neutrino distributions 
between the inertial and comoving frames.  
This causes changes of neutrino energies by scatterings 
in the inertial frame even for the iso-energetic scattering 
in the comoving frame.  
Therefore, the consideration of energy change 
in the collision term becomes mandatory.  
Studies by the developments mentioned above 
will be reported as the following articles 
after the current study in series.  

Further generalization of the 3D numerical code 
would require supercomputers in Exa-scale.  
In principle, one has to include the energy-changing 
reactions such as electron-neutrino scatterings 
in the collision term.  
Considering the energy coupling makes the size of 
block matrix larger and the computation formidable.  
The non-linear treatment of the pair processes 
requires iterations of the whole solution.  
The general relativistic treatment of the whole framework 
makes the computation further challenging.  
Therefore, the full 3D simulations with the extensions 
mentioned above awaits the next generation supercomputers 
to perform.  
The current study proves that the calculation of the neutrino transfer 
in 3D is now feasible, marking the first step toward this direction.  


\acknowledgments

We are grateful to 
K. Nakazato, K. Kotake, K. Kiuchi, N. Ohnishi and H. Suzuki 
for the fruitful collaborations and the profitable discussions 
on supernova simulations.  
We thank 
T. Takiwaki for providing the data of profiles 
from his 3D simulations of core-collapse supernovae and 
H. Nagakura and S. Furusawa for collaborative works.  
K. S. is grateful to S. Nasu and T. Sato for the investigation 
on the evaluation of neutrino reactions.  
We express our gratitude to T.-H. Janka, E. M\"uller and W. Hillebrandt 
for the hospitality at Max Planck Institut f\"ur Astrophysik 
where the developement was initiated during our stay.  
K. S. expresses his thanks to C. Ott and T. Mezzacappa for valuable comments 
on the numerical methods of neutrino transfer in 2D and 3D.  
K. S. thanks the advice on parallel computing from H. Matsufuru, 
A. Imakura and T. Sakurai as well as 
the support by S. Hashimoto, S. Aoki, M. Shibata and H. Toki 
toward the full simulation of 3D supernovae by supercomputing 
beyond the peta-scale.  
K. S. thanks also H. Monden-Torihata for the technical assistance 
on 3D graphics.  

The numerical computations in this work were performed on the supercomputers 
at Research Center for Nuclear Physics (RCNP) in Osaka University, 
The University of Tokyo, 
Yukawa Institute for Theoretical Physics (YITP) in Kyoto University, 
Japan Atomic Energy Agency (JAEA) and 
High Energy Accelerator Research Organization (KEK).  

A part of this work has been done during the research workshops: 
"Microphysics in Computational Relativistic Astrophysics" 
at The Perimeter Institute for Theoretical Physics in Ontario, Canada and 
"Extreme Computing and its Implications 
for the Nuclear Physics / Applied Mathematics / Computer Science Interface" (INT-11-2a) 
at The Institute for Nuclear Theory, University of Washington, USA.  
K. S. would like to express his gratitude for their supports 
and the discussions exchanged with the participants during the stay.  

This work is partially supported by 
the Grant-in-Aid for Scientific Research on Innovative Areas (Nos. 20105004, 20105005) 
and 
the Grant-in-Aid for the Scientific Research (Nos. 19104006, 21540281, 22540296) 
from the Ministry of Education, Culture, Sports, Science and Technology (MEXT) in Japan.  

This numerical study on core-collapse supernovae using the supercomputer 
facilities is supported by the HPCI Strategic Program of MEXT, Japan.  



\appendix

\section{Appendices}\label{appendix}

\subsection{Neutrino Direction}\label{direction}

We define the unit vectors 
at the position ($r$, $\theta$, $\phi$) as

\begin{equation}
\mathbf{e}_{r} = {\rm sin}~\theta ~{\rm cos}~\phi ~\mathbf{i}
               + {\rm sin}~\theta ~{\rm sin}~\phi ~\mathbf{j}
               + {\rm cos}~\theta                ~\mathbf{k} ,
\end{equation}
\begin{equation}
\mathbf{e}_{\theta} = {\rm cos}~\theta ~{\rm cos}~\phi ~\mathbf{i}
                    + {\rm cos}~\theta ~{\rm sin}~\phi ~\mathbf{j}
                    - {\rm sin}~\theta                 ~\mathbf{k} ,
\end{equation}
\begin{equation}
\mathbf{e}_{\phi} = - {\rm sin}~\phi ~\mathbf{i}
                    + {\rm cos}~\phi ~\mathbf{j} .
\end{equation}
The reversed transformation can be found in \citep{pom73}.  
The neutrino direction in the inertial frame is defined 
by two angles, $\theta_{\nu}$ and $\phi_{\nu}$, 
with respect to these unit vectors.  
The polar angle between the neutrino direction, {\boldmath $n$}$^{in}$, and 
the radial coordinate, $\mathbf{e}_{r}$, 
is denoted by $\theta_{\nu}$.
The azimuthal angle around the radial coordinate 
from the $\mathbf{e}_{\theta}$-direction is denoted by $\phi_{\nu}$.  
The three components of {\boldmath $n$}$^{in}$ are given by
\begin{equation}
n^{in}_{r}= {\rm cos}~\theta_{\nu},
\end{equation}
\begin{equation}
n^{in}_{\theta}= {\rm sin}~\theta_{\nu}~{\rm cos}~\phi_{\nu},
\end{equation}
\begin{equation}
n^{in}_{\phi}= {\rm sin}~\theta_{\nu}~{\rm sin}~\phi_{\nu}.
\end{equation}
They can be expressed in terms of the angle variable as 
\begin{equation}
n^{in}_{r}= \mu_{\nu},
\end{equation}
\begin{equation}
n^{in}_{\theta}= (1 - \mu_{\nu}^2)^{\frac{1}{2}}~{\rm cos}~\phi_{\nu},
\end{equation}
\begin{equation}
n^{in}_{\phi}  = (1 - \mu_{\nu}^2)^{\frac{1}{2}}~{\rm sin}~\phi_{\nu}.
\end{equation}

\subsection{Computational Grids}\label{scheme-grid}

We arrange the radial, polar and azimuthal coordinates 
by $N_r$, $N_{\theta}$ and $N_{\phi}$ grid points.  
We cover the neutrino energy and angles 
by $N_{\varepsilon}$, $N_{\theta_{\nu}}$ and $N_{\phi_{\nu}}$ grid points.  
Hereafter, we use lowercase subscripts (e.g. $i$, $j$) for the mesh centers 
and uppercase subscripts (e.g. $I$, $J$) to index the location.  
We assign the integer index from 1 to $N_{d}$ for the cell centers 
and the integer index from 0 to $N_{d}$ for the cell interfaces, 
where $N_{d}$ is the number of grid points 
for the coordinates ($N_{r}$, $N_{\theta}$ and $N_{\phi}$).  
The $i$-th cell center is located between the ($I-1$)th and $I$th interfaces 
\citep[e.g.][]{yam97}.  

The radial grid points at the cell interface are denoted by $r_I$, 
which covers from $r_{I=0}$ to $r_{I=N_{r}}$.  
We define the radial grid points at the cell center by 
\begin{equation}
\label{eqn:rdc}
r_i = \left[~\frac{r_{I-1}^{3} + r_{I}^{3}}{2}~\right]^{\frac{1}{3}} ~.  
\end{equation}
We determine the grid points for the neutrino angle, $\mu_{\nu}=\cos \theta_{\nu}$, 
by the points and weights obtained from the formula of the Gaussian quadrature.  
The $\mu_{\nu}$-grid points for the cell center, ${\mu_{\nu}}_{j}$, 
are given by the quadrature points. 
The $\mu_{\nu}$-grid points for the cell interface are determined by 
\begin{equation}
\label{eqn:munu}
{\mu_{\nu}}_{J} = {\mu_{\nu}}_{J-1} + {d\mu_{\nu}}_{j} ~, 
\end{equation}
where ${d\mu_{\nu}}_{j}$ are the quadrature weights 
for the cell containing ${\mu_{\nu}}_{j}$.
Starting with ${\mu_{\nu}}_{J=0}=-1$, we can determine 
the $\mu_{\nu}$-grid points up to ${\mu_{\nu}}_{J=N_{\theta_{\nu}}}=1$.  
The angle factor at the cell interface, 
which is necessary for the evaluation of the $\mu_{\nu}$-advection, 
is obtained by 
\begin{equation}
\label{eqn:munu2}
(1-{\mu_{\nu}}^2)_{J} = (1-{\mu_{\nu}}^2)_{J-1} - 2 {\mu_{\nu}}_{j} {d\mu_{\nu}}_{j} ~.  
\end{equation}
The angle factor at the boundary is set as $(1-{\mu_{\nu}}^2)_{J=0}=0$ 
and those at other grid points are determined accordingly 
up to $(1-{\mu_{\nu}}^2)_{J=N_{\theta_{\nu}}}=0$.   
We use indices $i$ and $I$ for spatial coordinates and 
$j$ and $J$ for neutrino angle coordinates, hereafter.  

We arrange the grid points for the polar coordinate, $\theta$, 
in terms of $\mu=\cos \theta$ 
in the same manner as the $\mu_{\nu}$-grid points described above.  
We set the grid for the cell center, $\mu_{i}$, and 
the corresponding weights, $d\mu_{i}$, from the Gaussian quadrature points.  
The grid points for the cell interface are determined by 
\begin{equation}
\label{eqn:mu}
{\mu}_{I} = {\mu}_{I-1} + {d\mu}_{i} ~,
\end{equation}
to cover from ${\mu}_{I=0}$ to ${\mu}_{I=N_\theta}$.  
The angle factor at the cell interface for the polar advection 
is given by 
\begin{equation}
\label{eqn:mu2}
{(1-{\mu}^2)^{\frac{1}{2}}}_{I} = {(1-{\mu}^2)^{\frac{1}{2}}}_{I-1} 
- \frac{{\mu}_{i}}{(1-{\mu_i}^2)^{\frac{1}{2}}} {d\mu}_{i} ~, 
\end{equation}
starting from the value at the boundary as 
${(1-{\mu}^2)^{\frac{1}{2}}}_{I=0} = (1-{\mu_{I=0}}^2)^{\frac{1}{2}}$.  
Similarly, the $\phi_{\nu}$-grid points at the cell center are set 
by the Gaussian quadrature with ${\phi_{\nu}}_{i}$ and $d{\phi_{\nu}}_{i}$.  
The grid points at the cell interface are determined by 
\begin{equation}
\label{eqn:phinu}
{\phi_{\nu}}_{J} = {\phi_{\nu}}_{J-1} + {d\phi_{\nu}}_{j} ~, 
\end{equation}
from ${\phi_{\nu}}_{J=0}=0$ to $\phi_{\nu}=\pi$.  
The angle factor at the cell interface used for the  $\phi_{\nu}$-advection 
is given by 
\begin{equation}
\label{eqn:phinu2}
(\sin \phi_{\nu})_{J} = (\sin \phi_{\nu})_{J-1} 
+ {\cos {\phi_{\nu}}_j}~{d\phi_{\nu}}_{j} ~.  
\end{equation}
starting with $(\sin \phi_{\nu})_{J=0}=0$.  
For three dimensional calculations, we need to cover $\phi_{\nu}$ from 0 to 2$\pi$.  
We repeat the above procedure from $\phi_{\nu}=\pi$ to $\phi_{\nu}=2\pi$ 
after dividing the whole angle range into two parts.  
This is to avoid the $\phi_{\nu}$-grid being asymmetric around the radial direction.   

The $\phi$-grid is given in a similar manner 
to $\phi_{\nu}$-grid.  
The $\phi$-grid points for the cell center are denoted by $\phi_{i}$ 
and those for the cell interface are given by 
\begin{equation}
\label{eqn:phi}
{\phi}_{I} = {\phi}_{I-1} + {d\phi}_{i} ~.  
\end{equation}
The points, ${\phi}_{i}$, and weights, ${d\phi}_{i}$, are taken 
from the Gaussian quadrature points.  

Finally, the energy grid points at the cell center are defined by 
\begin{equation}
\label{eqn:eps}
{\varepsilon}_{k} = ({\varepsilon}_{K}{\varepsilon}_{K-1})^{1/2} ~, 
\end{equation}
from the energy grid points at the cell boundary, ${\varepsilon}_{K}$, 
covering the energy range 
from ${\varepsilon}_{K=0}$ to ${\varepsilon}_{K=N_{\varepsilon}}$.  
The energy integration is done by the cell volume for the energy phase space, 
$\varepsilon^2 d\varepsilon = d(\varepsilon^3 /3) $, as 
\begin{equation}
\label{eqn:deps}
d({\varepsilon}_{k}^3 / 3) = 
({\varepsilon}_{K}^3 -{\varepsilon}_{K-1}^3 ) / 3 ~.
\end{equation}

\subsection{Numerical Scheme for Advection}\label{scheme-adv}

We describe a discretized form for the advection terms in our scheme.  
The current approach is an extension of the method by \citet{mez93} 
adopting an interpolation between first- and second-order finite 
difference representation of the neutrino advection.  
We pay attention to the evaluation of quantities at the cell interfaces 
for a wide range of opacities.  
We assume the evaluation is done at the time step $n+1$, 
although we drop here the index.  

First, we explain the advection terms for $r$ and $\mu_{\nu}$ 
which are the two basic variables used under spherical symmetry.  
The radial advection is expressed by 
\begin{equation}
\label{eqn:advection-radial}
\left[ \frac{\mu_{\nu}}{r^{2}} \frac{\partial}{\partial r} (r^{2} f) \right]
=
\left[       \mu_{\nu}         \frac{\partial}{\partial (r^3 / 3)} (r^{2} f) \right]
= {\mu_{\nu}}_{j} ~ \frac{3}{r_{I}^{3} - r_{I-1}^{3}} 
~ ( r_{I}^{2} ~ f_{I} - r_{I-1}^{2} ~ f_{I-1} ), 
\end{equation}
where $f_{I-1}$ and $f_{I}$ are the neutrino distributions 
at the cell interfaces for $f_{i}$.  
The value of ${\mu_{\nu}}_{j} f_{I}$ at the cell boundary is evaluated by 
\begin{equation}
\label{eqn:fnu-radial}
{\mu_{\nu}}_{j} f_{I} = 
  \frac{ {\mu_{\nu}}_{j} - | {\mu_{\nu}}_{j} | }{2} 
\{ ( 1 - \beta_{I} ) f_{i} +       \beta_{I}   f_{i+1} \}
+ \frac{ {\mu_{\nu}}_{j} + | {\mu_{\nu}}_{j} | }{2} 
\{         \beta_{I} f_{i} + ( 1 - \beta_{I} ) f_{i+1} \}, 
\end{equation}
for both inward and outward directions of neutrinos, 
depending on the sign of ${\mu_{\nu}}_{j}$.  
For example, the expression for outward neutrinos 
(${\mu_{\nu}}_{j} \geq 0$) becomes 
\begin{equation}
\label{eqn:fnu-radial2}
{\mu_{\nu}}_{j} f_{I} = {\mu_{\nu}}_{j}
\{         \beta_{I} f_{i} + ( 1 - \beta_{I} ) f_{i+1} \} ~, 
\end{equation}
where $\beta_{I}$ is a weighting factor 
to bridge the upwind and central differencing.  
We note that the expression, $f_{I}=f_{i}$, is obtained 
for the upwind differencing by setting $\beta_{I}=1$, 
whereas the expression, $f_{I}=(f_{i} + f_{i+1}) / 2$, is obtained 
for the central differencing by setting $\beta_{I}=1/2$.  
The value of $\beta_{I}$ is determined by a smooth function 
to connect the diffusion ($\beta_{I}=1/2$) and 
free-streaming ($\beta_{I}=1$) regimes.  
We employ the following expression 
\begin{equation}
\label{eqn:beta-radial}
\beta_{I} = 1 - \frac{1}{2} 
\frac{\alpha \Delta r_{I} / \lambda_{I}}{1 + \alpha \Delta r_{I} / \lambda_{I}} ~, 
\end{equation}
based on the formula by \citet{mez93}.  
Here the interval of radial grid points is defined 
by $\Delta r_{I} = r_{i+1} - r_{i}$ and 
the average of mean free paths is defined by 
$\lambda_{I}^{-1} = ( \lambda_{i+1}^{-1} + \lambda_{i}^{-1} ) / 2$.  
An adjustable factor, $\alpha$, is set to be 100 to ensure 
$\beta_{I}=1/2$ for the diffusion regime even when $\Delta r_{I} \sim \lambda_{I}$.  

The advection term for $\mu_{\nu} = \cos \theta_{\nu}$ is given by 
\begin{equation}
\label{eqn:advection-munu}
\left[ \frac{1}{r} 
  \frac{\partial}{\partial \mu_{\nu}} [~(1-\mu_{\nu}^{2}) f~] \right] 
= \frac{3}{2} ~ \frac{r_{I}^{2} - r_{I-1}^{2}}{r_{I}^{3} - r_{I-1}^{3}}
~ \frac{1}{{d \mu_{\nu}}_{j}} 
~ \left[ (1-{\mu_{\nu}}^2)_{J} f_{J} - (1-{\mu_{\nu}}^2)_{J-1} f_{J-1} \right], 
\end{equation}
where we take the upwind differencing and set simply $f_{J} = f_{j}$.  
We use the angle factor given by Eq. (\ref{eqn:munu2}).  
The relation of Eq. (\ref{eqn:munu2}) must be used to guarantee 
the steady state for infinite homogeneous medium (constant $f_i$) 
in the neutrino transfer under spherical symmetry 
by adopting the advection terms with Eqs. 
(\ref{eqn:advection-radial}) and (\ref{eqn:advection-munu})
as described in \citet{mez93}.  

Next, we describe the advection terms for $\theta$ and $\phi_{\nu}$, 
which are essential for the description under axial symmetry.  
We rewrite the $\theta$-advection term by using the variable $\mu$ 
and discretize the term by 
\begin{eqnarray}
\label{eqn:advection-polar}
\left[ \frac{\sqrt{1-\mu_{\nu}^{2}}~{\rm cos}~\phi_{\nu}}{r {\rm sin}~\theta} 
  \frac{\partial}{\partial \theta} ({\rm sin}~\theta f) \right] 
=
\left[ - \frac{\sqrt{1-\mu_{\nu}^{2}}~{\rm cos}~\phi_{\nu}}{r } 
  \frac{\partial}{\partial \mu} [~(1-\mu^{2})^{\frac{1}{2}} f~] \right] \nonumber \\
= - \frac{3}{2} ~ \frac{r_{I_r}^{2} - r_{I_r-1}^{2}}{r_{I_r}^{3} - r_{I_r-1}^{3}} 
(1-{\mu_{\nu}}_{j_{\theta}}^{2})^{\frac{1}{2}} {\cos \phi_{\nu}}_{j_{\phi}}~ \frac{1}{d \mu_{i_{\theta}}} 
~ \left[ (1-{\mu}^2)^{\frac{1}{2}}_{I_{\theta}}   f_{I_{\theta}} 
       - (1-{\mu}^2)^{\frac{1}{2}}_{I_{\theta}-1} f_{I_{\theta}-1} \right] ~.  
\end{eqnarray}
We use the angle factor given by Eq. (\ref{eqn:mu2}).  
The factor, $(1-{\mu_{\nu}}_{j_{\theta}}^{2})^{\frac{1}{2}} {\cos \phi_{\nu}}_{j_{\phi}}$, 
determines the direction of advection and 
the evaluation of $f_{I_{\theta}}$ at the cell interface.  
Depending on the sign of $\cos \phi_{\nu}$, the value at the interface is given by 
\begin{eqnarray}
\label{eqn:fnu-polar}
{\cos \phi_{\nu}}_{j_{\phi}} f_{I_{\theta}} = 
  \frac{ {\cos \phi_{\nu}}_{j_{\phi}} + | {\cos \phi_{\nu}}_{j_{\phi}} | }{2} 
\{ ( 1 - \beta_{I_{\theta}} ) f_{i_{\theta}} +       \beta_{I_{\theta}}   f_{i_{\theta}+1} \} \nonumber \\
+ \frac{ {\cos \phi_{\nu}}_{j_{\phi}} - | {\cos \phi_{\nu}}_{j_{\phi}} | }{2} 
\{       \beta_{I_{\theta}}   f_{i_{\theta}} + ( 1 - \beta_{I_{\theta}} ) f_{i_{\theta}+1} \} ~.  
\end{eqnarray}
This expression is adopted to connect the diffusion and free-streaming limits 
smoothly by the factor, $\beta_{I_{\theta}}$, in the same way 
as the one used in the radial advection.  
The form of $\beta_{I_{\theta}}$ is given in a similar form as Eq. (\ref{eqn:beta-radial}) 
by replacing the width of grids and the average mean free path 
to the ones in polar direction.  

The advection term for $\phi_{\nu}$ is given by 
\begin{eqnarray}
\label{eqn:advection-phinu}
\left[ - \frac{\sqrt{1-\mu_{\nu}^{2}}}{r} 
  \frac{{\rm cos}~\theta}{{\rm sin}~\theta}
  \frac{\partial}{\partial \phi_{\nu}} ({\rm sin}~\phi_{\nu} f) \right] 
=
\left[ - \frac{\sqrt{1-\mu_{\nu}^{2}}}{r} 
  \frac{\mu}{\sqrt{1-\mu^{2}}}
  \frac{\partial}{\partial \phi_{\nu}} ({\rm sin}~\phi_{\nu} f) \right] \nonumber \\
= - \frac{3}{2} ~ \frac{r_{I_r}^{2} - r_{I_r-1}^{2}}{r_{I_r}^{3} - r_{I_r-1}^{3}} 
(1-{\mu_{\nu}}_{j_{\theta}}^{2})^{\frac{1}{2}} ~ \frac{\mu_{i_{\theta}}}{(1-{\mu_{i_{\theta}}^{2})^{\frac{1}{2}}}}
~ \frac{1}{{d \phi_{\nu}}_{j_{\phi}}} 
\left[ (\sin \phi_{\nu})_{J_{\phi}} f_{J_{\phi}} - (\sin \phi_{\nu})_{J_{\phi}-1} f_{J_{\phi}-1} \right] ~, 
\end{eqnarray}
using the variable, $\mu$.  
The angle factor is given by Eq. (\ref{eqn:phinu2}).  
The sign of $\mu_{i_{\theta}} (\sin \phi_{\nu})_{J_{\phi}}$ determines 
the direction of advection.  
We take the upwind differencing for 
the evaluation of $f_{J_{\phi}}$ at the cell interface through the relation as 
\begin{eqnarray}
\label{eqn:fnu-phinu}
\mu_{i_{\theta}} (\sin \phi_{\nu})_{J_{\phi}} f_{J_{\phi}} = 
\frac{ \mu_{i_{\theta}} (\sin \phi_{\nu})_{J_{\phi}} + | \mu_{i_{\theta}} (\sin \phi_{\nu})_{J_{\phi}} | }{2} 
f_{j_{\phi}+1} \nonumber \\
+
\frac{ \mu_{i_{\theta}} (\sin \phi_{\nu})_{J_{\phi}} - | \mu_{i_{\theta}} (\sin \phi_{\nu})_{J_{\phi}} | }{2} 
f_{j_{\phi}} ~.  
\end{eqnarray}
In the same manner as the spherical case, 
the combination of Eqs. (\ref{eqn:mu2}) and (\ref{eqn:phinu2}) 
guarantees the steady state for infinite homogeneous medium 
in the neutrino transfer under axial symmetry 
when the advection terms of $\theta$ and $\phi_{\nu}$ are expressed 
by Eqs. (\ref{eqn:advection-polar}) and (\ref{eqn:advection-phinu}) 
as described in \citet{cas04}.  

Finally, we describe the advection term for $\phi$, 
which is necessary for three dimensional calculations.  
The advection term is expressed in terms of $\mu$ as 
\begin{eqnarray}
\label{eqn:advection-azimuthal}
\left[ \frac{\sqrt{1-\mu_{\nu}^{2}}~{\rm sin}~\phi_{\nu}}{r {\rm sin}~\theta} 
  \frac{\partial f}{\partial \phi} \right] 
=
\left[ \frac{\sqrt{1-\mu_{\nu}^{2}}~{\rm sin}~\phi_{\nu}}{r \sqrt{1-\mu^{2}}} 
  \frac{\partial f}{\partial \phi} \right] \nonumber \\
= \frac{3}{2} ~ \frac{r_{I_r}^{2} - r_{I_r-1}^{2}}{r_{I_r}^{3} - r_{I_r-1}^{3}} 
(1-{\mu_{\nu}}_{j_{\theta}}^{2})^{\frac{1}{2}} ~ \frac{{\sin \phi_{\nu}}_{j_{\phi}}}{(1-{\mu_{i_{\theta}}^{2})^{\frac{1}{2}}}}
~ \frac{1}{{d \phi}_{i_{\phi}}} \left[ f_{I_{\phi}} - f_{I_{\phi}-1} \right] ~.
\end{eqnarray}
The evaluation of $f_{I_{\phi}}$ is made by 
\begin{eqnarray}
\label{eqn:fnu-azimuthal}
{\sin \phi_{\nu}}_{j_{\phi}} f_{I_{\phi}} = 
\frac{ {\sin \phi_{\nu}}_{j_{\phi}} + | {\sin \phi_{\nu}}_{j_{\phi}} | }{2} 
\{      \beta_{I_{\phi}}  f_{i_{\phi}} + (1 - \beta_{I_{\phi}}) f_{i_{\phi}+1}\} \nonumber \\
+
\frac{ {\sin \phi_{\nu}}_{j_{\phi}} - | {\sin \phi_{\nu}}_{j_{\phi}} | }{2} 
\{ (1 - \beta_{I_{\phi}}) f_{i_{\phi}} +      \beta_{I_{\phi}}  f_{i_{\phi}+1}\} ~, 
\end{eqnarray}
depending on the sign of ${\sin \phi_{\nu}}_{j_{\phi}}$.  
The form of $\beta_{I_{\phi}}$ is given by a smooth function, 
which is similar to $\beta_{I_{r}}$ and $\beta_{I_{\theta}}$.  
This form of $\phi$-advection fulfills the steady state in infinite homogenous matter 
and the advection vanishes when $f_{i_{\phi}}$ is constant.  

\subsection{Moments of Energy and Angle}\label{moments}

We define the moments of energy and angle of the neutrino distributions.  
All evaluations here are done using the quantities in the inertial frame, 
though we dropped superscripts for the compactness.  
First of all, the neutrino density is evaluated by
\begin{equation}
\label{eqn:density}
n_{\nu} = 
\int \frac{d \varepsilon~\varepsilon^2}{(2 \pi)^3} \int d \Omega ~
f(\varepsilon, \Omega).  
\end{equation}
The first moment of energy is defined by 
\begin{equation}
\label{eqn:enu}
\langle \varepsilon \rangle = \frac{E_{\nu}}{n_{\nu}},
\end{equation}
where the energy density of neutrinos is given by 
\begin{equation}
\label{eqn:energy}
E_{\nu} = 
\int \frac{d \varepsilon~\varepsilon^2}{(2 \pi)^3} \int d \Omega ~
\varepsilon
f(\varepsilon, \Omega).  
\end{equation}

The first moment of angle is defined by 
\begin{equation}
\label{eqn:angle}
\langle \mu_{\nu} \rangle = \frac{f^{r}_{\nu}}{n_{\nu}}, 
\end{equation}
where the radial number flux is given by 
\begin{equation}
\label{eqn:flux}
f^{r}_{\nu} = 
\int \frac{d \varepsilon~\varepsilon^2}{(2 \pi)^3} \int d \Omega ~
n_{r}
f(\varepsilon, \Omega).  
\end{equation}
The polar and azimuthal fluxes are obtained by
\begin{equation}
\label{eqn:flux2}
f^{\theta}_{\nu} = 
\int \frac{d \varepsilon~\varepsilon^2}{(2 \pi)^3} \int d \Omega ~
n_{\theta}
f(\varepsilon, \Omega), 
\end{equation}
\begin{equation}
\label{eqn:flux3}
f^{\phi}_{\nu} = 
\int \frac{d \varepsilon~\varepsilon^2}{(2 \pi)^3} \int d \Omega ~
n_{\phi}
f(\varepsilon, \Omega).  
\end{equation}
The radial luminosity of neutrinos is defined by 
\begin{equation}
\label{eqn:lnu}
L_{\nu} = 4 \pi r^2 F_{\nu}, 
\end{equation}
where the energy flux is given by
\begin{equation}
\label{eqn:eflux}
F_{\nu} = 
\int \frac{d \varepsilon~\varepsilon^2}{(2 \pi)^3} \int d \Omega ~
\varepsilon n_{r}
f(\varepsilon, \Omega).  
\end{equation}

The second moment of angle is defined by 
\begin{equation}
\label{eqn:angle2}
\langle \mu_{\nu}^2 \rangle = \frac{1}{n_{\nu}}
\int \frac{d \varepsilon~\varepsilon^2}{(2 \pi)^3} \int d \Omega ~
n_{r}n_{r}
f(\varepsilon, \Omega).  
\end{equation}
We evaluate the elements of the Eddington tensor 
as defined by
\begin{equation}
\label{eqn:Eddington}
k^{ij} = \frac{P_{\nu}^{ij}}{E_{\nu}}, 
\end{equation}
where the elements of the pressure tensor are defined by 
\begin{equation}
\label{eqn:pressure}
P_{\nu}^{ij} = 
\int \frac{d \varepsilon~\varepsilon^2}{(2 \pi)^3} \int d \Omega ~
\varepsilon
n_{i}
n_{j}
f(\varepsilon, \Omega).  
\end{equation}
The subscripts, $i$ and $j$, denote one of 
the three components of $r$, $\theta$ and $\phi$.  
The diagonal elements in the spherical coordinate are 
$k^{rr}$, $k^{\theta\theta}$ and $k^{\phi\phi}$.  
The non-diagonal elements, 
$k^{r\theta}$, $k^{r\phi}$ and $k^{\theta\phi}$, 
may be non-zero in 3D calculations, 
while they are zero under spherical symmetry.  
For the treatment with multi-energy group, 
we utilize 
the integrand of Eqs. (\ref{eqn:energy}) and (\ref{eqn:pressure}) 
with the angle average.  
We evaluate the Eddington tensor for each energy zone by 
\begin{equation}
\label{eqn:Eddington-multienergy}
k^{ij}(\varepsilon_{k}) = 
\frac{\int d \Omega ~ n_{i} n_{j} f(\varepsilon_{k}, \Omega)}
     {\int d \Omega ~             f(\varepsilon_{k}, \Omega)} ~, 
\end{equation}
dropping off the common factor of the energy phase space.  

\subsection{Computing Size}\label{CompSize}
We describe briefly the size of memory and computational load 
necessary for the current simulations as well as larger simulations 
in future.  
The typical size of memory requirement for 
the numerical calculations in \S \ref{2Dconfig} and \S \ref{3Dconfig} 
is $\sim$30GB in the case of $N_r=200$, $N_{\theta}=9$ and $N_{\phi}=9$ 
with $N_{\varepsilon}=14$, $N_{\theta_{\nu}}=6$ and $N_{\phi_{\nu}}=12$.  
This includes the matrices for the equations and the vectors 
for the neutrino distributions for three species.  
We need $\sim$130 MB to store the neutrino distribution for each species.  
It takes $\sim$100 sec to proceed one time step on 1 node (32cpu) of 
Hitachi SR16000.  
Our largest calculations in the current report is the case of 
the diffusion of the 3D Gaussian packet in \S \ref{diffusion}.  
It costs the memory of $\sim$900 GB on NEC SX9 (8cpu) 
for the case of a fine mesh.  
Note that this specification of the computational speed is obtained 
within the basic optimization and the automatic parallelization.  
A parallel version of the numerical code for massive parallel architectures 
is ready for tests.  
In future, we would need the full coverage of the sphere 
with a high resolution by 
$N_r=400$, $N_{\theta}=64$ and $N_{\phi}=128$, for example, for spatial grid.  
The memory requirement would be $\sim$6 TB for the program 
and $\sim$26GB for the neutrino distribution, 
which are available on the recent supercomputers.  
We plan to perform such large scale simulations 
after we optimize the numerical code with the parallelization.  








\clearpage


\clearpage

\clearpage

\clearpage

\clearpage

\clearpage

\clearpage

\clearpage

\clearpage

\clearpage

\clearpage

\clearpage

\clearpage

\clearpage

\clearpage

\clearpage

\clearpage

\clearpage

\clearpage

\clearpage

\clearpage

\clearpage

\clearpage

\clearpage

\clearpage

\clearpage

\clearpage

\clearpage

\clearpage

\clearpage

\clearpage

\clearpage

\clearpage

\clearpage

\clearpage

\clearpage

\clearpage

\clearpage

%

\end{document}